\documentclass[prl,twocolumn,floatfix]{revtex4}
\usepackage{ifpdf}
\usepackage{graphicx}
\usepackage{fancyhdr}
\usepackage{extramarks}
\usepackage{wasysym}
\usepackage{pstricks}
\usepackage{bbm}
\usepackage{epsfig}
\usepackage{natbib}
\usepackage{chapterbib}
\usepackage{psfrag}
\usepackage[percent]{overpic}

\input  babarsym

\newcommand{\EE}[1] {\ensuremath{\times 10^{#1}}\xspace}

\def\ze#1   {\ensuremath{\zeta_{#1}}\xspace}

\def\pepii    {\pep2 \xspace}

\def\superb   {Super$B$\xspace}

\def\Bz   {\ensuremath{B_{z}}\xspace}

\def\CO2  {$\mathrm{CO}_2$\xspace}

\long\def\inst#1{\par\nobreak\kern 4pt\nobreak
    {\it #1}\par\vskip 10pt plus 3pt minus 3pt}

\def\NS {\ensuremath{\Upsilon{(nS)}}\xspace}
\def\qq{\mathbbmss{q}}

\renewcommand{\headrulewidth}{0pt}
\renewcommand{\footrulewidth}{0pt}

\newcommand{\beq}{\begin{equation}}
\newcommand{\eeq}{\end{equation}}
\def\be{\begin{equation}}
\def\bea{\begin{eqnarray}}
\def\eea{\end{eqnarray}}
\def\nnb{\nonumber}
\def\btolnu     {\ensuremath{B^\pm\to \ell^\pm \nu}}
\def\btoenu     {\ensuremath{B^\pm\to e^\pm \nu}}
\def\btomunu     {\ensuremath{B^\pm\to \mu^\pm \nu}}
\def\btotaunu     {\ensuremath{B^\pm\to \tau^\pm \nu}}

\def\eg{\ensuremath {e.g.}}
\def\ie{\ensuremath {i.e.}}

\makeatletter
\DeclareRobustCommand\bfseries{%
  \not@math@alphabet\bfseries\mathbf
  \fontseries\bfdefault\selectfont\boldmath}

\makeatother

\renewcommand\thesection{\arabic{section}}

\makeatletter
\@latex@info{//// style change: modifying section title style}
\def\@hangfrom@section#1#2{#1#2}%
\def\section{%
  \@startsection%
    {section}%
    {1}%
    {\z@}%
    {0.8cm \@plus1ex \@minus .2ex}%
    {0.5cm}%
    {%
      \normalfont\small\larger\larger\bfseries
      \centering
    }%
}%
\makeatother

\def\tauhhp{\ensuremath{\tau_{hh}^+}}
\def\tauhhm{\ensuremath{\tau_{hh}^-}}
\def\tauhh{\ensuremath{\tau_{hh}}}
\def\tauKpi{\ensuremath{\tau_{K\pi}}}
\def\yCP{\ensuremath{y_{C\!P}}}
\def\deltaY{\ensuremath{\Delta Y}}

\def\DZ{\ensuremath{{D}^{\rm{0}}}}
\def\DZB{\ensuremath{\overline{{D}}^{\rm{0}}}}

\begin{document}

\pagestyle{empty}

\onecolumngrid
\begin{center}
{\boldmath{
\phantom{\Huge{I}}
\phantom{\Huge{I}}
\phantom{\Huge{I}}
\phantom{\Huge{I}}
\phantom{\Huge{I}}

{\huge\bf
{Proceedings\\
of \\
\vskip 12pt
Super$B$ Workshop VI}}\\
 \ \phantom{\Huge{I}}\\%
{\huge\bf{
New Physics\\
at the\\
\vskip 12pt
Super Flavor Factory}}\\
\phantom{\Huge{I}}
{\Large Valencia, Spain}\\
\vskip 8pt
{\Large  January 7-15, 2008}\\
\phantom{\Huge{I}}
\phantom{\Huge{I}}
}}
\end{center}
\bigskip
\begin{center}
\large \bf Abstract
\end{center}
The sixth \superb\ Workshop was convened in response to questions posed by the INFN Review Committee,
evaluating the \superb\ project at the request of INFN. The working groups addressed the capability of a high-luminosity flavor factory that can gather a data sample of 50 to 75 ab$^{-1}$ in five years to elucidate New Physics phenomena unearthed at the LHC. This report summarizes the results of the Workshop.

\vfill\clearpage

{
%
\author{{\phantom{            }}}
\affiliation{{\phantom{       }\\ }}
\author{D.G.~Hitlin}
\author{C.H.~Cheng}
\affiliation{California Institute of Technology, Pasadena, California 91125, USA}

\author{D.M.~Asner}
\affiliation{Carleton University, Ottawa, Ontario, Canada K1S 5B6}

\author{T.~Hurth}
\author{B.~McElrath}
\affiliation{CERN, CH-1211 Geneve 23, Switzerland}

\author{T.~Shindou}
\affiliation{DESY, D-22603 Hamburg, Germany}

\author{F.~Ronga}
\affiliation{ETH, CH-8093 Zurich, Switzerland}

\author{M.~Rama}
\affiliation{INFN, Laboratori Nazionali di Frascati, Frascati, Italy}

\author{S.~Tosi}
\affiliation{Universit\`a di Genova, Dipartimento di Fisica and INFN, I-16146 Genova, Italy}

\author{G.~Simi}
\affiliation{University of Maryland, College Park, Maryland 20742, USA}

\author{S.~Robertson}
\affiliation{McGill University, Montr\`eal, Qu\'ebec, Canada H3A 2T8}

\author{P.~Paradisi}
\affiliation{Technische Universit\"at Munchen,
D-85748 Garching, Germany}

\author{I.~Bigi}
\affiliation{University of Notre Dame, Notre Dame, Indiana 46556, USA}

\author{A.~Stocchi}
\author{B.~Viaud}
\affiliation{Laboratoire de lÕAcc\'el\`erateur Lin\`eaire, IN2P3/CNRS et Universit\'e de Paris-Sud XI,
Centre Scientifique dÕOrsay, F-91898 Orsay Cedex, France}

\author{F.~Domingo}
\author{E.~Kou}
\affiliation{Laboratoire de Physique Theorique,
Universit\'e de Paris-Sud XI,
F-91405 Orsay Cedex, France}

\author{M.~Morandin}
\affiliation{Universit\`a di Padova, Dipartimento di Fisica and INFN, I-35131 Padova, Italy}

\author{G.~Batignani}
\author{A.~Cervelli}
\author{F.~Forti}
\author{N.~Neri}
\author{J.~Walsh}
\affiliation{Universit\`a di Pisa, Dipartimento di Fisica, Scuola
Normale Superiore and INFN, I-56127 Pisa, Italy}

\author{M.~Giorgi}
\author{G.~Isidori}
\author{A.~Lusiani}
\affiliation{Universit\`a di Pisa, Dipartimento di Fisica, Scuola
Normale Superiore and INFN, I-56127 Pisa, Italy\\ and INFN, Pisa,
 Italy}

\maketitle
}

\onecolumngrid\twocolumngrid\hbox{}

{
\author{A.~Bevan}
\affiliation{Queen Mary, University of London, E1 4NS,~United Kingdom}

\author{R.~Faccini}
\author{F.~Renga}
\affiliation{Universit\`a di Roma "La Sapienza" and INFN Roma, I-00185 Roma, Italy}

\author{A.~Polosa}
\author{L.~Silvestrini}
\author{J.~Virto}
\affiliation{INFN Roma, I-00185 Roma, Italy}

\author{M.~Ciuchini}
\affiliation{INFN Roma Tre, I-00146 Roma, Italy}

\author{S.~Heinemeyer}
\affiliation{Universitat de Cantabria, E-39005 Santander, Spain}

\author{M.~Carpinelli}
\affiliation{Universit\`a di Sassari and INFN, Sassari, Sardinia, Italy}

\author{P.~Gambino}
\affiliation{ Univerist\`a di Torino and INFN, I-1015 Torino, Italy}

\author{V.~Azzolini}
\author{J.~Bernab\'eu}
\author{F.~Botella}
\author{G.~C.~Branco}
\author{N.~Lopez~March}
\author{F.~Martinez~Vidal}
\author{A.~ Oyanguren}
\author{A.~Pich}
\author{M.~A.~Sanchis~Lozano}
\author{J.~Vidal}
\author{O.~Vives}
\affiliation{IFIC, Universitat de Valencia-CSIC, E-46071 Valencia, Spain}

\author{S.~Banerjee}
\author{J.~M.~Roney}
\affiliation{University of Victoria, Victoria, British Columbia, Canada V8W 3P6}

\author{T.~Gershon}
\affiliation{University of Warwick, Coventry, CV4 7AL, United Kingdom}

\maketitle
}

\clearpage
\onecolumngrid\twocolumngrid\hbox{}

\clearpage

\pagestyle{fancyplain}

\fancyfoot{} 
\fancyfoot[LE,RO]{\it{Proceedings of SuperB Workshop VI,  Valencia, Jan 7-15, 2008}}
\fancyhead{} 
\renewcommand{\sectionmark}[1]%
                  {\markright{#1}}
\rhead[\fancyplain{}{\bf Introduction}]%
      {\fancyplain{}{\bf\thepage}}

\clearpage
{\centerline{\rule[0in]{0.9\columnwidth}{2pt}}
\vspace {-0.7cm}
\part*{\centerline{Introduction}}
\vskip -5pt
{\centerline{\rule[ 0.15in]{0.9\columnwidth}{2pt}}

\bigskip
\smallskip



The Sixth \superb\ Workshop, held at the IFIC in Valencia, Spain from January 7-15, 2008,  was convened to update our understanding of the physics capabilities of the \superb\ project, proposed for construction on the campus of Rome University Tor Vergata. In particular, the Workshop addressed several questions posed by members of the International Review Committee appointed by INFN to review the project. 
The workshop was organized into several working groups; this document comprises the reports from these groups. It is not intended as a comprehensive review of the physics capability of \superb; rather, it should be read as a supplement to the physics section of the \superb\ Conceptual Design Report (CDR)\cite{Bona:2007qt}.

The motivation for undertaking a new generation of \epem experiments is, of course, to measure effects of New Physics on the decays of heavy quarks and leptons. A detailed picture of the observed pattern of such effects will be crucial to gaining an understanding of any New Physics found at the LHC. As detailed herein, much of the study of the capability of the LHC to distinguish between, for example, models of supersymmetry breaking have emphasized information accessible at high $p_{\rm T}$.  Many of the existing constraints on models of New Physics, however, come from flavor physics. Improving limits and teasing out new effects in the flavor sector will be just as important in constraining models after New Physics has been found as it has been in the construction of viable candidate models in the years before LHC operation.

In confronting New Physics effects on the weak decays of $b$, $c$ quarks and $\tau$ leptons it is crucial to have the appropriate experimental sensitivity. The experiment must measure \CP asymmetries in very rare decays, rare branching fractions and interesting kinematic distributions to sufficient precision to make manifest the expected effects of New Physics, or to place constraining limits.  There is a strong consensus in the community that doing so requires a data sample corresponding to an integrated luminosity of 50 to 100 \invab.  There is also a consensus that a reasonable benchmark for obtaining such a data sample is of the order of five years of running. Meeting both these constraints requires a collider luminosity of \hbox{$10^{36}$ cm$^{-2}$s$^{-1}$} or more, yielding 15 \invab/Snowmass Year of 1.5$\times 10^7$ seconds. It is these boundary conditions that set the luminosity of \superb.

Reaching this luminosity with a collider design extrapolated from \pepii or KEKB, such as SuperKEKB, is difficult; beam currents and thus power consumption are very high, and the resulting detector backgrounds are formidable. The low emittance, crabbed waist design of \superb\ provides an elegant solution to the problem; \superb\ can reach unprecedented luminosity with beam currents and power consumption comparable to those at \pepii.  A test of the crabbed waist concept is underway at Frascati; it is proceeding very well, producing impressive increases in the specific luminosity at DA$\Phi$NE. More remains to be done, but the results are very encouraging.

It is important that results with sensitivity to New Physics be obtained in a timely way, engendering a ``conversation'' with the LHC experiments. \superb\ can confidently be expected to produce a very large data sample before the end of the next decade. The more gradual SuperKEKB approach to achieving high peak luminosity cannot produce comparable data samples until close to the end of the following decade \cite{Ohnishi_Atami}.


 $\tau$ physics will likely assume great importance as a probe of physics beyond the Standard Model. \superb\ includes in the baseline design an 85\% longitudinally polarized electron beam and spin rotators to facilitate the production of polarized $\tau$ pairs. This polarization is the key to the study of the structure of lepton-flavor-violating couplings in $\tau$ decay, as well as the search for a $\tau$ EDM, or for $\CP$ violation in $\tau$ decay. SuperKEKB does not incorporate a polarized beam.

The recent observation of large \DzDzb mixing raises the exciting possibility of finding $\CP$ violation in charm decay, which would almost certainly indicate physics beyond the Standard Model. \superb can attack this problem in a comprehensive manner, with high luminosity data sample in the \FourS region and at the $\psi(3770)$ resonance, as the collider is designed to run at lower center-of-mass energies, at reduced luminosity. With very short duration low energy runs, a data sample an order of magnitude greater than that of the final BES-III sample can readily be obtained. SuperKEKB cannot run at low energies.


The following is a brief resum\'e of the capabilities of \superb. In some instances, comparisons are made between physics results that can be obtained with the five year, \hbox{75 \invab} \superb\ sample and a \hbox{10 \invab} sample such as could perhaps be obtained in the first five years of running of SuperKEKB. More detailed discussions will be found in the ensuing sections.
\vskip-18pt

\subsection*{$B$ Physics}
\label{sec:B:UKEKB}
\vskip-8pt



$B$ physics remains a primary objective of \superb. With \babar\ and Belle having clearly established the ability of the CKM phase to account for \CP-violating asymmetries in tree-level $b \rightarrow c {\bar c} s$ decays, the focus shifts to the study of very rare processes. With a SUSY mass scale below 1 TeV, New Physics effects in \CP-violating asymmetries, in branching fractions and kinematic distributions of penguin-dominated decays and in leptonic decays can indeed be seen in the five-year \superb\ data sample. 

\begin{table}[!htb]

\caption{\label{tab:ukekb-comp}
    Comparison of current experimental sensitivities with a 10 \invab sample
    and the five year \superb\  75 \invab sample.
    Only a small selection of observables are shown.
    Quoted sensitivities are relative uncertainties if given as a percentage,
    and absolute uncertainties otherwise. An ``X'' means that the quantity is not measured at this integrated
    luminosity.
    For more details, see text and Refs.~\cite{Bona:2007qt,Browder:2007gg,Browder:2008em}.
    }
  \begin{tabular}{lccc}
    \hline\hline
    Mode & \multicolumn{3}{c}{Sensitivity} \\
    & Current & $10 \ {\rm ab}^{-1}$ &$75 \ {\rm
      ab}^{-1}$  \\
    \hline
    ${\cal B}(B \to X_s \gamma)$ & 7\% & 5\% & 3\% \\
    $A_{\CP}(B \to X_s \gamma)$ & 0.037 & 0.01 & 0.004--0.005 \\
    ${\cal B}(B^+ \to \tau^+ \nu)$ & 30\% & 10\% & 3--4\% \\
    ${\cal B}(B^+ \to \mu^+ \nu)$ & X & 20\% & 5--6\% \\
    ${\cal B}(B \to X_s l^+l^-)$ & 23\% & 15\% & 4--6\% \\
    $A_{{\rm FB}}(B \to X_s l^+l^-)_{s_0}$ & X & 30\% & 4--6\% \\
    ${\cal B}(B \to K \nu \overline{\nu})$ & X & X & 16--20\% \\
    $S(\KS \pi^0 \gamma)$ & 0.24 & 0.08 & 0.02--0.03 \\
    \hline\hline
  \end{tabular}
\end{table}


Table~\ref{tab:ukekb-comp} shows a quantitative comparison of the two
samples for
some of the important observables that will be
measured at \superb, including all the so-called ``golden processes'' of
Table~\ref{tab:golden} (see the following section).
We list below some additional comments on the entries of
Table~\ref{tab:ukekb-comp}

\begin{itemize}
\item
  The measurements of ${\cal B}(B \to X_s \gamma)$ and \hbox{${\cal B}(B^+ \to \ell^+ \nu)$ }
  are particularly important in minimal flavor violation scenarios.
  It is crucial to be able to search for small deviations from the Standard Model value.
  Therefore the improvement is sensitivity provided by \superb is highly
  significant (see Figure~\ref{fig:btaunu}).
\item
A 10 \invab sample is not sufficiently large to take advantage of
  the theoretical cleanliness of several inclusive observables, such as
  the zero-crossing of the forward-backward asymmetry in $b \to s\ell^+\ell^-$.
  Results with 10 \invab would not match the precision from the
  exclusive mode $B \to K^* \mu^+\mu^-$, which will be measured by LHC$b$.
  Furthermore, these exclusive channel measurements will be limited by hadronic uncertainties.
  \superb\  can provide a much more precise and theoretically clean measurement using inclusive modes.
\item
  Several interesting rare decay modes, such as \hbox{$B \to K \nu \bar{\nu}$,}
  cannot be observed with the statistics of 10 \invab,
  unless dramatic and unexpected New Physics enhancements are present.
  Preliminary studies are underway on several other channels in this category, such as $B \to \gamma\gamma$ and $B \to$ invisible
  decays which are sensitive to New Physics models with extra-dimensions.
\item
  Another area for comparison is the phenomenological analysis
  within the MSSM with generic mass insertion discussed in the \superb\ CDR.
  Fig. ~\ref{fig:d13} shows how well the $(\delta_{13})_{LL}$ can be
  reconstructed at \superb\ and with 10 \invab. Improvements in lattice QCD
  performance, as discussed in the Appendix of the CDR, are assumed in both
  cases. The remarkable difference in sensitivity stems mainly from the different performance
  in measuring the CKM parameters $\bar\rho$ and $\bar\eta$.

\end{itemize}
\vspace{-5mm}
\begin{figure}[!htb]
  \begin{center}
  \hspace{+0.1cm}\includegraphics[width=0.35\textwidth]{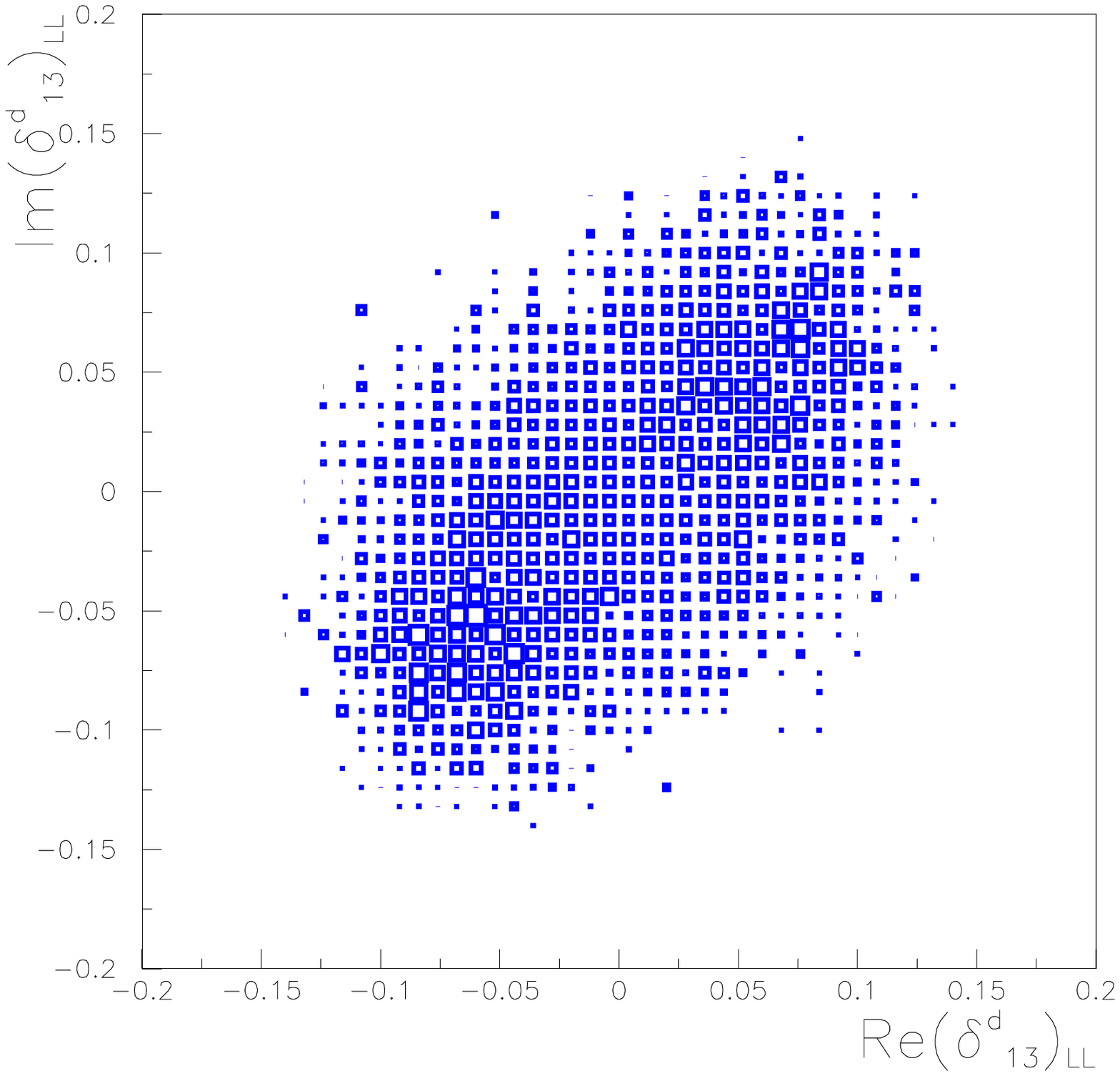}
    \hspace{-0.2cm}\includegraphics[width=0.35\textwidth]
{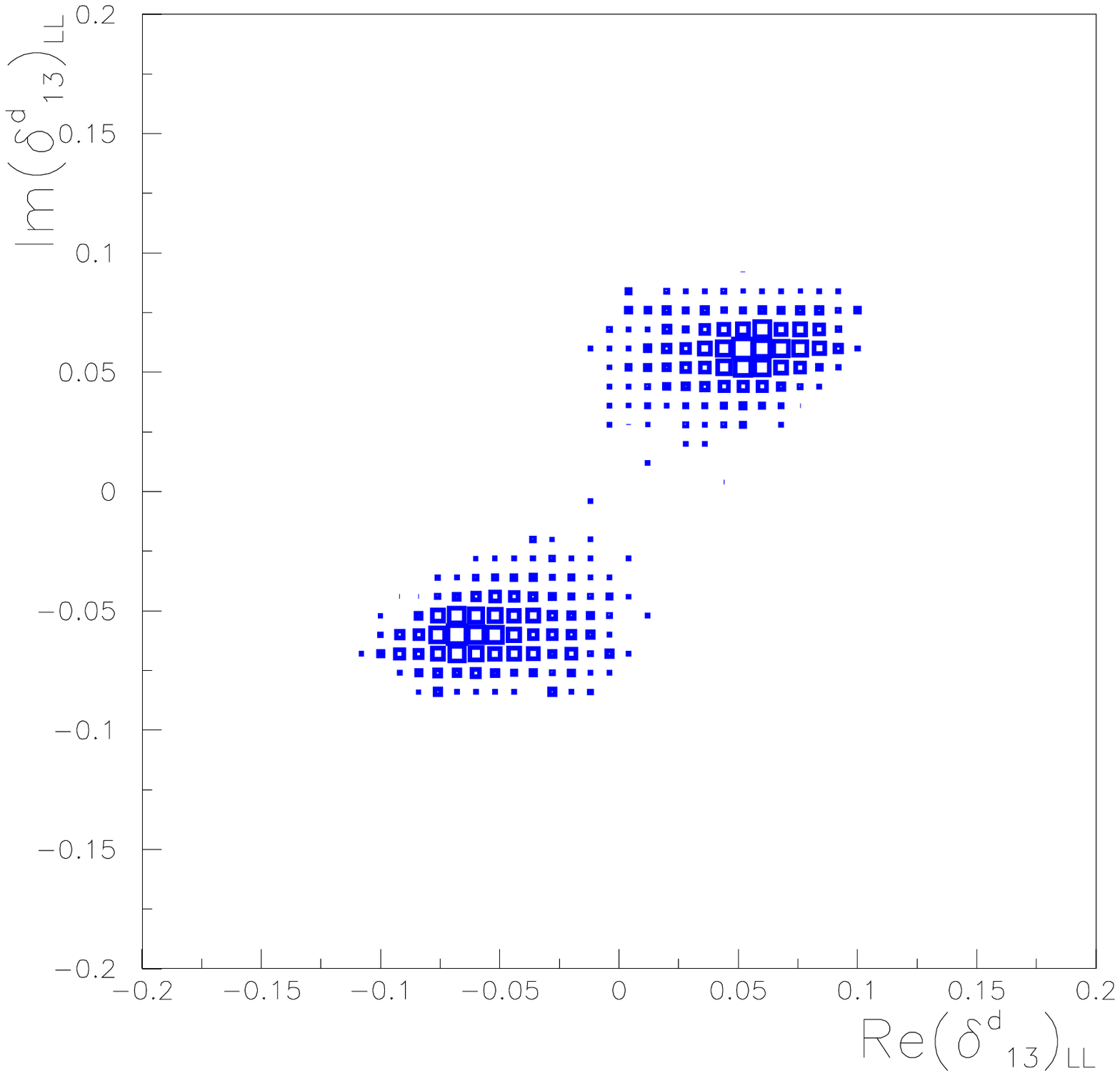}
    \caption{\label{fig:d13}
     Determination of the SUSY mass-insertion parameter $(\delta_{13})_{LL}$
     with a 10 \invab sample (top) and with \superb\  (bottom).
    }
  \end{center}
\end{figure}

\vspace{-10mm}

\bigskip\bigskip
\subsection*{Charm Physics}
\vskip-8pt
The influence of New Physics on the charm sector is often
overlooked. Constraints on flavor-changing neutral currents from new physics
in the up quark sector are much weaker than in the down quark sector. Thus high sensitivity studies of rare charm decays offer the possibility of isolating New Physics effects in $\DzDzb$ mixing, in \CP violation and in rare decay branching fractions.

The recent observation of substantial $\DzDzb$ mixing raises the very exciting possibility of measuring \CP violation in charm decays. Many of the most sensitive measurements remain statistics limited even with \superb\-size data samples, providing a substantial motivation for gathering 75 \invab.

In several specific cases, \CP violation in mixing can be studied more precisely by taking advantage of the clean environment provided by exclusive $\DzDzb$ production at the $\psi(3770)$ resonance.  We have therefore included in the \superb\ design the unique capability of running at this center-of mass-energy. Long data-taking runs are not required; a run of two months duration at the $\psi(3770)$ would yield a data sample an order of magnitude larger than the total BES-III sample at that energy.

\subsection*{Tau Physics}
\vskip-8pt
It is not unlikely that the most exciting results on New Physics in the flavor sector at \superb\ will be found in $\tau$ decays. With 75 ab$^{-1}$ \superb\ can cover a significant portion of the parameter space of most New Physics scenarios predictions for lepton flavor violation (LFV) in tau decays.

The sensitivity in radiative processes such as ${\cal B}(\tau\rightarrow\mu\gamma)$ ($2\times 10^{-9}$) and in ${\cal
B}(\tau\rightarrow\mu\mu\mu)$ decays ($2\times 10^{-10}$) gives \superb\ a real chance to observe these LFV decays. These
measurements are complementary to searches for $\mu\rightarrow e\gamma$ decay. In fact, the ratio
$\BR(\tau\rightarrow\mu\gamma)/\BR(\mu\rightarrow e\gamma)$ is an important diagnostic of SUSY-breaking scenarios.
If LFV decays such us $\tau\rightarrow\mu\gamma$ and $\tau\rightarrow\mu\mu\mu$ are found, the polarized electron beam of \superb provides us with a means of determining the helicity structure of the LFV coupling, a most exciting prospect. The polarized beam also provides a novel additional handle on backgrounds to these rare processes.

The longitudinally polarized high energy ring electron beam, which is a unique feature of \superb, is also the key to searching for \CP violation in tau production or decay. An asymmetry in production would signal a $\tau$ EDM, with a sensitivity of $\sim 10^{-19}\ e$cm, while an unexpected \CP-violating asymmetry in decay would be a clear signature of New Physics.

The polarized beam and the ability to procure a data sample of sufficient size to find lepton flavor-violating events, as opposed to setting limits on LFV processes are unique to \superb.

\vspace{-5mm}

\bigskip\bigskip
\subsection*{Spectroscopy}
\vskip -6pt
One of the most surprising results of the past decade has been the plethora of new states with no ready quark model explanation by the $B$ Factories and the Tevatron. These states clearly indicate the existence of exotic combinations of quarks and gluons into hybrids, molecules or tetraquarks.

These studies, which promise to greatly enhance our understanding of the non-perturbative regime of QCD, are at an early stage. Many new states have been found. These may be combinations involving light quarks or charmed quarks, but only in the case of the $X$(3872) have there been observations of more than a single decay channel. It is crucial to increase the available statistics by of the order of one hundred-fold in order to facilitate searches for additional decay modes. In the case of the $X$(3872) state, for example, it is particularly critical to observe both decays to charmonium and to $D$ or $D_s^+$ pairs, the latter having very small branching fractions. It is also important to provide enhanced sensitivity to search for additional states, such as the neutral partners of the $Z$(4430).

Bottomonium studies are quite challenging, since the expected but not yet observed states are often broad and have many decay channels, thus requiring a large data sample. Leptonic decays of bottomonium states also provide, through lepton universality tests, a unique window on New Physics.

Data samples adequate for these studies, which in some cases require dedicated runs of relatively short duration, in both the 4 and 10 GeV regions, are obtainable only at \superb.

\onecolumngrid\twocolumngrid\hbox{}



\onecolumngrid\twocolumngrid\hbox{}

\onecolumngrid\twocolumngrid\hbox{}

\setcounter{section}{0}
\rhead[\fancyplain{}{\bf $B$ Physics}]%
      {\fancyplain{}{\bf\thepage}}
\clearpage

{\centerline{\rule[0in]{0.9\columnwidth}{2pt}}
\vspace {-0.7cm}
\part*{\centerline{$B$ Physics}}
\vskip -8pt
{\centerline{\rule[ 0.15in]{0.9\columnwidth}{2pt}}



\def\version{0.0.1}

\def\mySpecialText{DRAFT version \version \ \today}
\def\myspecial#1{\special{#1}}       

\def\be{\begin{equation}}
\def\ee{\end{equation}}
\def\bea{\begin{eqnarray}}
\def\eea{\end{eqnarray}}
\def\nnb{\nonumber}
\def\dps{\displaystyle}
\def\lk{\left(}
\def\rk{\right)}
\def\lek{\left[}
\def\rek{\right]}
\def\bbuildrel#1_#2^#3{\mathrel{\mathop{\kern 0pt#1}\limits_{#2}^{#3}}}
\def\slash#1{\setbox0=\hbox{$#1$}#1\hskip-\wd0\dimen0=5pt\advance
       \dimen0 by-\ht0\advance\dimen0 by\dp0\lower0.5\dimen0\hbox
         to\wd0{\hss\sl/\/\hss}}
\def\gev{{\rm GeV}}
\def\mev{{\rm MeV}}

\def\todo       {{\Large \bf TO DO}}
\def\btolnu     {\ensuremath{B^\pm\to \ell^\pm \nu}}
\def\btoenu     {\ensuremath{B^\pm\to e^\pm \nu}}
\def\btomunu     {\ensuremath{B^\pm\to \mu^\pm \nu}}
\def\btotaunu     {\ensuremath{B^\pm\to \tau^\pm \nu}}
\label{sec:B:intro}

\bigskip
\smallskip

%

The physics case for \superb\ has been
discussed in some detail in the \superb\ Conceptual Design Report (henceforth CDR)~\cite{Bona:2007qt}.
In the CDR, and in the following, we consider the discovery potential of \superb\
in two scenarios: whether or not the LHC finds evidence for New Physics. \\

\noindent
{\it LHC discovers new particles} \\
\par
\vskip - 9pt
\noindent
  If the LHC finds physics beyond the Standard Model, the essential, and unique, role of \superb\ will be to determine the flavor structure of the New Physics.  In that sense, measurements from \superb\ that
  are consistent with the Standard Model are as valuable as those that show
  significant deviations -- in either case these measurements provide information
  about the New Physics flavor structure that cannot be provided by other experiments.
  In this context, the measurement of  theoretically  clean rare  decays,
  even when found to be Standard Model-like, will yield valuable
  insights into the structure of New Physics models, providing
 information complementary to LHC results.

  It is, of course, generally regarded as more valuable to find deviations from Standard Model predictions than to
  find a result that agrees with the Standard Model.
  In fact,
  many New Physics flavor structures do produce measurable
  effects. As shown in the discussion on benchmark points below, there are also
  scenarios in which flavor effects can be very small, and perhaps barely visible, even
  with \superb. The great precision reached at \superb\ can still provide positive
  information on the underlying theory, even in a Standard Model-like flavor scenario.
  Indeed, we emphasize that measurement of the New Physics flavor couplings are the primary discovery goal
  of \superb;  results from both LHC and \superb\ are required to reconstruct the New Physics Lagrangian.\\

\noindent
{\it There is no New Physics discovery at LHC} \\
\par
\vskip -9pt
\noindent
  If evidence for New Physics does not readily appear at LHC, the goal of \superb\ would
  then be to emphasize measurement precision to search for deviations in
  flavor observables.  In this scenario, finding such small effects could provide the first
  evidence of New Physics!  The absence of knowledge about the New Physics scale from LHC would
  make it impossible to reconstruct the New Physics Lagrangian, but a New Physics discovery at
  \superb\ would provide a solid indication that the New Physics scale is only slightly
  above the reach of LHC. \\

The chapter is organized as follows.
We first present a description of work done since
the writing of the CDR~\cite{Bona:2007qt}, concentrating on some particularly
interesting channels that were only partially covered or not covered at all.
We then update the phenomenological studies presented in the CDR,
including a classification of golden modes, performance at LHC benchmark points,
the impact of \superb\ on explicit models of SUSY breaking, and a brief discussion on
the interplay of flavor and high $p_{T}$ physics.
We concentrate on $B$ physics at the \FourS,
since we have little to add to previous studies of
the potential for $B_s$ physics at the \FiveS~\cite{Baracchini:2007ei}.

\vspace {-3mm}

\section{Studies of selected $B$ decay channels}
\label{sec:B:channels}

In this section we present new studies on a selected set of
$B$ meson decay channels,
updating the determination of the following processes: \\
\par\noindent
{{\it The CKM matrix element $|V_{ub}|$.} This measurement, crucial to the
model-independent determination of the CKM matrix,
can only be done at an \epem  machine. We update the calculation of the \superb\ reach, as
suggested by the International Review Committee.\\
\par\noindent
{\it The rare branching fractions  \hbox{${\cal B}(B \to X_s \gamma)$,}} {\hbox{\it ${\cal B}(B \to X_s \ell^+ \ell^-)$.}}
These channels were not thoroughly studied in the CDR, as they are limited by experimental
and theoretical systematic uncertainties. In the CDR we concentrated on
other observables, such as the photon polarization and $\CP$ and isospin asymmetries.
However ${\cal B}(B \to X_s \gamma)$, at present  one of the most powerful New Physics probes,
remains a powerful constraint, even in the Minimal Flavor Violation case. We have therefore reassessed
the experimental and theoretical sensitivities for these modes at \superb. We have also
done a preliminary sensitivity study for $B \to K^{(*)} \tau^+ \tau^-$.\\
\par\noindent
{{\it  The branching fraction ${\cal B}(B \to X_s \nu \overline{\nu})$.}} A new detailed
study has been performed on this mode, evaluating the possibility of measuring the
branching fraction with the full \superb\  data sample. This information complements
the measurements of $B \to X_s \gamma$ and $B \to X_s \ell^+ \ell^-$ in accessing
New Physics that can contribute to $\Delta B=1$ box, photon penguin, and $Z^0$ penguin diagrams.\\
\par\noindent
{{\it  Leptonic decay modes.}} The precise measurement of ${\cal B}(B \to \ell \nu$)
is particularly interesting in New Physics scenarios with a charged
Higgs at high tan$\beta$. Following the suggestion of the IRC, we discuss
possible improvements in signal efficiency and systematic uncertainties at \superb.
We also present a new study of radiative leptonic decays and some discuss considerations
relevant to LFV modes.\\

\noindent
\subsection*{Precise determination of the CKM element $|V_{ub}|$}
\label{sec:B:channels:Vub}

The precise measurement of $|V_{ub}|$ is a crucial ingredient in
the determination of the CKM parameters $\bar{\rho}$ and $\bar{\eta}$
in the presence of New Physics. At the time \superb\ commences operation,
LHCb will have already provided precise measurements of $\sin2\beta$ and
$\gamma$. This will allow for an improved determination of CKM parameters within the
Standard Model.  However, in the presence of generic New Physics contributions,
this information alone is not sufficient to obtain the same precision. As precise information
on CKM parameters is essential for any New Physics flavor analysis
in the $K$ and $B$ sectors, an improved determination of $|V_{ub}|$ turns out to be quite important
in New Physics searches.

%

The precise study of both inclusive and exclusive $B$
semileptonic branching fractions is a unique feature of \superb. \\
\par\noindent
{{\it Inclusive decays}}\\
\vskip -9pt
\noindent
The current $5$--$10\%$ theoretical error on the inclusive determination
of $|V_{ub}|$ is due mainly to uncertainties in the $b$ quark mass,
in weak annihilation (WA) contributions, in missing higher order perturbative
corrections, and in the modeling of the shape functions.

At the time \superb\ takes data, new
calculations should decrease the perturbative error,
and  lattice calculations, together with
improved analyses of $e^+e^- \to \ {\rm hadrons}$ and
measurements of the moments of semileptonic and radiative $B$ decay spectra should
provide better determinations of $m_b$;
a precision of $20 \ {\rm MeV}$ on $m_b$ is possible.
Weak annihilation contributions are relevant only at high $q^2$, and
can be efficiently constrained by studying the $q^2$ spectrum.
The shape functions can also be better-constrained by studies of the
$B \to X_u \ell \nu$ spectra, but their importance will
decrease as the measurements become increasingly more inclusive.
A pioneering analysis in this area has
recently been published by \babar~\cite{Aubert:2006qi}.
In this analysis the $M_X$ cut is raised to values for which
the shape function sensitivity becomes negligible.
Such measurements are not competitive now,
but the situation will be quite different at \superb.

As a result, we expect the theoretical uncertainty on the inclusive
determination of $|V_{ub}|$ to eventually be dominated by the uncertainty in the $b$ quark mass.
In this respect, it should be stressed that in current analyses, $|V_{ub}|$
depends quite strongly on the precise value of $m_b$.
Typically, for a cut of $M_X < 1.7 \ {\rm GeV}$,
the relative error on $V_{ub}$ scales as  $4(\delta m_b)/m_b$.
Currently, with $\delta m_b = 40 \ {\rm MeV}$, the error induced on $V_{ub}$ is about $3.5\%$.
If the error on $m_b$ were halved, $|V_{ub}|$ extracted in this way would
have a parametric uncertainty below $2\%$.
However, the presence of the $M_X$ cut increases the sensitivity to $m_b$,
because the distribution functions also strongly depend on $m_b$.
Increasing the $M_X$ cut (as mentioned above) reduces the sensitivity to $m_b$.
Indeed, the total rate is proportional to $m_b^5$,
and for a totally inclusive measurement one has
$\delta V_{ub}/V_{ub} \simeq 2.5 (\delta m_b)/m_b$.
Therefore, if one could measure the total $B \to X_u \ell \nu$ rate,
the uncertainty induced by $\delta m_b = 20 \ {\rm MeV}$ on $|V_{ub}|$
would be only $1\%$.

A promising way to deal with the large $B \to X_c \ell \nu$ backgrounds with no cut
on the inclusive $B \to X_u \ell \nu$ decays phase space is to reconstruct the semileptonic decays in the recoil
against the other  $B$ fully reconstructed in a hadronic final state in $e^+e^- \to \FourS \to B\bar{B}$
events (the so-called ``hadronic tag technique'').
This technique provides full knowledge of the event, including the flavor of the $B$,
and allows the precise reconstruction of the neutrino four-momentum,
significantly improving background rejection against, for example.
events with several neutrinos or with one or more $K_L$ mesons.
At present, these measurements
are limited by low signal efficiency, and have large statistical uncertainty.
At \superb\, however, the statistical uncertainty will be less than $\sim1\%$.
The leading systematic errors will also be reduced: those due to detector effects
could reach 2$\%$ using the large data control samples available.
The current analyses have uncertainties due to $B \to X_c \ell \nu$ background
(branching fractions and form factors) as low as $4\%$ --
it is possible to reduce this by a factor of two.
Indeed, higher statistics and improvements in the detector and analysis will yield better measurements of these quantities.
Moreover, the enhanced hermeticity and superior vertexing capability of the \superb\ detector
will further improve background rejection through more precise neutrino
reconstruction and the detection of the displaced $D$ meson vertex.
A total experimental uncertainty on $|V_{ub}|$ of approximately $2$--$3\%$
can thus be achieved with this method.

Combined with the theoretical uncertainty discussed above,
an overall precision of $3\%$ on the determination of $|V_{ub}|$
using inclusive $B \to X_u \ell \nu$ decays at \superb\ will be possible.\\
\par\noindent
{{\it Exclusive decays}}\\
\vskip -9 pt
\noindent
The measurement of $|V_{ub}|$ using exclusive decays
is presently limited by theoretical
uncertainties on the form factors (about $12\%$).
Lattice calculations are expected to improve significantly
in the next five years, mainly due to an increase of available computing power.
Results from these calculations will decrease the uncertainty to approximately
2--3\% in the case of the most promising decay $B \to \pi \ell \nu$ (see
the Appendix of the CDR~\cite{Bona:2007qt}).

Using the hadronic tag approach (as in Refs.~\cite{Aubert:2006ry,Abe:2006gb}),
the statistical uncertainty on $|V_{ub}|$ will be below $1\%$.
This measurement being almost background-free, the systematic uncertainties
are dominated by detector effects, and should be of the order of $2\%$.
A total experimental uncertainty of 2--3\% on $|V_{ub}|$ can thus be achieved,
leading to an overall precision as good as 3--4\%.

These figures basically confirm the sensitivities presented in the CDR
for the measurement of $|V_{ub}|$.

\noindent
\subsection*{Rare radiative decays}
\label{sec:B:channels:rare}



\subsubsection*{The branching fraction ${\cal B}(B \to X_s \gamma)$}



The inclusive branching fraction ${\cal B}(B\to X_s \gamma)$ has been
measured at the $B$ factories
~\cite{Chen:2001fja,Abe:2001hk,Koppenburg:2004fz,Aubert:2005cu,Aubert:2006gg};
the current experimental world average is~\cite{Barberio:2006bi}:
\[
  \label{eq:btosgamma:HFAG}
    {\cal B}(B\rightarrow X_s \gamma)|_{E_\gamma >1.6 \ \rm{GeV}} =
  (3.55 \pm 0.26) \times 10^{-4}\,.
\]
The $7\%$ error on the branching fraction is a mixture of statistical,
systematic and theoretical contributions, where the latter comes
primarily from extrapolating the partial branching fraction,
typically measured for photon energies above $1.9 \ {\rm GeV}$, down to
the value of $1.6 \ {\rm GeV}$ used for the theoretical prediction.

Several different experimental approaches have been pursued to make a
measurement of the inclusive $B\rightarrow X_s \gamma$ branching fraction.
The approach that yields the most precise measurement depends on
the available statistics.
Untagged inclusive analyses, in which only the high-momentum photon is
reconstructed, have been carried out at $B$ factories, but are limited by
systematic errors that will make them uncompetitive in the \superb\ era.
Similarly, the semi-inclusive approach, which attempts to reconstruct as
many exclusive modes as possible, and then applies a correction
due to the missing rate, is already limited by the $X_s$ fragmentation
properties, {\it i.e.}, by uncertainty in the estimate of the fraction of the
total rate that is not reconstructed.
This systematic uncertainty amounts to about $15\%$
on the branching ratio~\cite{Aubert:2005cu}. More detailed studies are
needed to evaluate how much this systematic could be reduced with the
statistics available at \superb.

The most promising approaches for \superb\ are those that make use of
recoil analysis, in which the ``other $B$'' in the $B\bar B$ event is
tagged in either a semileptonic or hadronic decay.
This allows backgrounds to be reduced to acceptable levels
without putting constraints on the $X_s$ system.
The most recent semileptonic tag analysis~\cite{Aubert:2006gg} currently has
comparable statistical and systematic uncertainties (about $8\%$ each), but a
sizable portion of the systematic uncertainty is actually statistical in
nature, since it depends on the size of control samples derived from the data.
The current systematic uncertainty of the hadronic tagged
analysis~\cite{Aubert:2007my} is larger, but it seems probable that
refinements to this relatively new technique will be able significantly to reduce the
systematic error.

With the data sample of \superb, all approaches will
be systematics-limited.
We estimate that the hadronic and semileptonic tagged analyses
will be able to reduce systematic uncertainties to about $4$--$5\%$.
Since the systematics are mostly uncorrelated, the combined branching fraction
can be expected to have a systematic error of around $3\%$.\\

The Standard Model prediction of ${\cal B}(B\to X_s \gamma$) for $E_{\gamma} > 1.6 \ {\rm GeV}$ is
\[
  \label{eq:btosgamma:NNLL}
  {\cal B}( B \to X_s\gamma)|_{E_\gamma >1.6 \ \rm{GeV}} =
  \left\{
    \begin{array}{ll}
      (3.15  \pm 0.23) \times 10^{-4}   & \hbox{\cite{Misiak:2006zs}} \\
      (2.98  \pm 0.26) \times 10^{-4}   & \hbox{\cite{Becher:2006pu}}.
    \end{array}
  \right.
\]
The two predictions differ in their use of resummation of log-enhanced terms which are included
in the result of \cite{Becher:2006pu}. There is no consensus on the consistency of the resummed result~\cite{misiaktalk}. We therefore quote both predictions pending clarification.
For both results, the overall uncertainty consists of non-perturbative ($5\%$),
parametric ($3\%$), higher-order ($3\%$) and $m_c$-interpolation ($3\%$), which
have been added in quadrature.

There are other perturbative NNLL corrections that are not yet included
in the present NNLL estimate, but are expected to be smaller than the current
uncertainty, producing a shift of the central value of about 1.6$\%$.

While the uncertainties due to the input parameters and due to the
$m_c$~interpolation could be further reduced, the perturbative error
of $3\%$ will remain until a new major effort to compute the NNNLO
is carried out.
However, the theoretical prediction has now reached the non-perturbative
boundaries. The largest uncertainty is presently due to nonperturbative
corrections that scale with $\alpha_s \Lambda_{\rm QCD}/m_b$.
A local expansion is not possible for these contributions;
it is not clear if the corresponding uncertainty of $5\%$
(based on a simple dimensional estimate) can be reduced.
Recently, a specific piece of these additional nonperturbative corrections
has been estimated~\cite{Lee:2006wn}, and found to be consistent with the
dimensional estimate. It is also included in the prediction of Ref.~\cite{Becher:2006pu}.\\

Two explicit examples should demonstrate the stringent constraints that
can, with these uncertainties, be derived from the measurement of the $B \rightarrow X_s \gamma$
branching fractions.

Fig.~\ref{fig:MHc} shows the dependence of ${\cal B}(B \to X_s \gamma)$ on the
charged Higgs mass in the 2-Higgs-doublet model (2HDM-II) ~\cite{Misiak:2006zs}.
\begin{figure}[!htb]
  \begin{center}
    \includegraphics[width=8cm,angle=0]{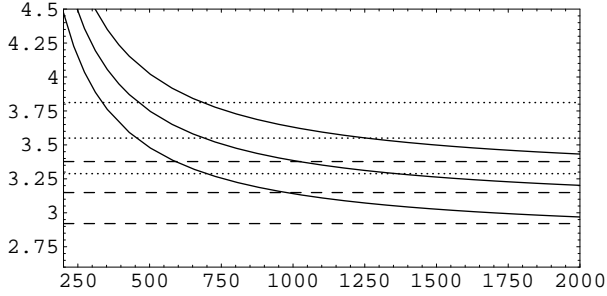}
  \end{center}
  \caption{
    ${\cal B}(B \to X_s \gamma) \times 10^{-4}$ as a function of the charged
    Higgs boson mass $M_{H^{+}}$ (GeV) in the 2HDM~II for $\tan\beta=2$ (solid lines).
    Dashed and dotted lines show the Standard Model and experimental results,
    respectively.
    \label{fig:MHc}
  }
\end{figure}
The bound on $M_{H^{+}}$ = 295 GeV at 95$\%$~CL, shown in Fig.~\ref{fig:MHc}, is currently
the strongest available lower limit on the charged Higgs mass.

Similarly, the bound on the inverse compactification radius of the minimal
universal extra dimension model (mACD) derived from ${\cal B}(B \rightarrow X_s \gamma)$~\cite{Haisch}
is $1/R > 600$ GeV at $95\%$ confidence level, as shown in Fig.~\ref{fig:ACDLO}.

\begin{figure}[!htb]
\begin{center}
\vspace{2mm}
\makebox{
\begin{psfrags}
\newcommand{\psfragtextscale}{1}
\providecommand{\psfragtextscale}{1}
\providecommand{\psfragmathscale}{\psfragtextscale}
\providecommand{\psfragnumericscale}{\psfragtextscale}
\providecommand{\psfragtextstyle}{}
\providecommand{\psfragmathstyle}{}
\providecommand{\psfragnumericstyle}{}

\psfrag{x}[cc][cc][1.1][0]{$1/R~[{\rm TeV}]$}
\psfrag{y}[bc][bc][1.1][0]{${\cal B} (\bar{B} \to X_s \gamma)~[10^{-4}]$}
\makebox{\hspace{-1.1cm} \includegraphics[width=3.5in,height=2.25in]{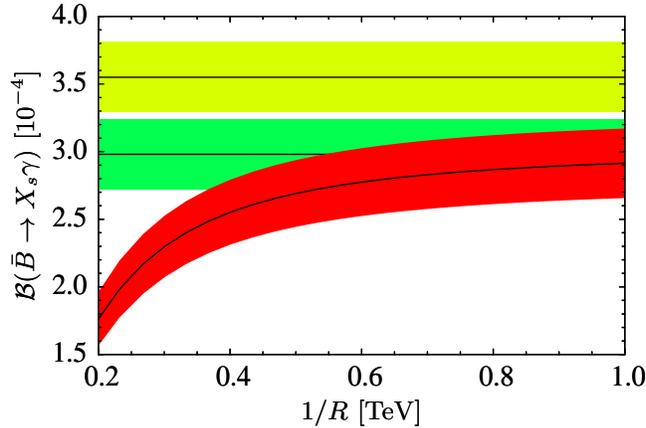}}
\end{psfrags}
}
\end{center}
\vspace{-4mm}
\caption{Branching fraction for $E_0 = 1.6 \, GeV$ as a function of $1/R$.
  The red (dark gray) band corresponds to the LO mUED result.
  The $68 \%$ CL range and central value of the
  experimental/Standard Model result is indicated by the yellow/green
  (light/medium gray) band underlying the straight solid line.}
\label{fig:ACDLO}
\end{figure}

\subsubsection*{$B \rightarrow X_s \ell^+ \ell^-$ decay modes}


The decay $B \rightarrow X_s \ell^+\ell^-$ is particularly important to the
\superb\ physics programme, due to the sensitivity to New Physics effects on
kinematic observables, such as the dilepton invariant  mass spectrum and the
forward--backward asymmetry $A_{\rm FB}$.

In the $B \rightarrow X_s \ell^+\ell^-$ system, one has to remove contributions from $c \bar c$
resonances that appear as large peaks in the dilepton invariant mass spectrum, by appropriate  kinematic cuts.
It is conventional to define ``perturbative windows'' with $s=q^2/m_b^2$ away from charmonium resonances,
namely the low dilepton-mass region $ 1 \ {\rm GeV} < q^2 < 6 \ {\rm GeV}$
and the high dilepton-mass region with $q^2 > 14.4 \ {\rm GeV}$.
In these windows theoretical predictions for the invariant mass spectrum
are dominated by the perturbative contributions;
a theoretical precision of order $10\%$ is, in principle, possible.

In the following, we collect the most accurate predictions for
observables in $B \rightarrow X_s \mu^+ \mu^-$ decay. Formulae for
the electron case should be modified to take into account
the experimental resolution for collinear photons.

The value of the dilepton invariant mass $q^2_0$,
for which the differential asymmetry $A_{\rm  FB}$ vanishes,
is one of the most precise predictions in flavor physics,
with a theoretical uncertainty of order $5\%$~\cite{Huber:2007vv}:
\begin{eqnarray}
(q_0^2)_{\mu\mu} & = &
\Big[ 3.50 \pm 0.10_{\rm scale} \pm 0.002_{m_t} \pm 0.04_{m_c,C}\nnb \\
&& \hspace*{-15pt}
\pm 0.05_{m_b} \pm 0.03_{\alpha_s(M_Z)} \pm 0.001_{\lambda_1} \pm 0.01_{\lambda_2}
\Big] \, \gev^2 \nnb \\
& = & ( 3.50 \pm 0.12) \, \gev^2\,. \label{muonzero}
\end{eqnarray}
This accuracy cannot be reached with the analogous exclusive
observable in $B \rightarrow K^* \ell^+\ell^-$, due to the unknown
$\Lambda_{\rm QCD}/m_b$ corrections.

The latest update of the dilepton mass spectrum, integrated
over the low and the high dilepton invariant mass region in the muonic case,
leads respectively to~\cite{Huber:2005ig}:
\bea
{\cal B}_{\mu\mu}^{\rm low} & = &
\Big[ 1.59 \pm 0.08_{\rm scale} \pm 0.06_{m_t} \pm 0.024_{ C,m_c } \nnb \\
&& \hspace*{-40pt}
\pm 0.015_{m_b} \pm 0.02_{\alpha_s(M_Z)} \pm 0.015_{\rm CKM} \pm 0.026_{{\rm BR}_{sl}}
\Big] \times 10^{-6} \nnb \\
& = & (  1.59  \pm 0.11 ) \times 10^{-6} \, \label{muonBRlow}
\eea

and

\bea
{\cal B}_{\mu\mu}^{\rm high} & = &
2.40 \times 10^{-7} \Big(
1 + \left[ {}^{+0.01}_{-0.02} \right]_{\mu_0} + \left[ {}^{+0.14}_{-0.06} \right]_{\mu_b}
\pm 0.02_{m_t} \nnb \\
&& \hspace*{-25pt}
+ \left[ {}^{+0.006}_{-0.003} \right]_{{C,m_c}} \pm 0.05_{m_b} + \left[ {}^{+0.0002}_{-0.001} \right]_{\alpha_s} \nnb \pm 0.002_{\rm CKM} \\
&& \hspace*{-25pt}
\pm 0.02_{{\rm BR}_{sl}} \pm 0.05_{\lambda_2} \pm 0.19_{\rho_1} \pm 0.14_{f_s} \pm 0.02_{f_u} \Big)  \nnb \\
&=& 2.40 \times 10^{-7} \; (1\,^{+0.29}_{-0.26} ) \,. \label{muonBR}
\eea

In the high ${s}$ region, the uncertainties are larger, due to the breakdown of
the heavy-mass expansion at the endpoint. However the uncertainties can be
significantly reduced by considering quantities normalized to the semileptonic
$b\to u \ell \nu$ rate integrated over the same $s$ interval~\cite{Ligeti:2007sn}:

\begin{equation}
  {\cal R}(s_{\rm min}) = \frac{
    \int_{s_{\rm min}}^1 {\rm d}s \,
    \frac{{\rm d}\Gamma(B\to X_s \ell^+\ell^-)}{{\rm d}s}
  }{
    \int_{s_{\rm min}}^1 {\rm d}s \,
    \frac{{\rm d}\Gamma(B\to X_u \ell \nu)}{{\rm d}s}
  } \, .
\end{equation}

The numerical analysis shows that the uncertainties due ${\cal O}(1/m_b)$
power corrections which correspond to the parameters $\lambda_2$, $\rho_1$,
$f_u^0+f_s$ and $f_u^0-f_s$ are now under control~\cite{Huber:2007vv}:
\bea
{\cal R}(s_{\rm min})_{\mu\mu}^{\rm high} & = &
2.29 \times 10^{-3} \Big(
1 \pm 0.04_{\rm scale} \pm 0.02_{m_t} \nnb \\
&& \hspace*{-25pt}
\pm 0.01_{{C,m_c}} \pm 0.006_{m_b} \pm 0.005_{\alpha_s} \pm 0.09_{\rm CKM}
\nnb \\
&& \hspace*{-25pt}
\pm 0.003_{\lambda_2} \pm 0.05_{\rho_1} \pm 0.03_{f_u^0+f_s} \pm 0.05_{f_u^0-f_s} \Big) \nnb \\
&=& 2.29 \times 10^{-3} ( 1 \pm 0.13 ) \,. \label{muonR}
\eea
The largest remaining source of error is now $|V_{ub}|$, which will be further reduced
with the precise CKM determination at \superb.
As in the $B \rightarrow X_s \gamma$ case, additional uncertainties, such as the still
unknown non-perturbative corrections that scale with $\alpha_s\Lambda_{\rm QCD}/m_b$,
are about $5\%$.
The cuts in the hadronic invariant mass spectrum lead to
additional uncertainties of order $5\%$, which correspond to the effects of
subleading shape functions~\cite{Lee:2005pk,Lee:2005pw}. \\

Published analyses
for $B \to X_s l^+ l^-$~\cite{Aubert:2004it,Iwasaki:2005sy}
have used a semi-inclusive approach ($X_s = 1K + n\pi, n \le 3$).
This technique is affected by large systematics arising from uncertainties
on the ratio used to extrapolate from the semi-inclusive to the inclusive
branching ratio. This type of analysis is expected to be systematics
dominated, with statistics around $1 \ {\rm ab}^{-1}$.

With larger statistics, a fully inclusive analysis using
semileptonic or hadronic tags is likely to be more sensitive.
Feasibility studies for such an analysis show that about 40 signal events per
${\rm ab}^{-1}$ can be expected with a signal-to-background ratio of $\sim 1.5$.
At \superb, a few percent statistical error on the inclusive branching ratio
can be achieved, well below the present theoretical error
(see Eqs.~\ref{muonBRlow} and~\ref{muonBR}).
No detailed studies are available for the systematic uncertainties,
but they are likely to become dominant over
experimental statistical uncertainties at this level of precision. \\

\subsubsection*{$B \to K^{(*)} \tau \tau$ decay modes}

The branching ratio of
$B \to X_s \tau^+ \tau^-$ is smaller by a factor of about 20,
with respect to $B \to X_s \ell^+ \ell^-$ ($\ell = e,\mu$),
in the low $q^2$ region,
but is expected to be about $2$--$3 \times 10^{-7}$,
comparable to $B \to X_s \ell^+ \ell^-$ (see Eq.~\ref{muonBR}),
in the high $q^2$ region.

An inclusive experimental determination is essentially impossible,
but an analysis of the exclusive decays $B \to K^{(*)} \tau \tau$ might be possible.
These decays are predicted to make up $50$--$60\%$ of the total inclusive
rate~\cite{Du:1993sh}.
Preliminary simulation studies using the hadronic tag technique indicate
that the Standard Model branching fractions could be measurable with the full \superb\
integrated luminosity. Other interesting measurements such as the polarization
asymmetry~\cite{Hewett:1995dk} are under study. \\

\subsubsection*{$B \to K^{(*)} \nu \overline \nu$ decay modes}

The rare decay $B \to K^{(*)} \nu \overline \nu$ is an
interesting probe for New Physics in $Z^0$ penguins~\cite{Buchalla:2000sk}, such as
chargino-up-squark contributions in a generic supersymmetric theory.
Moreover, since only the $b \to s$ +
missing energy process can be detected, the measured rate can be affected by
exotic sources of missing energy, such as light dark matter~\cite{Bird:2004ts}
or ``unparticle physics''~\cite{Georgi:2007ek,Aliev:2007gr}. Notice also that
New Physics effects can modify the kinematics of the decay, which implies that any
selection applied on kinematical variables has an impact on the theoretical
interpretation of the measured branching ratio. The best upper limit among the
exclusive decay channels is
${\cal B}(B^+ \to K^+\nu{\overline\nu}) < 14 \times 10^{-6}$~\cite{Chen:2007zk},
still far above the Standard Model branching fraction of $4 \times 10^{-6}$~\cite{Buchalla:2000sk}.

Due to the undetected neutrinos, it is not possible to reject
background by means of the usual kinematical constraints, so the search for
these decays must be performed using a recoil analysis.

In the $B^+ \to K^+ \nu \overline \nu$ analysis, only one track is required on
the signal side. A selection on the kaon momentum is usually applied.
A final selection is applied on the extra energy $E_{\rm extra}$, defined as
the sum of the energies of the neutral electromagnetic calorimeter clusters that are not associated with
the $B_{\rm tag}$ or the signal side. Current analyses employ a counting
technique, but a maximum likelihood (ML) fit to the $E_{\rm extra}$ distribution can be
used to improve performance. To be conservative, we assume the current
analysis technique. From toy MC simulations, combining the results from the
semileptonic and the hadronic recoil, the observation of the decay is expected with
between 10 and $20 \ {\rm ab}^{-1}$ with an expected error of $18\%$, in the most
conservative scenario, at $50 \ {\rm ab}^{-1}$.
The improvement in the precision as a function of luminosity is shown in
Fig.~\ref{fig:snunu}.

\begin{figure}[!htb]
  \begin{center}
    \includegraphics[width=\columnwidth]{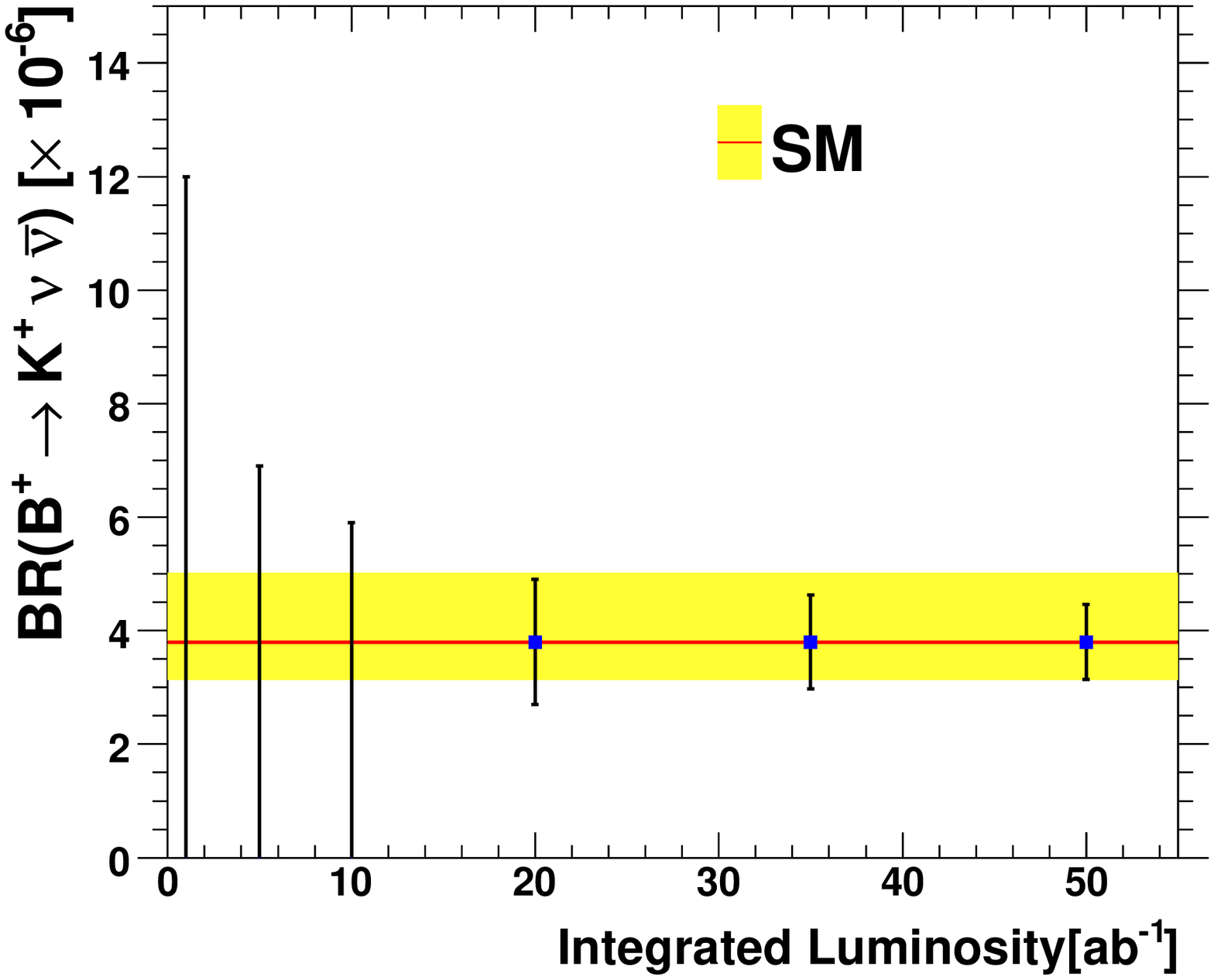}
    \includegraphics[width=\columnwidth]{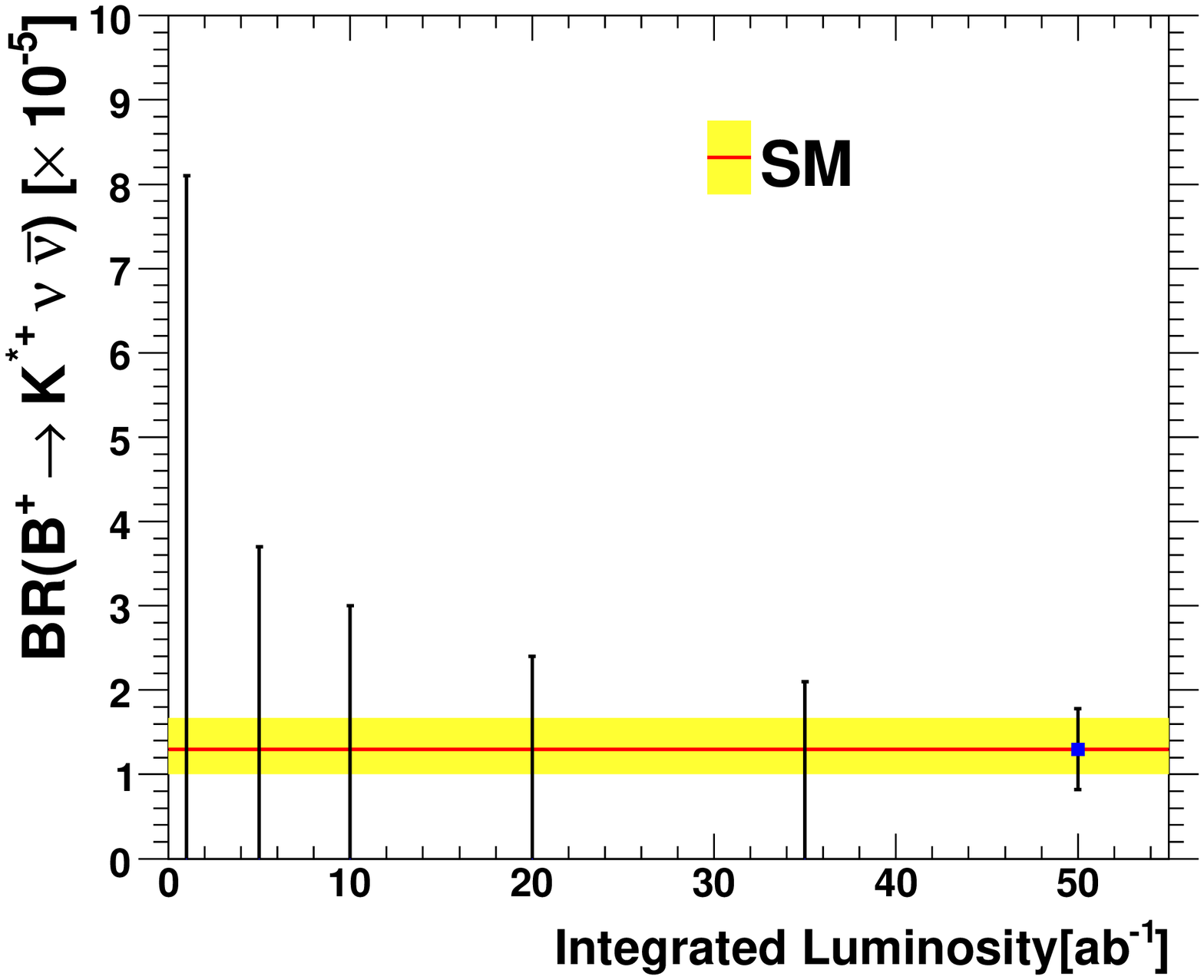}
    \caption{
      Expected precision of the measurements of the branching fractions of
      (top) $B^+ \to K^+ \nu \bar{\nu}$ and
      (bottom) $B^+ \to K^{*+} \nu \bar{\nu} \ (K^{*+} \to K_S \pi^+)$
      evaluated as a function of the integrated luminosity,
      assuming efficiencies and backgrounds as in the current \babar\ analyses.
      The bands indicate the range of the Standard Model predictions.
      \label{fig:snunu}
    }
    \end{center}
\end{figure}

In the $B^0 \to K^{*0} \nu \overline \nu$ analysis, the $K^{*0}$ is
reconstructed in the $K^{*0} \to K^+ \pi^-$ channel, with no cut on the
kinematical variables. A maximum likelihood fit is used to extract the signal yield from the
$E_{\rm extra}$ distribution. Observation of this decay is expected between 10
and $20 \ {\rm ab}^{-1}$ with an expected error of $20\%$, in the most conservative
scenario, at $50 \ {\rm ab}^{-1}$.

The same approach is adopted in the $B^\pm \to K^{*\pm} \nu \overline \nu$
analysis, where $K^{*\pm} \to K_S^0 \pi^\pm$ or $K^{*\pm} \to K^\pm \pi^0$.
The observation is expected around $40 \ {\rm ab}^{-1}$ with an expected error of  $25\%$,
in the most conservative scenario, at $50 \ {\rm ab}^{-1}$
(see Fig.~\ref{fig:snunu}).

An irreducible background contribution from $B \to \tau \nu$ decays
is expected in the $B \to K^{(*)\pm} \nu \overline \nu$ analyses.
However, the effect of this background can be controlled with
improvements in the analyses (such as using a maximum likelihood fit).
Moreover, the performance of the recoil technique will be improved by
the improved hermeticity of the \superb\ detector, making the cuts usually applied on
the track multiplicity of the signal side more effective. Preliminary studies have shown
that a $30\%$ reduction in the background contamination with the baseline
\superb\ design is possible.
For background dominated channels such as $B \to K^{(*)\pm} \nu \overline \nu$,
a reduction in background of $30\%$ can be shown to be roughly equivalent to
an increase in statistics of $1/0.7$, {\it i.e.} about $40\%$.
Therefore, such an improvement has a significant effect on the sensitivity.

If the background can be reduced sufficiently, it will be possible to do higher
multiplicity studies of $b \to s\nu\bar{\nu}$ decays such as
 $B^+ \to K_1^+ \nu\bar{\nu} \,, \ K_1^+ \to K^+ \pi^+\pi^-$.
This information could be used to make a semi-inclusive measurement of
${\cal B}(b \to s \nu \bar{\nu})$.
Further background rejection can come from an improved vertex detector,
that allows to apply vertexing requirements (poorly used now)
and secondary vertex information.
The semi-inclusive approximation may provide the best possible
analysis of $B \to X_s \nu \overline \nu$ decay.
Owing to the complete absence of any powerful constraint to be applied
on the signal side, the fully inclusive analysis appears to be difficult
in the face of large backgrounds.
If a fully inclusive analysis could be performed at \superb, it may be possible
to make a test of the theoretically clean Standard Model prediction~\cite{Buras:2000dm}
\begin{equation}
\frac{
  {\cal B}(B \to X_d \nu \bar{\nu})
}{
  {\cal B}(B \to X_s \nu \bar{\nu})
} = \left| \frac{V_{td}}{V_{ts}} \right|^2 \, .
\end{equation}
Studies of corresponding ratios using exclusive modes are less theoretically
clean, however the prospects for measuring $B \to \pi \nu \bar{\nu}$ at
\superb\ look good~\cite{Bona:2007qt}.


\noindent
\subsection*{Leptonic $B$ decays}
\label{sec:B:channels:leptonic}

\noindent
{\underline{The branching fraction of $B \to \ell \nu$}}

The decays \btolnu\ can be used to constrain the Standard Model mechanism of
quark mixing.  New Physics contributions can enhance the branching fractions
of \btolnu, as described in the \superb\ CDR~\cite{Bona:2007qt}.
Precision measurements of the branching fraction
of \btolnu\ where $\ell = e, \mu, \tau$ can be used to constrain New Physics.

Recent measurements have provided evidence for
$B \to \tau \nu$~\cite{Ikado:2006un,Aubert:2007bx,Aubert:2007xj}
These measurements rely on recoil analyses in which fully (partially)
reconstructed $B$ meson decays to hadronic (semileptonic) final states
of the non-signal $B$ in the event ($B_{\rm tag}$) are used to help reduce
background for the partially reconstructed signal.
This approach is required for the $B \to \tau \nu$ analysis,
in which there are at least two missing neutrinos in the final state.
For $B \to \mu \nu$ and $B \to e \nu$~\cite{Satoyama:2006xn,Aubert:2004yu,Aubert:2006at},
on the other hand, the high momentum lepton alone provides a characteristic
signature.
Nevertheless, recoil analyses appear preferable also for these channels, due
to the additional kinematic constraints and the reduction in background.

A number of possible improvements to the \btolnu\ analyses
are being explored.
These include
\begin{enumerate}
\item All existing \babar\ measurements rely on reconstructing $B_{\rm tag}$
  modes with a $D$ or $D^*$ in the final state.  However it is also possible
  to increase the signal efficiency by including charmonium decay modes
  with a \jpsi\ in the $B_{\rm tag}$ final state.
\item The $B_{\rm tag}$ categories all have different purities.
  It is a natural extension of the existing analyses to investigate the gain
  in precision that one can obtain by subdividing the data according to the
  $B_{\rm tag}$ purity in a multi-dimensional maximum likelihood fit, and if
  necessary, exclude any $B_{\rm tag}$ category in which systematic uncertainties
  are not under control.
  (Similar strategies have been successfully employed in time-dependent \CP\
  asymmetry measurements at the $B$ factories.)%
\item In the case of \btotaunu, $B_{\rm sig}$ has contributions from several
  reconstructed $\tau$ decay channels that have different purities; so one
  should subdivide the data according to $B_{\rm tag}$ and $B_{\rm sig}$ purity.%
\item Existing analyses rely heavily on a variable constructed from the sum of
  electromagnetic energy unassociated wither with the $B_{\rm sig}$ or
  $B_{\rm tag}$ to isolate signal ($E_{\rm extra}$).  In order to do this reliably,
  one has to understand, and accurately simulate, noise in the calorimeter as
  well as the geometric acceptance of the detector to backgrounds in which final
  state particles escape down the beam pipe, or into uninstrumented regions of
  the detector.  Not only does this rely on accurate accounting of material in
  the inner parts of the detector, but also in the calorimeter itself, and
 a finely-tuned understanding of the production mechanisms for all types of $B$
  and non-$B$ backgrounds.  It is not clear if the continual use of such a
  variable would facilitate a precision measurement of \btolnu\ branching
  fractions.  It would be possible to improve control of systematic
  uncertainties by limiting the analysis to high purity $B_{\rm tag}$ samples
  and/or  to $B_{\rm sig}$ channels only.  During the detector R\&D stage, one
  should also consider the effects of non-active material in the
  calorimeter, and material in front of the calorimeter, as it is critical that %
  this is correctly accounted for in GEANT simulations of \superb.%
\item The current analyses that extract the yield from a fit of the %
  $E_{\rm extra}$ distribution determine the shape of the signal PDF %
  using a control sample of semileptonic $B \to D^{(*)}\ell\nu$ decays %
  on the recoil of $B_{\rm tag}$.
  With \superb\ statistics it would be possible to use hadronic $B$
  decays for the control sample, which could lead to a reduction of systematic
  uncertainty.  This approach has been used as a systematic cross-check in one
  search for \btotaunu~\cite{Aubert:2006at}, and has also been employed by
  CLEO in the measurement of $f_{D_s}$ using
  $D_s^+ \to \tau^+ \nu_\tau$~\cite{Ecklund:2007zm}.%
\item There are alternatives to the $E_{\rm extra}$ variable that do not rely
  so critically on our understanding of the detector material, acceptance,
  response and details of the background kinematics.  Examples of such
  variables include the highest energy cluster unassociated with $B_{\rm sig}$ or
  $B_{\rm tag}$.%
\item Improvements in the detector hermeticity would, as well as increasing
  the signal efficiency, lead to smaller backgrounds due to particles that
  travel down the beam pipe.  Similarly, improvements in the efficiency with
  which $K_L$ mesons are detected would help to reduce the background.%
\end{enumerate}

The emphasis in these improvements is on increasing the signal efficiency, and
on better control of systematic uncertainties associated with measuring
\btolnu\ branching fractions. It must be emphasized that, while the Standard Model
expectation for the branching fraction of \btomunu\ is significantly lower
than that of \btotaunu, the experimental signature, a high momentum muon
with missing energy, is much cleaner than that of a $\tau$ lepton.
Therefore, at very high luminosities, \btomunu\ is expected to provide
a more precise branching fraction measurement, as it will not be systematics
limited.
Measurements of \btomunu\ and \btotaunu\ are central to the New Physics
search capability of \superb.
The phenomenological impact of these measurements is discussed in
Section~\ref{sec:B:pheno}. \\

\subsubsection*{Radiative leptonic decays}

Radiative leptonic decays, namely
$B_{u}\to \ell \nu\gamma$, $B_{d(s)}\to \ell\ell\gamma$ and $B_{d(s)}\to \gamma\gamma$,
do not contain any hadrons except the $B$ meson.
This simple observation drastically reduces theoretical uncertainties
originating from the strong interaction, such as final state interactions.
\superb\ may be able to observe these extremely rare processes,
due to its good efficiency for reconstruction of the radiative photon.

It has been shown that, in the Standard Model, the strong interaction factorizes at the
large $m_b$ limit, making it possible to describe these three processes in
terms of an universal non-perturbative
form-factor~\cite{DescotesGenon:2002ja}.
Rough estimates of the branching ratios yield
${\cal B}(B_{u}\to \ell \nu\gamma)�\sim {\cal O}(10^{-6})$,
${\cal B}(B_{d(s)}\to \ell\ell\gamma) \sim {\cal O}(10^{-10(-9)})$ and
${\cal B}(B_{d(s)}\to \gamma\gamma) \sim {\cal O}(10^{-8(-6)})$.
It should be emphasized that the helicity suppression,
which diminishes the branching ratio of the
pure-leptonic processes corresponding to the first two channels,
$B_u\to \ell \nu$ and $B_{d (s)} \to \ell\ell$,
does not occur here, due to the additional photon.
As a result, one can take advantage of all three final states
with $\ell = e, \mu, \tau$, which have similar decay rates.

A strategy to search for New Physics with these channels would be to first determine
the form factor through
the tree level $B^+ \to \ell \nu\gamma$ process~\cite{Korchemsky:1999qb}
and then use it to extract New Physics effects from the loop level
$B_{d(s)}\to \ell\ell\gamma$ and $B_{d(s)}\to \gamma\gamma$ processes.
In the former, the most recent experimental results~\cite{Aubert:2007yh} are
already close to the Standard Model expectation.
\superb\ can make a precise measurement of this decay; theoretical
uncertainties due to the restricted phase space used in the analysis
(necessary to reduce backgrounds from final state radiation photons) may then
become a limiting factor.
The current experimental upper limits on $B_d\to \ell\ell\gamma$ are
at the $10^{-7}$ level~\cite{Aubert:2006ft}; since these are not background-limited,
\superb\ can improve the limits to close to the Standard Model level.
Once observed, kinematical distributions in these processes provide additional New Physics sensitivity.
New Physics effects on the branching ratio and the forward-backward asymmetry
of the $B_{d(s)}\to \ell\ell\gamma$ process have been investigated,
e.g. in~\cite{Yilmaz:2004tr,Choudhury:2005rz}.
For example, those effects could come from an anomalous $bd(s)Z$ coupling, that
could be also seen the in $B\to K^{(*)} \ell\ell$ and $B_{d(s)}\to \ell\ell$
processes.

On the other hand, the new physics effect to $B_{d(s)}\to \gamma\gamma$ process
could come from two kinds of short-distance contributions: anomalous  $bs\gamma$ coupling
and the $bs\gamma\gamma$ coupling. In particular, the later contribution has not been
explored yet and \superb\ sensitivity will reveal these couplings for the first time.
It should be noted that this contribution can be also studied in
$B\to K\gamma\gamma$~\cite{Hiller:2004wc}.
Detailed investigations of the supersymmetric contributions to
$B_s\to\gamma\gamma$ and $B\to X_s\gamma\gamma$
have been performed~\cite{Bertolini:1998hp}.
As discussed in the CDR~\cite{Bona:2007qt}, $B_s\to\gamma\gamma$ could be
observed at \superb\ after accumulating about $1 \ {\rm ab}^{-1}$ at the
\FiveS.
Extrapolating from existing upper limits on the $B_d \to\gamma\gamma$
decay~\cite{Aubert:2001fm,Abe:2005bs}, \superb\ could probe down to the Standard Model
level of this New Physics-sensitive decay.

%
%
%

\section{Phenomenology}
\label{sec:B:pheno}

\subsection*{Golden processes}
\label{sec:B:pheno:gold}

\begin{table*}[!htb]
 \caption{
    \label{tab:golden}
     Golden modes in different New Physics scenarios. A ``X'' indicates the golden channel of a given scenario.
     An ``O'' marks modes which are not the ``golden'' one of a given scenario but can still display a measurable deviation from the Standard Model.
     The label $CKM$ denotes golden modes which require the high-precision determination of the CKM parameters achievable at \superb. }
   \begin{tabular}{lcccccc}
    \hline\hline
                                      &      $H^+$      &   ~Minimal~  &   ~Non-Minimal~   &  ~Non-Minimal~    &     NP            & ~Right-Handed~   \\
                                      & high tan$\beta$ &     FV       &     FV (1-3)      &   FV (2-3)        & ~Z-penguins~      &   currents       \\
    \hline
    \BR$(B \to X_s \gamma)$            &                 &     X        &                   &       O           &                   &      O           \\
    $A_{\CP}(B \to X_s \gamma)$        &                 &              &                   &       X           &                   &      O           \\
    \BR$(B \to \tau\nu)$               &      X-$CKM$    &              &                   &                   &                   &                  \\
    \BR$(B \to X_s l^+l^-)$            &                 &              &                   &       O           &      O            &      O           \\
    \BR$(B \to K \nu \overline{\nu})$  &                 &              &                   &       O           &      X            &                  \\
    $S(K_S \pi^0 \gamma)$              &                 &              &                   &                   &                   &      X           \\
    $\beta$                           &                 &              &      X-$CKM$      &                   &                   &      O           \\
    \hline\hline
  \end{tabular}
\end{table*}

At \superb, a golden channel is any channel that is very well known in the Standard Model.  This
includes ``null tests'' (observables that are zero, at least approximately, in
the Standard Model) but also other channels predicted with small errors.  This places more emphasis on
inclusive modes than on exclusive decays.  While there are probably
specific channels that can be selected in charm and in $\tau$ physics, in $B$
physics there are so many golden channels that selecting one or two risks
missing the point. In addition processes that are golden (\ie\ display a
measurable deviation from Standard Model) for given New Physics scenario could be uninteresting
in a different scenario.
The rationale for building \superb\ based on the New Physics-sensitivity of any individual
channel can certainly be challenged -- the motivation is the large range of golden
channels for which \superb\ has unsurpassed sensitivity.
We will nonetheless, in response to the IRC, select some specific channels for which \superb\ has unique potential. However, the argument given above makes it clear that golden modes are defined only in the context of a limited
and non-orthogonal set of New Physics scenarios.
We thus want to stress once more that one of the most sensitive searches for
New Physics will be the 1$\%$ determination of CKM parameters; the possibility
of performing such a precise determination in the presence of New Physics is a unique
feature of \superb. The precision measurements required to achieve this goal are $|V_{ub}|$ and
the CKM angles. In the spirit of indicating the golden modes, we select
$|V_{ub}|$ and $\alpha$, being $\beta$ and $\gamma$ precisely measured at LHCb.
In the following, we denote by $CKM$ those places in which the improvements
on the CKM parameters achieved by \superb\ are crucial to the corresponding New Physics
searches. We do not include rare kaon decays in which a precise CKM measurement is also extremely
important. Notice that whenever a high precision CKM determination is required,
progress on Lattice QCD calculations, as discussed in the Appendix of the CDR,
is needed.
In Table \ref{tab:golden} we show the result of our selection of golden modes in
different New Physics scenarios. For each scenario, ``X'' marks the golden channel while ``O'' marks
those modes which can display measurable deviation from the Standard Model.

A few comments are in order on this selection. Notice first that ${\cal B}(B \to X_s \gamma)$ is important in several scenarios,  in
particular in the MFV scenarios, and therefore we put it in the list, even though at \superb\ it is limited by theoretical errors,
unless a major breakthrough in non-perturbative calculations of power suppressed corrections is achieved. In some
of the scenarios considered, of course, this list is far from complete; many other measurements are expected to show deviations from their Standard Model
values. For example, in the case of non-minimal flavor violation in the transitions between third and second generations, the entire cohort
of $b \to s$ penguins-dominated non-leptonic modes could show a deviation in the measured value of time-dependent \CP\  asymmetries
compared to those measured in $b \to c \bar{c} s$ transitions.

\noindent
\subsection*{Benchmarks}
\label{sec:B:pheno:benchmarks}

The problem of defining proper benchmarks for \superb\ has not been addressed yet. In fact benchmarks
for flavor physics clearly require the specification of the New Physics flavor structure, which is not needed
(at least at first approximation) for high-$p_T$ physics. Nonetheless, stimulated by the IRC, we estimate
the relevant flavor observable measured at \superb\ within the mSUGRA models at the SPS$1a$, SPS$4$ and
SPS5 benchmark points defined for the LHC in ~\cite{Allanach:2002nj} .
The purpose of this exercise is to evaluate the deviation from the Standard Model of flavor observables in a
MFV scenario where LHC can reconstruct a large part of the SUSY spectrum. We consider a set of measurements
which are likely to be affected in the MFV model under consideration.

In terms of the fundamental parameters at the high scale, the SPS considered points are
defined as:
\begin{eqnarray}
{\rm SPS}1a: &&\quad m_0 = 100 {\rm GeV}, \quad m_{1/2} = 250 {\rm GeV}, \\ \nonumber
             &&\quad A_0 = -100 {\rm GeV}, \quad \tan\beta = 10, \quad \mu > 0 \\ \nonumber
{\rm SPS}4:  &&\quad m_0 = 400 {\rm GeV}, \quad m_{1/2} = 300 {\rm GeV}, \\ \nonumber
             &&\quad A_0 = 0, \quad \tan\beta = 50, \quad \mu > 0 , \\ \nonumber
{\rm SPS}5:  &&\quad m_0 = 150 {\rm GeV}, \quad m_{1/2} = 300 {\rm GeV}, \\ \nonumber
             &&\quad A_0 = -1000,\quad \tan\beta = 5, \quad \mu > 0 . \nonumber
\end{eqnarray}

Note that SPS$1a$, a ``typical'' mSUGRA scenario with intermediate tan$\beta$, is extremely good for LHC and indeed the most studied -
the pattern of sparticle masses allows them all to be measured with very good accuracy~\cite{Weiglein:2004hn}.
By contrast, the relatively high squark masses and the low value of $\tan\beta$ suppress effects on flavor observables.
SPS4 is an mSUGRA scenario with large $\tan\beta$. Unfortunately, no detailed studies are available at
LHC for this point. Nevertheless we roughly estimated the LHC performance by studying the decay chain starting
from the computed SUSY spectrum. We found a single study at SPS5~\cite{Borjanovic:2005tv}, a parameter configuration with relatively
light stop quark and low $\tan\beta$. Here again the LHC performance in measuring the SUSY spectrum is rather good.

Based on these studies, and using the tools developed at the recent CERN-Workshop ``Flavour in the LHC Era''
\cite{Buchmueller:2007zk} we produced the predictions presented in Table \ref{tab:spspoints}.

\begin{table*}[!h]
 \caption{
    \label{tab:spspoints}
    Predictions of flavor observables based on expected measurements
    from LHC in mSUGRA at SPS1a, SPS4, SPS5 benchmark points. Quantities denoted ${\cal R}$ are the ratios of
    the branching fractions to their Standard Model values. Quoted uncertainties (when available) come from the
    errors on the measurement of the New Physics parameters at LHC. Uncertainties on the Standard Model predictions of flavor
    observables are not included. For the SPS4 benchmark point the sensitivity study at LHC are not available.
  }
  \bigskip
   \begin{tabular}{lccc|c}     
    \hline\hline %
                                                       &        SPS1a             &       SPS4            &     SPS5           \\ \hline
    ${\cal R}(B \to X_s\gamma)$                          &     0.919  $\pm$ 0.038   &  0.248  & 0.848  $\pm$ 0.081 \\
    ${\cal R}(B \to \tau\nu)$                          &     0.968  $\pm$ 0.007   &  0.436  & 0.997  $\pm$ 0.003 \\
    ${\cal R}(B \to X_s l^+l^-)$                       &     0.916  $\pm$ 0.004   &  0.917  & 0.995  $\pm$ 0.002 \\
    ${\cal R}(B \to K \nu \overline{\nu})$             &     0.967  $\pm$ 0.001   &  0.972  & 0.994  $\pm$ 0.001 \\
    ${\cal B}(B_d \to \mu^+\mu^-)/10^{-10}$            &     1.631  $\pm$ 0.038   &  16.9   & 1.979  $\pm$ 0.012 \\
    ${\cal R}(\Delta m_s)$                             &     1.050  $\pm$ 0.001   &  1.029  & 1.029  $\pm$ 0.001 \\
    ${\cal B}(B_s \to \mu^+\mu^-)/10^{-9}$             &     2.824  $\pm$ 0.063  &   29.3   & 3.427  $\pm$ 0.018 \\
    ${\cal R}(K \to \pi^0 \nu \overline{\nu})$         &     0.973  $\pm$  0.001  &  0.977  & 0.994  $\pm$ 0.001 \\  \hline\hline %
  \end{tabular}
\bigskip
\end{table*}

\begin{figure*}[!h]
  \begin{center}
    \includegraphics[width=4.cm,angle=-90]{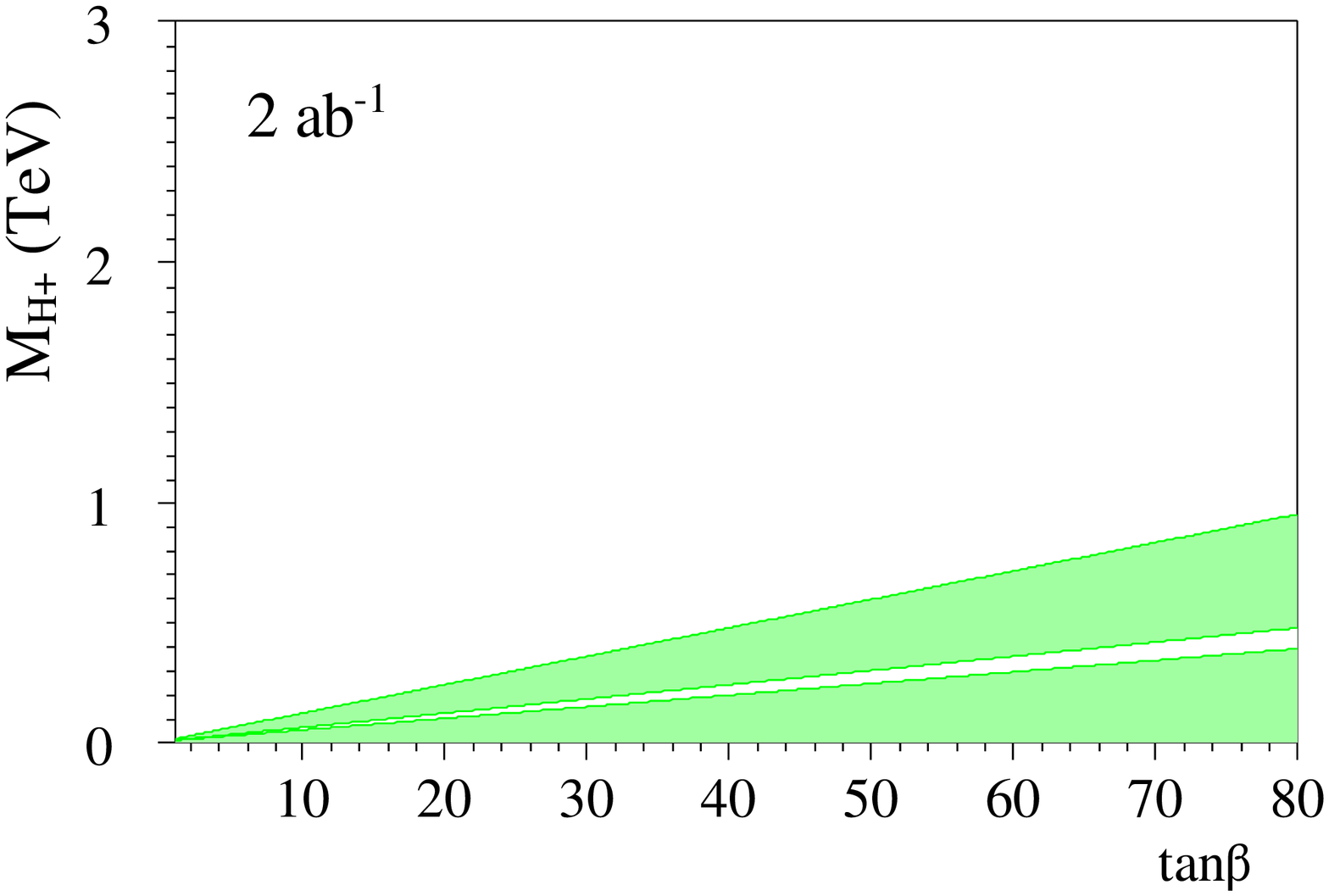}
    \includegraphics[width=4.cm,angle=-90]{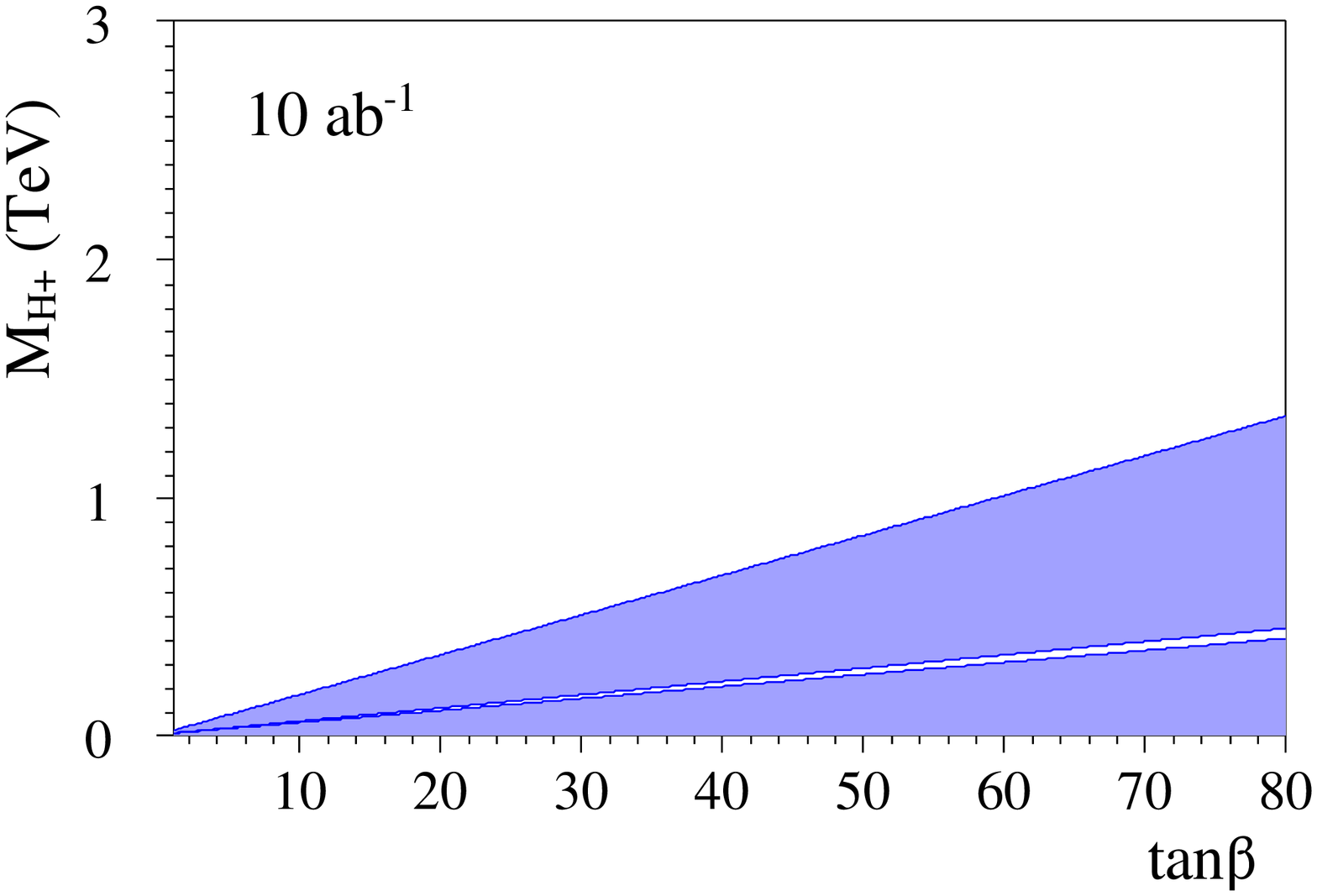} \\

    \includegraphics[width=4.cm,angle=-90]{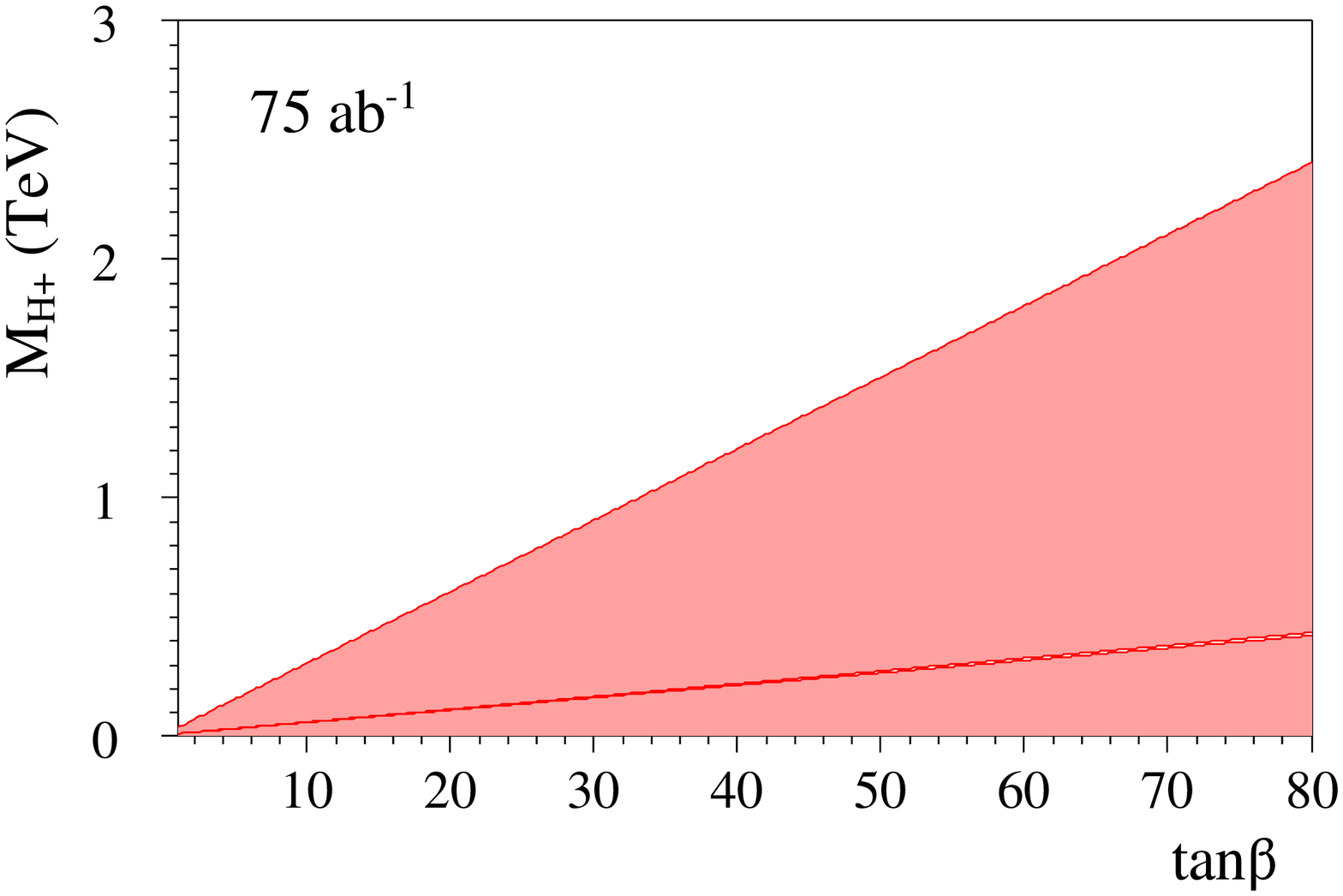}
    \includegraphics[width=4.cm,angle=-90]{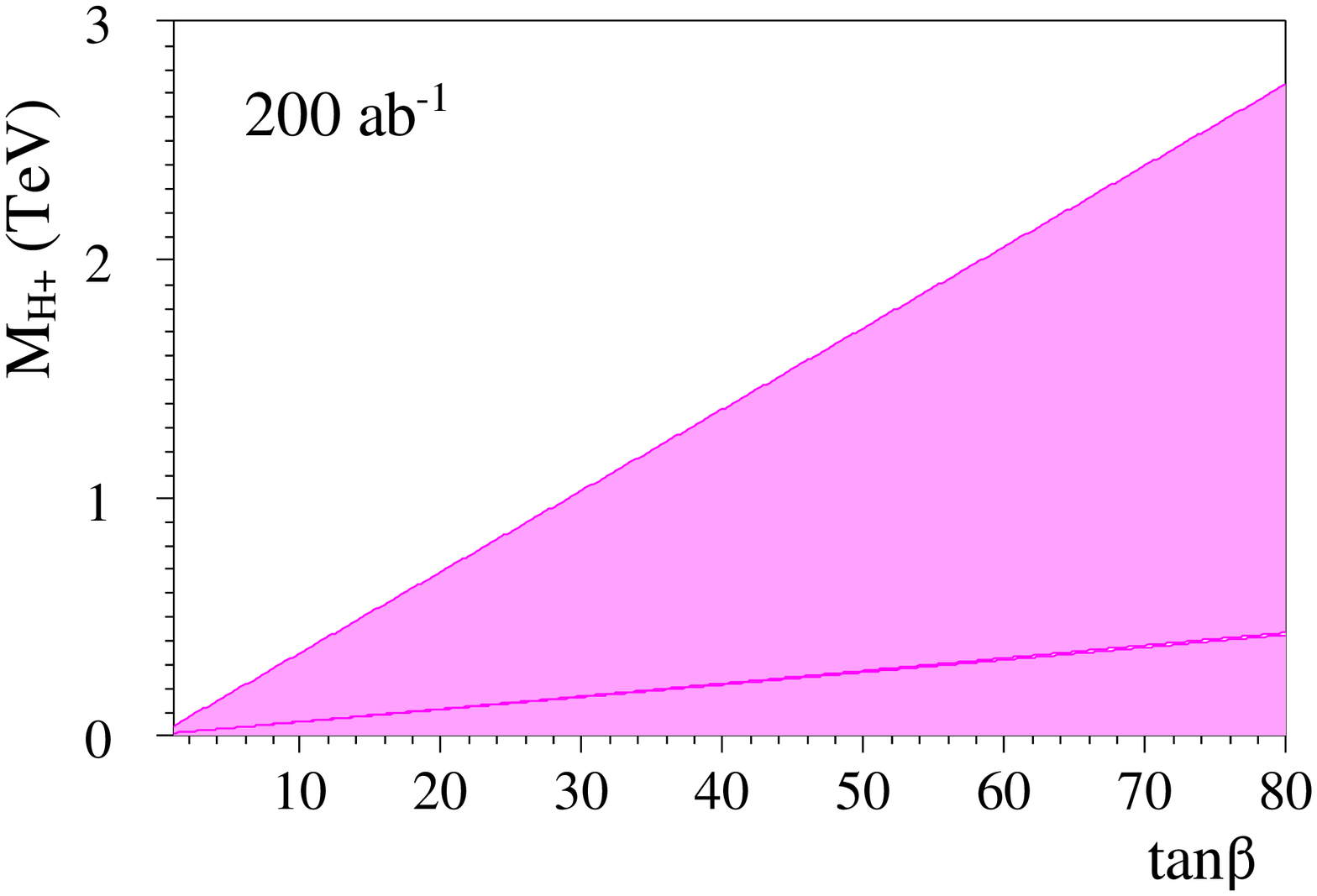}
    \caption{\label{fig:btaunu}
      Exclusion regions in the $m(H^+)$--$\tan\beta$ plane arising from the
      combinations of the measurement of ${\cal B}(B \to \tau \nu)$ and
      ${\cal B}(B \to \mu \nu)$ using 2 ab$^{-1}$ (top left), 10 ab$^{-1}$ (top right)
      75 ab$^{-1}$ (bottom left) and 200 ab$^{-1}$ (bottom right). We assume that
      the result is consistent with the Standard Model.
    }
  \end{center}
\end{figure*}

\begin{figure*}[!h]
  \begin{center}
    \includegraphics[width=4.cm,angle=-90]{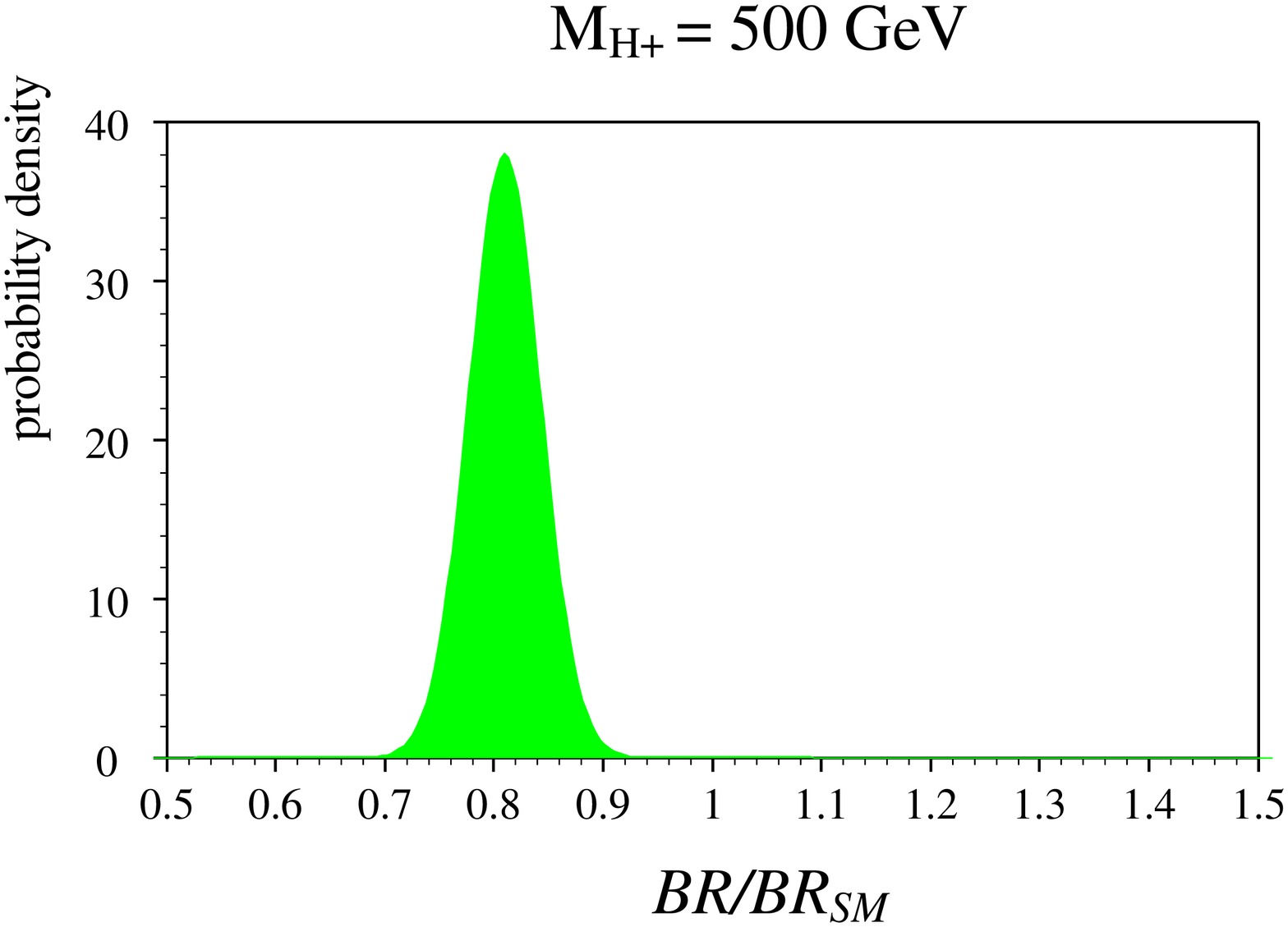}
    \caption{\label{fig:btaunu-sig}
      Distribution of $R= {\cal B}(B \to \ell \nu)/{\cal B}_{SM}(B \to\ell \nu)$
      in 2HDM, using $m(H^+)$=500GeV and $\tan\beta$=30 as it would be measured
      in 5 years at \superb.
    }
  \end{center}
\end{figure*}

The most striking feature of this result is that SPS4 is already ruled out by the present measurement
of ${\cal B}(B \to s\gamma)$ with high significance, showing the impact of flavor observables on the SUSY parameter space even in a MFV case. Indeed, from Eqs.~(\ref{eq:btosgamma:HFAG}) and (\ref{eq:btosgamma:NNLL}) one obtains R$^\mathrm{exp}(B\to X_s\gamma)=1.13\pm 0.12$. In the absence of a detailed analysis, we have not  attempted an estimate of the errors associated with the predictions of Table~\ref{tab:spspoints} at SPS4.
Nevertheless, even assuming an error of 50\%, much larger than the other points, R($B\to X_s\gamma$)=0.25 at SPS4  is more than 5$\sigma$ away from the present experimental value.
SPS5 is marginally compatible with present measurement of ${\cal B}(B \to s\gamma)$. Clearly this point will produced a measurable effect on ${\cal B}(B \to s\gamma)$ at \superb. Considering these results, it is not surprising that the recent MSSM analysis in \cite{Buchmueller:2007zk} found that the best fit to present data, using ${\cal B}(B \to s\gamma)$ among the constraints, resembles SPS$1a$.

SPS$1a$ is clearly the least favorable point from the flavor point of view. However, even here
\superb\ could see a definite pattern of 1-2 $\sigma$ deviations from the Standard Model in
${\cal R}(B \to \tau\nu)$, ${\cal R}(b \to s\gamma)$ and ${\cal R}(B \to X_s l^+l^-)$,
although this does depend, to some extent, on improvements in theory.
In any case, \superb\ flavor measurements are required to establish that
the New Physics flavor couplings are small as predicted by mSUGRA, since
LHC alone cannot establish which model is behind the measured SUSY spectrum.

One of the lessons of this exercise is that the benchmarks for flavor physics,
if needed, should mainly address the problem of defining a ``typical'' non-minimal flavor
structure with an economical number of parameters. A possible way to further investigate is using
models of SUSY-breaking as discussed below.

\noindent
\subsection*{Update on the $B \to \ell \nu$ predictions }
\label{sec:B:pheno:btaunu}

We update in this section the analysis of the decay $B \to \ell \nu$ in the
2HDM. The case of SUSY, discussed in the CDR, is very similar.

Figure~\ref{fig:btaunu} shows a comparison of the exclusion plot in the
$m(H^+)$--$\tan\beta$ plane coming from a measurement of ${\cal B}(B \to \tau
\nu)$ with different data samples, 2 ab$^{-1}$, 10 ab$^{-1}$, 75 ab$^{-1}$ and
200 ab$^{-1}$, assuming that the result is consistent with the Standard Model.

Note that moving from 10 ab$^{-1}$ to 75 ab$^{-1}$ the channel $B \to \mu \nu$
begins to give a significant contribution to the average, and the scale is then
larger than the naive statistical gain. With further increases in the integrated
luminosity beyond 75 ab$^{-1}$, $B \to \tau \nu$ become systematics-dominated
but $B \to \mu \nu$ still scales with statistics.

To give an example of a positive signal as seen at \superb,
Figure~\ref{fig:btaunu-sig}
shows the deviation of ${\cal B}(B \to \ell \nu)$ with
respect to its Standard Model value computed in the 2HDM for $m(H^+)$=500GeV and
$\tan\beta$=30 as it would be measured with a sample of 75 ab$^-1$. It's clear that the deviation is established with very high significance.

\subsection*{SUSY-breaking models}
\label{sec:B:pheno:SUSY-b}

Within supersymmetric extensions of the Standard Model, the flavor structure is directly
linked to the crucial question of the supersymmetry-breaking mechanism.
Indeed,
the bulk of soft SUSY-breaking terms is given by the sfermion bilinear and
trilinear couplings, which are matrices in flavor space.
Thus, once some SUSY particles have been found, the measurement of the flavor
sector can provide important information for distinguishing among models of
supersymmetry. This is a manifestation of the complementary nature of flavor
physics and collider physics. At the LHC direct searches for supersymmetric
particles are essential in establishing the existence of new physics. On the
other hand, there are a variety of possibilities for the origin of SUSY
breaking
and of flavor structures within supersymmetry. Flavor physics provides an
unique tool with which fundamental questions, such as how supersymmetry is
broken, can be addressed.

\begin{figure}[!t]
  \begin{center}
    \hspace{-0.1cm}\includegraphics[width=0.45\textwidth]{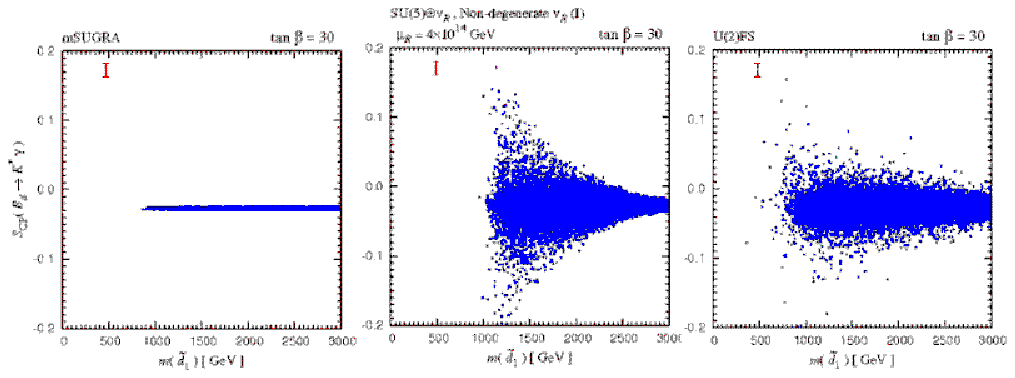}
    \hspace{-0.5cm}\includegraphics[width=0.45\textwidth]{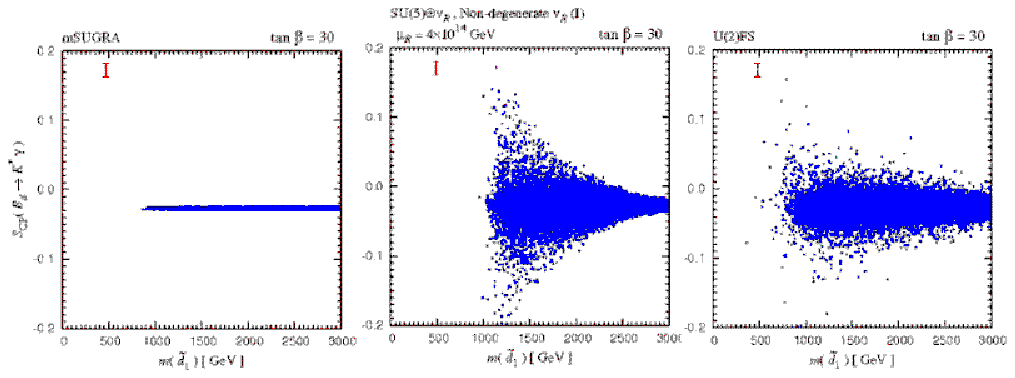}
    \caption{
      Time-dependent \CP\ asymmetry of $B \rightarrow  K_s \pi^0 \gamma$
      and the difference between the time-dependent asymmetries of
      $B \to \phi K_S$ and $B \to \jpsi\,K_S$ modes
      for three SUSY breaking scenarios:
      mSUGRA(left), SU(5) SUSY GUT with right-handed neutrinos
      in non-degenerate case (middle),
      and MSSM with U(2) flavor symmetry (right).
      The \superb\ sensitivities are also shown.
    }
    \label{Okada1}
  \end{center}
\end{figure}

\begin{figure}[!b]
  \begin{center}
     \includegraphics[width=0.5\textwidth]{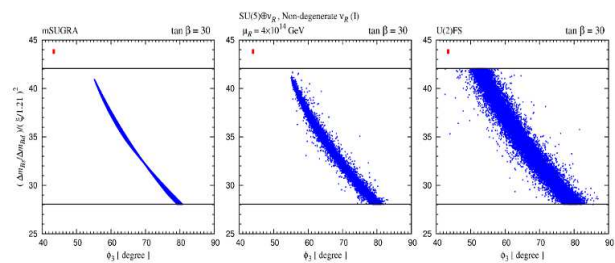}
    \caption{
      Correlation of $\Delta m_s/\Delta m_d$ and $\gamma$ ($\phi_3$)
      for three SUSY breaking scenarios:
      mSUGRA(left), SU(5) SUSY GUT with right-handed neutrinos
      in non-degenerate case (middle),
      and MSSM with U(2) flavor symmetry (right).
    }
    \label{Okada2}
  \end{center}
\end{figure}

A comprehensive analysis of the flavor patterns generated in
SUSY models with different SUSY-breaking sector has been
recently presented in Ref.~\cite{Goto:2007ee}. The models under study are
mSUGRA, MSSM with U(2) flavor symmetry, MSSM with right-handed neutrinos,
and SU(5) SUSY-GUT with right-handed neutrinos. Different scenarios for
the neutrino mass spectrum and Yukawa couplings have also been considered.
For our purpose, it is sufficient to consider a few examples. We refer the
reader
to the original publication for all the details.

Figs.~\ref{Okada1}--\ref{Okada2} from Ref.~\cite{Goto:2007ee} are examples of the power of \superb\ in
discriminating
different SUSY-breaking scenarios.  Additional information can certainly be
obtained
from a systematic study of correlations among flavor observables. It is
interesting to
notice that the plot in Fig.~\ref{Okada2} calls for a determination of
$\gamma$ ($\phi_3$)
with a sub-degree precision, which could be obtained at \superb\ with 100
$ab^{-1}$.

%
%
%

\section{Interplay of flavor and high~$p_T$ physics}
\label{sec:B:interplay}

In this section we want to report some result of the recent
workshops ``Flavour in the LHC era''~\cite{delAguila:2008iz,Buchalla:2008jp,Raidal:2008jk} from the
perspective of \superb.

We have already commented on the complementarity of the
physics goals of flavor and high~$p_T$ physics, which are
both necessary to identify the structure of the New Physics models.

Three analyses out of these reports should demonstrate the importance of
the interplay in our future new physics search:

In the context of this workshop the study of several SUSY-breaking models,
along the same lines of the previous section, have been presented to show
the capability of combined flavor and high~$p_T$ data in identifying the
SUSY-breaking mechanism.

Another study that started at the workshop concerns the effects of
flavor violation on direct searches at LHC, which are often not fully
taken into account.
It has been shown that flavor violation could, in some cases,
change the decay chains used at LHC to reconstruct the New Physics mass
spectrum, possibly making the analysis more involved~\cite{delAguila:2008iz,Hurth:2003th}.

The workshop result most relevant to Super$B$ physics comes from
a first attempt at combining of flavor and high~$p_T$ physics on the
same New Physics parameter space.
Based on existing flavor physics and high-energy computer codes,
a so-called master tool was developed which combines calculations from
both low-energy and electroweak observables in one common code.
The details of the analysis presented at the workshop can
be found in~\cite{Buchalla:2008jp}.

The complementarity of flavor physics and high~$p_T$ physics is
shown in Figure ~\ref{fig:tanbeta_a}. It is clearly demonstrated that,
without the inclusion of  both the flavor and
electroweak constraints, the parameters  $\tan\beta$ and $M_A$ are much
less well-determined. It can be seen, as well, that LHC mainly constrains the
mass, whereas flavor physics constrains the flavor coupling (\ie\
~the $\tan\beta$-enhanced Yukawa coupling).
Even in a model such as CMSSM with only a few New Physics parameters, both constraints
are required to effectively bound the parameter space.

\begin{figure}[!htb]
\begin{center}
\epsfig{file=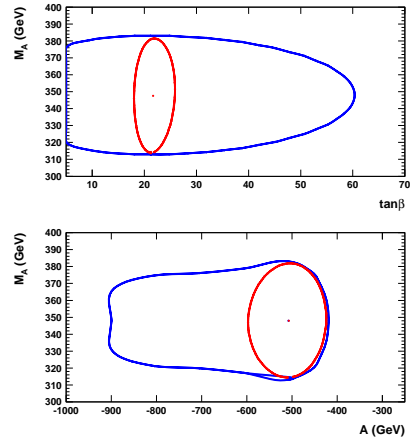, scale=0.3}
\caption{The red (clear) contour corresponds to the LHC scenario that includes the low-energy and
electroweak constraints, while the blue (darker) contour makes the same
assumptions about the assumed LHC discoveries, but does not include any
external constraints.}
\label{fig:tanbeta_a}
\end{center}
\end{figure}

A working group on the ``Interplay between high-$p_T$ and flavor physics''
has been set up~\cite{interplay1}; the first meeting was held at CERN in December 2007~\cite{interplay2}.




\onecolumngrid\twocolumngrid\hbox{}

\setcounter{section}{0}
\rhead[\fancyplain{}{\bf Charm Physics}]%
      {\fancyplain{}{\bf\thepage}}
\clearpage
{\centerline{\rule[0in]{0.9\columnwidth}{2pt}}
\vspace {-0.7cm}
\part*{\centerline{Charm Physics}}
\vskip -8pt
{\centerline{\rule[ 0.15in]{0.9\columnwidth}{2pt}}



\bigskip
\smallskip

New Physics, in general, generates flavor-changing neutral currents (FCNC). Those could be much less suppressed in the up-type than
the down-type quark sectors. Among the up-type quarks, only charm  allows the {\em full} range of probes for FCNC, and thus
New Physics, in oscillation phenomena, in particular those involving $\CP$ violation. The Standard Model makes nontrivial predictions for $\CP$ violation
in charm transitions: {\em direct} $\CP$ violation should occur
only  in Cabibbo-suppressed
modes at an observable level $\sim$ ${\cal O}(10^{-3})$.

The recent evidence for $\DzDzb$ oscillations -- with $x_D$, $y_D$ $\simeq$
0.005--0.01 --  does not prove the presence of New Physics. However it greatly
widens the stage on which $\CP$ violation can appear as a manifestation of New
Physics. Within the Standard Model, time dependent $\CP$ asymmetries could
reach the $10^{-5}$ [$10^{-4}$] level in Cabibbo-allowed and
once [doubly]-suppressed modes, whereas New Physics could enhance these asymmetries by almost three orders of magnitude. A search for New Physics
should then aim at sensitivity levels
of ${\cal O}(10^{-3})$ or better and ${\cal O}(10^{-2})$ or better
in Cabibbo-allowed or once-suppressed nonleptonic channels and in doubly Cabibbo-suppressed or
wrong-sign semileptonic modes, respectively. Signals for New Physics might actually be clearer in
$D$ than in $B$ decays: for while {\em conventional} New Physics scenarios tend to create larger effects in the latter than the former,
those signals must also contend with a much larger Standard Model ``background'' in the latter than the former.

These searches can be done at the $\FourS$ using $D^*$ tagging and tracking of the $D$ production
and decay vertices. Relatively short runs in the charm threshold region can provide unique and important information on strong phases needed
for a proper interpretation of results obtained in $\FourS$ runs.
They might reveal significantly enhanced effects that can be seen only in $e^+e^- \to \DzDzb$
exclusive production.



\section{New Physics in charm decays: mainly \CP violation}
\label{sec:charm-intro}

\subsection*{The landscape}

New Physics in general generates flavor changing neutral currents (FCNC). The Standard Model had to be crafted carefully to suppress them in the
strangeness sector down to the observed level. Those FCNC could actually be much less suppressed in the up-type than the down-type quark
sectors. Among the up-type quarks, only charm allows the {\em full} range of probes for FCNC, and thus, New Physics in oscillation
phenomena, in particular those involving $\CP$ violation: (i) Top quarks decay before they can hadronize;
without top hadrons $T^0$ oscillations cannot occur. Furthermore the sheer size of phase space in top decays greatly reduces
the coherence between different amplitudes needed to make
direct $\CP$ violation observable. (ii) Hadrons built with $u$ and $\bar u$ quarks, like the $\pi^0$ and
$\eta$, are their own antiparticle; thus there can be no $\pi^0 - \pi^0$ etc. oscillations as a matter
of principle. They also decay very rapidly. In addition, they possess so few decay channels that
$C\!P T$ invariance largely rules out $\CP$ asymmetries in their decays.


Strong evidence for $\DzDzb$ oscillations has been recently found~\cite{ref:hfag}. The most recent averages for the mixing parameters are
\begin{eqnarray}
x_D &\equiv& \frac{\Delta M_D}{\bar \Gamma_D} = 0.0097^{+0.0027}_{-0.0029}~, \\
y_D &\equiv& \frac{\Delta \Gamma_D}{2\bar \Gamma_D} = 0.0078^{+0.0018}_{-0.0019}~.
\end{eqnarray}
According to our present understanding -- or lack thereof -- these quantities could be produced by Standard Model dynamics, yet $x_D$ could still
harbour substantial contributions from New Physics. It will require a
theoretical breakthrough to resolve this ambiguity in the interpretation of the data.

We will be on much firmer ground in interpreting $\CP$ asymmetries. For on one hand, $\DzDzb$
oscillations greatly widen the stage on which $\CP$ violation can appear as a manifestation of New Physics; on the
other hand, the Standard Model makes nontrivial predictions for $\CP$ violation in charm transitions. In CKM dynamics there is a weak
phase in $\Delta C =1$ transitions
entering (in the Wolfenstein representation) through $V_{cs}$, yet it is highly diluted:
\beq
V_{cs} \simeq 1 - \frac{1}{2} \lambda ^2 - i \eta A^2 \lambda ^4 \simeq 0.97 - 6\cdot 10^{-4} i~.
\eeq
Furthermore two different, yet coherent, amplitudes must contribute to the same channel to produce a direct $\CP$ asymmetry.
Within the Standard Model this can happen at an observable level only in
Cabibbo-suppressed modes -- even in these channels, $\CP$ asymmetries can be no more than ${\cal O}(10^{-3})$. This means that any
observation of direct $\CP$ violation in Cabibbo-allowed or doubly-suppressed channel establishes the
intervention of New Physics. The only exception to this general rule is provided by modes like
$D^{\pm} \to K_S\pi^{\pm}$, where one becomes sensitive to (i) the interference between
$D^+ \to \bar K^0 \pi^+$ and $D^+ \to K^0 \pi^+$ and (ii) the slight $\CP$ impurity in the $K_S$ state. The latter effect dominates,
inducing a $\CP$ asymmetry of $3.3 \cdot 10^{-3}$.

With $x_D$, $y_D \sim$ 0.005 -- 0.01 the possibilities for $\CP$ asymmetries proliferate. In addition
to the aforementioned direct $\CP$ violation one can encounter {\em time dependent} $\CP$ asymmetries.
The latter can be induced by $\CP$ violation in $\Delta C=2$ dynamics, or even by $\CP$-conserving
contributions to the latter that can make the weak phase in a $\Delta C=1$ amplitude observable.
In both cases an educated Standard Model guess points to time-dependent $\CP$ asymmetries of order
$10^{-3} \sim x_D \sim 10^{-5}$.

\subsection*{The menu}

There are three classes of $\CP$ asymmetries:
\begin{enumerate}
\item
Direct $\CP$ violation can lead to a difference in the rates for $D \to f$ and $\bar D \to \bar f$:
\beq
|A_f| \equiv |A(D \to f)| \neq |\bar A_{\bar f}| \equiv |A(\bar D \to \bar f)|~.
\label{ANEQBARA}
\eeq
Strong phase shifts due to final state
interactions, are required to produce such asymmetries in partial widths. Since charm decays proceed in an environment populated by many resonances,
this requirement will not, in general, represent a limiting factor; it might make, however, the interpretation of signals a more complex task.
\item
Indirect $\CP$ violation -- {\it i.e.}, that which occurs only in $\Delta C=2$ transitions. One measure for it
is provided by
\beq
|q/p| \sim 1 + \frac{\Delta \Gamma_D}{\Delta M_D}{\rm sin}\phi_{\rm weak}  \neq 1 ~.
\label{QOVERP}
\eeq
The  same educated Standard Model guess mentioned above points to $|1 - |q/p|| \sim {\rm several}\times 10^{-4}$.
One should note here that the factor $\Delta \Gamma_D/\Delta M_D$ apparently is close to unity and thus provides no suppression
to this observable, unlike the case of $B^0$ mesons. Thus one has practically undiluted access to a weak phase due to the
intervention of New Physics in $\DzDzb$
oscillations. As discussed below,
such an asymmetry can be searched for cleanly in semileptonic decays of neutral $D$ mesons. While we already know the ratio of
wrong-sign leptons is small, their $\CP$ asymmetry could conceivably be as large as several percent!  While the rate of wrong-sign
leptons oscillates with time, the $\CP$ asymmetry does not.

\item
\CP violation in the interference between mixing and decay: In qualitative analogy to $B_d \to \jpsi \KS$, a time-dependent $\CP$ asymmetry can arise due to an interference between an oscillation
and decay amplitude:
\beq
{\phi_f} = {\rm arg}     \left(     \frac{q}{p}  \frac{ {\bar A}_{\bar f} }{A_f}        \right)      \ne 0 \; .
\label{INTER}
\eeq
A $\CP$ asymmetry generated by $\phi_f \neq 0$ is also proportional to sin$\Delta M_D t
\simeq x_D (t/\tau _D)$ and thus effectively bounded by $x_D$; {\ie}, the present lack of a signal for a
time-dependent $\CP$ asymmetry in $D^0 \to K^+K^-$ on about the 1\% level is not telling at all, in view
of $x_D \leq 1\%$. Yet any improvement in experimental sensitivity could reveal a genuine signal.
\end{enumerate}

Searching for $\CP$ violation in charm decays is not a ``wild goose chase''. We know that baryogenesis requires the presence of
$\CP$-violating New Physics. Signals for such New Physics might actually be clearer in
$D$ than in $B$ decays: for while {\em conventional} New Physics scenarios tend to create larger effects in the latter than the former,
those signals would also have to contend with a much larger Standard Model ``background'' in the latter than the former; \ie, the
theoretical ``signal-to-noise'' ratio could be better in charm decays.

The required searches can be undertaken very profitably in runs at the $\FourS$
by tagging the \Dz flavor at production time using $D^{*+} \to D^0 \pi^+$ decays
and reconstructing the proper decay time and its error.  This is done by tracking the $D$ production and decay vertices with
constraints provided by the position and size of the tight $e^+ e^-$ interaction region.
Relatively short runs in the charm threshold region, \eg, \psiprpr, can provide unique and important information on strong phases needed
for a proper interpretation of results obtained in $\FourS$ runs. In the latter \Dz flavor tagging exploits the
quantum correlations at \psiprpr; the poor proper time resolution (about the \Dz lifetime)
will make time-dependent measurements challenging.

{In summary}: Comprehensive and precise studies of $\CP$ invariance in charm decays provide sensitive probes for the presence of New Physics.
\begin{itemize}
\item
`Comprehensive' means that one analyses nonleptonic as well as semileptonic channels on all
Cabibbo levels in as many modes as possible; \ie, including final states containing neutrals.
\item
`Precise' means that one achieves sensitivity levels of $10^{-3}$ or better.

\end{itemize}

Charm decays provide another highly promising avenue towards finding $\CP$ violation, namely in final state distributions, rather
than in partial widths considered so far. This issue will be addressed separately
below.

\subsection*{Side remarks on rare decays}

The obvious motivation for measuring the branching fractions for \hbox{$D^+/D^+_s \to \mu^+ \nu$, $\tau^+\nu$} decays is to extract the decay
constants $f_D$ and $f_{D_s}$ in order to compare them with lattice QCD calculations and, hopefully, to validate
these calculations with high accuracy. A more ambitious goal is to probe for contributions from a charged Higgs field, as an indication of New Physics.

The mode $D^0 \to \mu^+\mu^-$ arises within the Standard Model mainly through a two photon intermediate
state -- $D^0 \to \gamma^* \gamma^* \to \mu^+\mu^-$ -- and can reach the $10^{-12}$ level. With the present
experimental upper bound of $1.3 \times 10^{-6}$ there is a search window for New Physics of six orders
of magnitude. Multi-Higgs models or SUSY models with $R$ parity breaking could conceivably induce a signal in
a range as ``large'' as few$\times 10^{-8}$ and $10^{-6}$, respectively.

Channels such as $D\to \gamma h$, $l^+l^-h$, $l^+l^-h_1h_2$, with $h$ denoting a hadron, receive
relatively sizable contributions within the Standard Model from long distance dynamics. Thus a search for New Physics contributions are
not very promising there, unless one can measure
precisely the lepton spectra in the final states.

One can probe a rather exotic variant of New Physics by searching for two-body modes
$D^+ \to K^+/\pi^+ f$; the charge neutral $f$ denotes a `familon', which could arise as the Nambu-Goldstone boson resulting
from the spontaneous breakdown of a global family symmetry. It has been searched for in $K^+$ and $B^+$ decays,
but apparently not yet in $D^+$ decays.


\section{\DzDzb mixing at \FourS and \psiprpr energies}
\label{sec:charm-mixing}

The parameters describing charm mixing can be measured in time-dependent
studies of $D$ mesons or with time-integrated observables of $D$ mesons
produced coherently near charm threshold.

The time-dependent $\DzDzb$ mixing formalism and a summary of recent
experimental results can be found in Ref.~\cite{pdg2008}.
Many different charm decay modes can be used to search for charm mixing.
\begin{itemize}
\item The appearance of ``wrong-sign'' kaons in semileptonic decays would provide
direct evidence for $\DzDzb$ oscillations (or another process of beyond Standard
Model origin).
\item The most precise limits are obtained by exploiting the time-dependence of
$D$ decays produced in $e^+e^-$ collision near 10~GeV.
\begin{itemize}
\item The wrong-sign hadronic decay $\DZ\to K^+\pi^-$ is sensitive to linear combinations of the mass and lifetime differences, denoted $x^{\prime 2}$ and $y^{\prime}$.
The relation of these parameters to $x_D$ and $y_D$ is controlled by a strong
phase difference $\delta_{K\pi}$.
\item Direct measurements of $x_D$ and $y_D$ independent of unknown
strong interaction phases can also be made using time-dependent studies of
amplitudes present in multi-body decays of the $\DZ$, for example,
$\DZ \to K^0_S\pi^+\pi^-$.
\item Direct evidence of $y_D$ can also appear through
lifetime differences between decays to $\CP$ eigenstates.
The measured quantity in
this case $\yCP$, is equivalent to $y_D$ in the absence of $\CP$ violation.
\end{itemize}
\item Another approach is to study quantum correlations near charm threshold
\cite{Asner:2005wf} in
$e^+e^- \to \DzDzb(\pi^0)$ and $e^+e^- \to \DzDzb\gamma(\pi^0)$ decays, which yield
$C$-odd and $C$-even $\DZ\DZB$ pairs, respectively. Taken together, the
time-integrated decay rate to semileptonic, $K\pi$, and $\CP$ eigenstates provide
sensitivity to $x_D$, $y_D$, and $\cos\delta_{K\pi}$.
\end{itemize}

Several recent results provide evidence that charm mixing is at the upper
end of the range of Standard Model predictions.

\babar~\cite{Aubert:2007wf} and CDF~\cite{Aaltonen:2007uc}
find evidence for oscillations in $\DZ \to K^+\pi^-$, with
3.9$\sigma$ $(\Delta {\rm Log} {\cal L})$ and 3.8$\sigma$ (Bayesian), respectively.
The most precise measurement is from Belle which
excludes $x^{\prime 2}=y^\prime=0$ at 2.1$\sigma$~\cite{Zhang:2006dp}
(Feldman-Cousins).

Belle~\cite{Staric:2007dt} and \babar~\cite{Aubert:2007en}
see 3.2$\sigma$ and 3$\sigma$ effects, respectively,
for $\yCP$ in $\DZ\to K^+K^-$.
The most precise measurement of $y_D$ is in $\DZ \to K^0_S\pi^+\pi^-$
from Belle~\cite{Abe:2007rd} and is only
1.2$\sigma$ significant. From the same analysis, Belle also reports a
2.4$\sigma$ significant result for $x_D$.
The current situation would greatly benefit from more precise
knowledge of the strong phase difference $\delta$; this would allow
one to unfold $x_D$ and $y_D$ from the $D^0 \to K^+\pi^-$ measurements of
$x^{\prime2}$ and $y^\prime$,
and directly compare them to the $D^0 \to K^0_S \pi^+\pi^-$ results.

All mixing measurements can be combined to obtain
world average (WA) values for $x$ and $y$. The Heavy Flavor
Averaging Group (HFAG) has done such a combination~\cite{Schwartz:2007fw,Schwartz:2008}.
The resulting $1\sigma$-$5\sigma$ contours
are shown in Fig.~\ref{fig:hfag_mixing_contour} and Fig.~\ref{fig:hfag_cpv_contour}.
The fits exclude the no-mixing
point ($x\!=\!y\!=\!0$) at $6.7\sigma$ for both the no $\CP$ violation scenario
and the case allowing for $\CP$ violation. One-dimensional
likelihood functions for parameters are obtained by allowing, for
any value of the parameter, all other fit parameters to take their
preferred values. The resulting likelihood functions give central
values, 68.3\% C.L. intervals, and 95\% C.L. intervals as listed
in Table~\ref{tab:HFAG}.

From the results of the HFAG averaging, we can conclude  the following:
\begin{itemize}
\item
The experimental data consistently indicates
that $D^0$ mesons undergo mixing. The effect is presumably dominated
by long-distance processes, and unless $|x|\gg |y|$, it may
be difficult to identify New Physics from mixing alone.
\item
Since $\yCP$ is positive, the $\CP$-even state is shorter-lived,
as in the $K^0$-$\overline K{^0}$ system. However, since $x$ appears to be
positive, the $\CP$-even state is heavier,
unlike in the $K^0$-$\overline K{^0}$ system.
\item
There is no evidence yet for $\CP$ violation in the $D^0$-$\overline D{^0}$
system.
\end{itemize}
\begin{table}[!htb]
\caption{\label{tab:HFAG}
HFAG Charm Mixing Averages.}
\begin{tabular}{clcc} \hline\hline
Fit & Parameter & HFAG Average & 95\% C.L. Interval\\
\hline
$C\!PV$ & $x(\%)$ & $0.97^{+0.27}_{-0.29}$ & (0.39:1.48) \\
 & $y(\%)$ & $0.78^{+0.18}_{-0.19}$ & (0.41:1.13) \\
 & $R_D(\%)$ &  $0.335\pm 0.009$ & (0.316:0.353) \\
  & $\delta_{K\pi}(^\circ)$ & $21.9^{+11.5}_{-12.5}$ & (-6.3:44.6) \\
 & $\delta_{K\pi\pi^0}(^\circ)$ &  $32.4^{+25.1}_{-25.8}$ & (-20.3:82.7) \\
 & $A_D(\%)$ &  $-2.2\pm 2.5$ & (-7.10:2.67) \\
 & $|q/p|$ & $0.86^{+0.18}_{-0.15}$ & (0.59:1.23) \\
 & $\phi(^\circ)$ & $-9.6^{+8.3}_{-9.5}$ & (-30.3:6.5) \\ \hline\hline
\end{tabular}
\end{table}


\begin{figure}[!htb]
\includegraphics[width=0.9\linewidth]{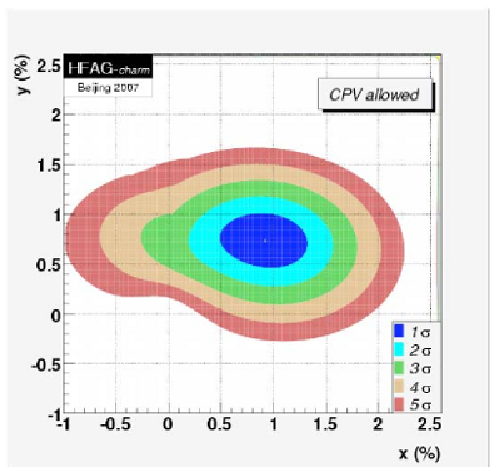}
\caption{\label{fig:hfag_mixing_contour}
Two-dimensional $1\sigma$-$5\sigma$ contours
for $(x,y)$, obtained from a global fit to the measured observables
for $x$, $y$, $|q/p|$, ${\rm Arg}(q/p)$,
$\delta_{K\pi}$, $\delta_{K\pi\pi^0}$, and $R_D$
from
measurements of $D^0\rightarrow K^+\ell\nu$, $D^0\rightarrow h^+h^-$,
$D^0\rightarrow K^+\pi^-$, $D^0\rightarrow K^+\pi^-\pi^0$,
\hbox{$D^0\rightarrow K^+\pi^-\pi^+\pi^-$,} and $D^0\rightarrow K^0_S\pi^+\pi^-$ decays,
and double-tagged branching fractions measured
at the $\psi(3770)$ resonance (from HFAG~\protect{\cite{hfag-charm}}).}
\end{figure}

\begin{figure}[!htb]
\includegraphics[width=0.9\linewidth]{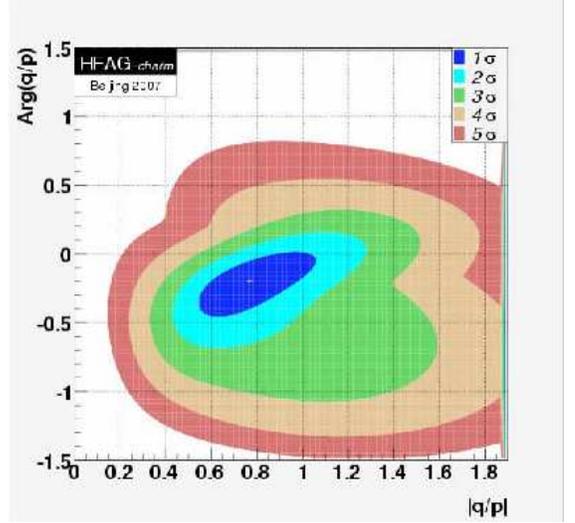}
\caption{
Two-dimensional $1\sigma$-$5\sigma$ contours
for $(|q/p|$, ${\rm Arg}(q/p))$, obtained from a global fit to the measured observables
for $x$, $y$, $|q/p|$, ${\rm Arg}(q/p)$,
$\delta_{K\pi}$, $\delta_{K\pi\pi^0}$, and $R_D$
from
measurements of $D^0\rightarrow K^+\ell\nu$, $D^0\rightarrow h^+h^-$,
$D^0\rightarrow K^+\pi^-$, $D^0\rightarrow K^+\pi^-\pi^0$,
$D^0\rightarrow K^+\pi^-\pi^+\pi^-$, and $D^0\rightarrow K^0_S\pi^+\pi^-$ decays,
and double-tagged branching fractions measured
at the $\psi(3770)$ resonance (from HFAG~\protect{\cite{hfag-charm}}).}
\label{fig:hfag_cpv_contour}
\end{figure}

{ The interpretation of the new results in terms of New Physics is inconclusive.
It is not yet clear whether the effect is caused by $x_D \neq 0$ or $y_D \neq 0$ or both, although the latter is favored,
as shown in Table~\ref{tab:HFAG}. Furthermore, there is no single 5$\sigma$
observation of charm mixing nor is one anticipated from the current $B$ Factories.
This situation will be remedied by results anticipated from \superb.
Table~\ref{tab:mixtab2} shows the sensitivity to mixing in $\DZ\to K^+\pi^-$, $K^+K^-$, and $K^0_S\pi^+\pi^-$ channels from the $\FourS$ data
is in excess of 5$\sigma$ if the lifetime and
mass differences in the $\DZ$ system lie at the upper end of the
range of Standard Model predictions.

Table~\ref{tab:mixtab2} also shows the sensitivity to mixing from two months of running at charm threshold. The sensitivity to the mixing parameters is comparable to
five years at $\FourS$, with different sources of systematic uncertainties.
The $\psiprpr$ data provides unique sensitivity to $\cos\delta_{K\pi}$.
Although
$\cos\delta_{K\pi}$
can be determined from a global fit to $\FourS$ results, the
direct measurement from $\psiprpr$ data allow $y^\prime$ and $x^{\prime 2}$
determined from $\DZ\to K^+\pi^-$ to contribute to the precision determination
of $x$ and $y$.

Although the $D$ mesons from $\psiprpr$ decay are produced nearly at rest in
the center-of-mass frame, the asymmetric $e^+e^-$ collisions make
time-dependent mixing analyses possible.
However, since the production rate of
charm during threshold running and $\FourS$ running is comparable, the
statistical power of the time-dependent analyses near threshold is small.

\begin{table*}[!htb]
  \caption{
    Expected precision ($\sigma$) on the measured
    quantities using methods described in the text for \superb\ with an integrated
    luminosity of 75~ab$^{-1}$ at \superb\ at 10~GeV, 300~fb$^{-1}$
    ($\sim$ two months) running at charm threshold with \superb, and
   LHC$b$ with 10~fb$^{-1}$\protect{\cite{LHCb_Spradlin}}.
  }
  \label{tab:mixtab2}
\begin{tabular}{ccccc}
\\ \hline\hline
Mode & Observable & $\FourS$  & $\psi(3770)$ & LHC$b$ \\
     &            & (75~ab$^{-1}$) &  (300~fb$^{-1}$) & (10~fb$^{-1}$) \\ \hline
$\DZ\!\to\! K^+\pi^-$ & $x^{\prime 2}$  & $3\times 10^{-5}$ &  &  $6\times 10^{-5}$\\
                  & $y^{\prime}$   & $7\times 10^{-4}$ & & $9\times 10^{-4}$\\
$\DZ\!\to\! K^+K^-$ & $\yCP$ & $5\times 10^{-4}$ & & $5\times 10^{-4}$\\
$\DZ\!\to\! K^0_S\pi^+\pi^-$ & $x$ & $4.9\times 10^{-4}$ & & \\
                        & $y$  &  $3.5\times 10^{-4}$ & & \\
                        & $|q/p|$  &  $3\times 10^{-2}$ & & \\
                        & $\phi$  &  $2^\circ$ & & \\
$\psi(3770) \!\to\! \DZ\DZB$ & $x^2$  & &  $(1\!-\!2)\times 10^{-5}$ & \\
                         & $y$ & & $(1\!-\!2)\times 10^{-3}$ & \\
                & $\cos\delta$ & & $(0.01\!-\!0.02)$ & \\
\hline\hline
\end{tabular}
\end{table*}

A serious limitation in the interpretation of charm oscillations in terms of New Physics is the theoretical uncertainty on the Standard Model prediction.
However, the recent evidence for oscillations opens the window to
searches for $\CP$ asymmetries that do provide unequivocal New Physics
signals. The sensitivity to these New Physics signals is shown in
Fig.~\ref{fig:hfag_cpv_projection}.

\begin{figure}[!htb]
\includegraphics[width=0.9\linewidth]{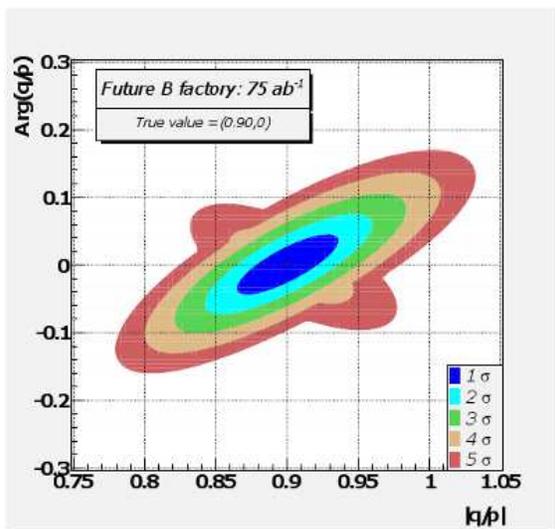}
\caption{\label{fig:hfag_cpv_projection}
Projected two-dimensional $1\sigma$-$5\sigma$ contours with 75 ab$^{-1}$
for $(|q/p|$, ${\rm Arg}(q/p))$, obtained from a global fit to the observables
for $x$, $y$, $|q/p|$, ${\rm Arg}(q/p)$,
$\delta_{K\pi}$, $\delta_{K\pi\pi^0}$, and $R_D$
from the sensitivity estimates in Table~\ref{tab:mixtab2}.
A ``true value'' of $|q/p|=0.90$, ${\rm Arg}(q/p)=0$ is assumed.}
\end{figure}





\section{\CP violation}
\label{sec:charm-CPV}

\subsection*{Direct $\CP$ violation}
\label{sec:charm-directCPV}

Searches for $\CP$ violation in $\Delta C=1$ transitions can be performed by measuring asymmetries in the partial widths or in final state distributions.

Golden modes
are the Cabibbo-suppressed decays $D^0 \to h^+ h^-$, $h=K,\pi$,
and the doubly Cabibbo-suppressed decay $D^0 \to K^+ \pi^-$.
These studies can be performed either time-integrated or by analyzing the time dependence
of the \Dz and \Dzb decay rates, although in both cases time-integrated asymmetries are measured.
Data at the \FourS provides the largest data sample
with excellent purities (as large as $\sim 99\%$). The contamination from \BB decays can be virtually eliminated by
imposing a $2.5$~\gevc cut on the $D$ momentum in the center-of-mass frame, which preserves more than 85$\%$ of signal events.

The most precise analysis to date~\cite{ref:babarDtoKKandDtoPiPi} compares time-integrated
$\Dz \to h^+ h^-$ and $\Dzb \to h^+ h^-$ rates,
$a_{\CP}^{hh} = [N_{\Dz} - N_{\Dzb}]/[N_{\Dz} + N_{\Dzb}]$, where $N_{\Dz}$ ($N_{\Dzb}$) is the
number of \Dz (\Dzb) mesons decaying into $h^+ h^-$ final state.
In this construction, all $\CP$ violation contributions, direct and
indirect are present. Direct $\CP$ violation in one or both modes would be signaled by a non-vanishing
difference between the asymmetries for $D^0 \to K^+ K^-$ and $D^0 \to \pi^+ \pi^-$, $a_{\CP}^{KK} \neq a_{\CP}^{\pi\pi}$ .
There are two main experimental challenges in these measurements.
Firstly, the experimental asymmetry in  \Dz flavor tagging.
This asymmetry is measured by determining the relative detection efficiency for soft pions in data, using
the Cabibbo-allowed decay $D^0 \to K^- \pi^-$ with (tagged) and without (non-tagged) soft-pion flavor tagging,
as a function of the pion-momentum  and the polar angle in the lab frame. For the azimuthal dependence, an
integrated scale factor is sufficient, since charm production is uniform in azimuth.
Since the reconstructed modes are $\CP$-even, this is the only detector asymmetry.
Secondly, the forward-backward (FB) asymmetry in $c\bar{c}$ production at \FourS,
a consequence of the $\gamma/Z^0$ interference and higher order QED corrections
(both at the percent level at this energy), coupled with the asymmetric acceptance of the detector,
which produces a difference in the number of reconstructed $D^0$ and $\bar{D}^0$ events.
This effect is directly measured by determining the number of \Dz and \Dzb events
(after soft pion asymmetry correction) as a function of $\cos\theta_D^{CM}$ and decomposing
these events  into even (representing the $\CP$ asymmetry and independent of $|\cos\theta_D^{CM}|$) and odd
(representing the FB production asymmetry) parts. The associated systematic uncertainties are therefore
not a limiting factor, and are mostly statistical in nature. Other potential
sources of uncertainty are highly suppressed because the
final states are reconstructed identically for $D^0$ and $\bar{D}^0$.
With a \superb\ data sample of 75 ab$^{-1}$, sensitivities at $3\times10^{-4}$ and $4\times10^{-4}$ level,
for $a_{\CP}^{KK}$ and $a_{\CP}^{\pi\pi}$ respectively, are foreseen.

A time-dependent $D$-mixing analysis of DCS (wrong sign) $\Dz \to K^+ \pi^-$ and $\Dzb \to K^- \pi^+$ decays can be
used to separate the contributions of DCS decays from $\DzDzb$ mixing, separately for \Dz and \Dzb.
A direct $\CP$ asymmetry can then be constructed from the difference of DCS \Dz and \Dzb decays,
$A_D = (R_{\Dz} - R_{\Dzb})/(R_{\Dz} + R_{\Dzb})$, where $R_{\Dz}(R_{\Dzb})$  is the \Dz(\Dzb) DCS rate.
The main experimental difficulties in this analysis are accurate proper time reconstruction and calibration,
together with asymmetry in the \Dz flavor tagging and the modeling of the differences between $K^+$ and $K^-$
absorption in the detector. At \superb, the much smaller luminous region and the
significantly enhanced vertexing capabilities provide proper time significances at the $10\sigma$ level
(3-4 times better than in \babar~\cite{ref:babarDtoKPi}, with decay length resolution of about 80~$\mu$m, $\sim 3\sigma$),
significantly reducing the systematic uncertainties associated with the modeling of the
long decay time component and possible biases.
Systematic uncertainties related to the asymmetry in the soft-pion tagging can be keep under
control using a similar procedure to that outlined above. Corrections 
due to the FB production asymmetry and kaon hadronic interactions can be performed
relying mainly on data, through untagged $\Dz \to K^- \pi^+$ and $\Dzb \to K^+ \pi^-$
decays measured as a function of $\cos\theta_D^{CM}$.
Scaling the statistical uncertainty from the \babar\ analysis to 75 ab$^{-1}$ we obtain a
sensitivity on $A_D$ of $4\times10^{-3}$. To reach or improve this sensitivity level, systematic uncertainties,
currently $15\times 10^{-3}$, will have to be reduced by a factor of five or better, which is feasible since the uncertainty of the systematic corrections scale with the size of the data sample.

For asymmetries in final state distributions, the simplest way is to compare $\CP$ conjugate Dalitz plots for 3-body decays.
Different regions of the Dalitz plot may exhibit $\CP$ asymmetries of varying signs that largely cancel out when one
integrates over the entire phase space, therefore subdomains of the Dalitz plot could contain significantly larger
$\CP$ asymmetries than the whole phase space. Since understanding the dynamics is not an easy goal to achieve,
one could try up to four strategies, three of which are model-independent. First, quantify differences
between the \Dz and \Dzb Dalitz plots in two dimensions. Secondly, look for differences in the angular moments
of \Dz and \Dzb intensity distributions. Thirdly, in a model-dependent approach, look for $\CP$ asymmetries
in the amplitudes describing intermediate states in the \Dz and \Dzb decays. Finally, look for the phase-space
integrated asymmetry. Asymmetries in the \Dz flavor assignment and FB production asymmetries only affect the last
method, and can be kept under control, as discussed above. From the pioneering \babar\ analysis using
$\Dz \to \pi^- \pi^+ \pi^0$ and $\Dz \to K^- K^+ \pi^0$~\cite{ref:babarDtoKKpi0andDtoPiPiPi0},
sensitivities at $3\times10^{-4}$ and $9\times10^{-4}$ level, respectively, are anticipated.

For more complex final states other probes have to be employed.
A golden example is discussed below.

%

\subsection*{Indirect $\CP$ violation at the  \FourS and \psiprpr}
\label{sec:charm-indirectCPV}

$\CP$ violation in mixing can be investigated from the data taken at the \FourS and at the \psiprpr resonances
in semi-leptonic transitions.
In both cases one measures an asymmetry from events in which the $D^0$ or $\bar D^0$, previously flavor tagged,
has oscillated (signaled as a wrong sign decay),
\begin{equation}
a_{SL} = \frac{N^{--}(t)-N^{++}(t)}{N^{--}(t)+N^{++}(t)} = \frac{|q|^4 - |p|^4}{|q|^4 + |p|^4}~,
\end{equation}
where $N^{--}$ ($N^{--}$) represents the number of
$\Dz \to \ell^- \nu X$ ($\Dzb \to \ell^+ \nu X$) decays when the other $D$ meson was tagged as \Dz (\Dzb) at production time.
Data at the \psiprpr benefit from a very clean environment with almost no background. Several decay
channels can be exclusively reconstructed
to increase the asymmetry sensitivity. Considering
the $D^0$ and $\bar D^0$ both decaying into $K^-\pi^+$, $K^-\pi^+\pi^0$, $K^-\pi^+\pi^+\pi^-$, $K^-e^+\nu$,
$K^-\mu^+\nu$, $K^{*-}e^+\nu$, $K^{*-}\mu^+\nu$, $K^{*-}e^+\nu$,  $\pi^- e^+\nu$, $\pi^-\mu^+\nu$, $K^-K^+$ and $\pi^-\pi^+$,
and using recent results for the $\DzDzb$ mixing parameters $x$ and $y$ \cite{ref:hfag}, a sensitivity to $\CP$
violation of 2.5$\%$ in one month of running at threshold is expected. The quantum correlation ensures that the
same-sign combinations
can only be due to mixing; thus hadronic modes can be treated like the semileptonic decays (no DCS contribution).
Control of systematic uncertainties is expected at the percent
level, dominated by channels with $\pi^0$ and $\nu$ particles~\cite{ref:cleoc_hbr,ref:cleoc_slbr}. Missing mass techniques with full reconstruction of
\psiprpr $\to D \bar D$ events, omitting one of the product particles, can be used to evaluate the accuracy in the reconstruction.
Large control samples of decay channels with unequivocal particle content like $D^0\to K^0_s\pi^+\pi^-$ and $D^+\to K^-\pi^+\pi^+$ will reduce
the uncertainty on PID efficiencies. Other sources of systematic uncertainties will also benefit from the precise measurement of the beam energy
and improved detector performance.

At the \FourS, the soft pion coming from $D^*$ decays ($D^{*+} \to D^0 \pi^+$)
can be used to tag the flavor of the $D^0$. The measurement of wrong sign leptons in
semileptonic decays then provides a clear signature of a mixed event. Data are taken from the continuum.
Background events from $B$ decays can be reduced by imposing a 2.5~\gevc cut on the $D$ momentum. 
With this method, the statistical sensitivity in the decay asymmetries would reach the $1\%$ level in one
year of data taking. Systematic uncertainties are foreseen to arise from the control of backgrounds and PID management
(mainly lepton identification), which will benefit from the vertex capabilities to suppress the background
and large control samples to study the PID.


\subsection*{$\CP$V in the interference of mixing and decay}
\label{sec:charm-interferenceCPV}





$\CP$ violation in the interplay of $\Delta C = 1,2$ dynamics can be searched for through time-dependent analyses
of $\Dz \to K^+ K^-$ and $\Dz \to \pi^+ \pi^-$ decays. $\CP$ violation and $\DzDzb$ mixing alter the decay time
distribution of $\Dz$ and $\Dzb$ mesons that decay into final states of specific $\CP$, and a time-dependent
analysis of the tagged \Dz and \Dzb intensities allows a measurement of the $\phi_f$.
To a good approximation, these decay time distributions can be treated as exponential, with effective
lifetimes \tauhhp and \tauhhm.
%

The effective lifetimes can be combined into the quantities $\yCP$ and $\Delta Y$:
\begin{displaymath}
\begin{array}{rcl}
\displaystyle y_{\CP}
= \frac{\tauKpi}{\langle\tauhh\rangle} - 1\;,
&&
\displaystyle\Delta Y = \frac{\tauKpi}{\langle\tauhh\rangle} A_\tau \;,
\label{eq:yCPcalc}
\end{array}
\end{displaymath}
where $\langle \tauhh \rangle = (\tauhhp+\tauhhm)/2$ and
$A_\tau = (\tauhhp-\tauhhm)/(\tauhhp+\tauhhm)$.
%
The golden mode is $\Dz \to \Kp\Km$, since the combinatorial background
is $\sim 10\times$ smaller than in the $\pi^+\pi^-$ channel, and the
selected sample is $\sim 2\times$ larger.  $\Dz \to \KS\phi$ instead has a
large ($\sim10\%$) contribution from $S$ wave, so it is better analyzed
using the Dalitz plot technique (see Sec.~\ref{sec:charm-3body}).

The \superb\ sensitivity to $\yCP$ and $\deltaY$ in the $KK$
and $\pi\pi$ modes can be extrapolated from the current \babar\
analysis~\cite{ref:babarDtoKKandDtoPiPi}, assuming that the systematic errors can be kept
under control. Provided that $\CP$ violation in mixing is small,
the sensitivity to the $\CP$-violating phase is dominated by
the first term in the expression for $\yCP$ and $\deltaY$.
\begin{eqnarray*}
2 y_{\CP}    = (|q/p|+|p/q|) y \cos\phi - (|q/p|-|p/q|) x \sin\phi, \\
2 \deltaY = (|q/p|-|p/q|) y \cos\phi - (|q/p|+|p/q|) x \sin\phi,
\end{eqnarray*}
therefore we can estimate the sensitivity as
$\delta(\cos \phi)\simeq \delta (y_{\CP}) /y \simeq  3\times 10^{-4} / y$,
$\delta(\sin \phi)\simeq \delta (\deltaY) / x \simeq  3\times 10^{-4} / x$.


Most of the systematic errors affecting the signal cancel in the
lifetime ratio. The errors associated with the background are
unrelated between \Dz and \Dzb and do not cancel; however they do improve
with statistics. In addition, the superior resolution of
the vertex detector will further reduce the systematic errors
associated with the position measurement. We therefore expect that the
systematic errors can be kept under control.

One underlying assumption in the recent \babar\
analysis~\cite{ref:babarDtoKKandDtoPiPi} is that the resolution bias
is the same for all the channels ($K\pi$, $KK$, $\pi\pi$) and does
not depend on the polar angle $\theta$. This could introduce a bias in
the measurements, because of the different polar angle acceptance in
the various channels. With a higher statistics sample, however, this
systematic effect can be overcome by splitting the sample into polar angle
(or other variable) intervals.  The production asymmetry is not
important with \babar\ statistics, but could become significant
at sensitivities of the order of few $\times 10^{-4}$. However this
can be handled using control samples, such as the untagged \Dz, which
have about 5 times more events (assuming \Dz and \Dstar have the same
asymmetry), as discussed in Sec.~\ref{sec:charm-directCPV}.

%



\subsection*{$T$ odd correlations}
\label{sec:Todd}


All $\CP$ asymmetries observed so far have surfaced in partial widths -- with one notable exception:
the forward-backward asymmetry $\langle A \rangle$ in the $\pi^+\pi^-$ and $e^+e^-$ planes in
$K_L \to \pi^+\pi^-e^+e^-$. $\langle A \rangle \simeq 14\%$ had been
predicted -- and confirmed by experiment -- as being driven by the indirect $\CP$ impurity $|\eta_{+-}| \simeq 0.23 \%$.
The reason for this magnification by two orders of magnitude is well understood:
$\langle A \rangle$ is induced by the interference between a $\CP$-violating and a $\CP$-conserving amplitude,
both of which are suppressed, albeit for different reasons. This explains why the enhancement of the $\CP$ asymmetry
comes at the expense of the branching ratio, which is about
$3\cdot 10^{-7}$; \ie, one has traded branching fraction for the size of the asymmetry.

It is possible that a similar effect and enhancement occurs in the analogous mode
$D_L \to K^+K^-\mu^+\mu^-$, where $D_L$ denotes the ``long-lived'' neutral $D$ meson. This mode
can be studied uniquely at \superb operating at the $\psi (3770)$ by
$\CP$-tagging the other neutral $D$ meson produced as a ``short-lived'' $D_S$:
\beq
e^+e^- \to \gamma^* \to \DzDzb \to [K^+K^-]_D D_L
\eeq

There is a more general lesson from the $\KL \to \pi^+\pi^-e^+e^-$ example, namely that
$\CP$ violation could surface in an enhanced  fashion in multi-body final states. This could turn
an apparent vice in charm decays -- the preponderance of multi-body final states -- into a virtue.
This issue will be addressed in detail in Sec.~\ref{sec:charm-3body}.

These considerations also apply to four-body modes, although less experience with such studies has been accumulated so far.
Some intriguing pilot studies have been performed on a comparison of
$D^0 \to f$ and $\Dzb \to f$, $f=K ^+K^- \pi^+\pi^-$ channels. Denoting by $\phi$ the angle between
the $\pi^+\pi^-$ and $K^+K^-$ planes, one has

\beq
\frac{d\Gamma}{d\phi}(D^0 \to f) =
\Gamma_1 {\rm cos}^2 \phi + \Gamma_2 {\rm sin}^2 \phi + \Gamma_3 {\rm cos} \phi {\rm sin}\phi~,
\eeq
\beq
\frac{d\Gamma}{d\phi}(\overline D^0 \to f) =
\overline \Gamma_1 {\rm cos}^2 \phi + \overline \Gamma_2 {\rm sin}^2 \phi - \overline \Gamma_3 {\rm cos} \phi {\rm sin}\phi~.
\eeq
%
Upon integrating over $\phi$, the $\Gamma_3$ and $\overline \Gamma_3$ terms cancel;
$(\Gamma_1, \Gamma_2) \neq (\overline \Gamma_1, \overline \Gamma_2)$ thus represents a
$\CP$ asymmetry in the partial widths. The $\Gamma_3$ and $\overline \Gamma_3$ terms can be projected out by integrating over two quadrants:
\begin{eqnarray}
\langle A \rangle = \frac{\int _0^{\pi/2}d\phi \frac{d\Gamma}{d\phi} - \int _{\pi/2}^{\pi}d\phi \frac{d\Gamma}{d\phi}}
{\int _0^{\pi}d\phi \frac{d\Gamma}{d\phi}}= \frac{2\Gamma_3}{\pi (\Gamma_1 + \Gamma_2)}~,\\
\langle \overline A \rangle =
\frac{\int _0^{\pi/2}d\phi \frac{d\overline \Gamma}{d\phi} -
\int _{\pi/2}^{\pi}d\phi \frac{d\overline \Gamma}{d\phi}}
{\int _0^{\pi}d\phi \frac{d\overline \Gamma}{d\phi}}=
\frac{2\overline \Gamma_3}{\pi (\overline \Gamma_1 + \overline \Gamma_2)}~.
\end{eqnarray}
While $\Gamma_3$ and $\overline \Gamma_3$ represent $T$-odd moments, they do not necessarily
signal $T$ violation, since they could be induced by strong final state interactions. Yet
\beq
\Gamma_3 \neq \overline \Gamma_3 \; \; \Longrightarrow \; \; { \CP \; {\rm violation}}.
\eeq
Such an analysis is theoretically clean, since the dependence on the angle $\phi$ is specifically predicted, which in
turn allows cross checks to control experimental systematics.

Alternatively, one can define another $T$-odd correlation among the pion and kaon momenta, namely
$C_T \equiv \vec p_{K^+} \cdot (\vec p_{\pi^+} \times \vec p_{\pi^-})$ for $D^0$  and
$\overline C_T \equiv \vec p_{K^-} \cdot (\vec p_{\pi^-} \times \vec p_{\pi^+})$ for $\Dzb$.
Similar to the previous case one has: $C_T \neq - \overline C_T \; \; \Longrightarrow$  \CP violation.
One can then construct $T$-odd moments
\begin{eqnarray}
A_T &=& \frac{\Gamma(C_T > 0 ) - \Gamma(C_T < 0 )}{\Gamma(C_T > 0 ) + \Gamma(C_T < 0 )}~, \\
\overline A_T &=& \frac{\Gamma(\overline C_T > 0 ) -
\Gamma(\overline C_T < 0 )}{\Gamma(\overline C_T > 0 ) + \Gamma(\overline C_T < 0 )} \;~,
\end{eqnarray}
and therefore
\beq
A_{\not T} = \frac{1}{2} (A_T - \overline A_T) \neq 0 \; \; \Longrightarrow \; \;
{\CP \; {\rm violation}}.
\eeq
A preliminary study based on 380 fb$^{-1}$ of \babar\ data suggests a sensitivity of $5.3 \times 10^{-3}$
in $A_{\not T}$ that
would extrapolate to $4 \times 10^{-4}$ for 75 ab$^{-1}$. With such a sample one can analyze even time slices of $A_{\not T}$.
These are very promising sensitivities.

Similar $\CP$ studies can be performed for other four-body modes, and one can also compare
$Y^0_L$ moments and even full amplitude analyses.

\subsection*{Charm baryon decays}

Charm baryons decays are sensitive only to direct $\CP$ violation.
Longitudinally polarized beams -- motivated mainly by $\CP$ studies in $\tau$ production and decays --
provide an intriguing handle for $\CP$ studies in charm baryon decays, since charm baryons would be produced with a net
longitudinal polarization that would allow the formation of novel $\CP$-odd correlations with the momenta of the particles in
the final state. The control of the sign of longitudinal polarization provides an excellent handle on systematics.


\section{Mixing and $\CP$V in 3-body decays}
\label{sec:charm-3body}

A Dalitz plot analysis of
 $\Dz \to \KS \pip \pim$ events provides a golden method for studying mixing and $\CP$ violation in mixing/decay/interference. If Dalitz plot model systematics can be kept under control, direct $\CP$-violation
 can also be investigated.
Present \babar\ data~\cite{Aubert:2005iz} show that at the \FourS, signal events from the decay chain $\Dstarp \to \Dz \pip$ with
 $\Dz \to \KS \pip \pim$ can be selected at a rate close to 1000/\invfb with a purity of 97.0\%, and a mistag probability
 of 0.1\%. \KS are reconstructed in the $\pip\pim$ final state; a requirement that the \KS proper time be $\le 8 \tau_S$
allows us to reduce \KL contamination to a level of $10^{-5}$.
Reconstructing the $\Dz \to \KS \pip\pim$ decay vertex, the \Dz proper time ($\tau_D$) can be measured with
  an average error of $\pm 0.2\ps$ in \babar\ and $\pm0.1\ps$ at \superb, to be compared with the \Dz
 lifetime of 0.4\ps.
\par
We use the invariant mass of $K\pi$ pairs: $m_+^2=m^2(\KS,\pip)$ and $m_-^2=m^2(\KS,\pim)$, and we define
the following Dalitz plot amplitudes ($f_D$) and probabilities ($p_D$), which
also depend on $t$:
\begin{eqnarray}
p_D(m_+^2, m_{-}^2,t) &\equiv& | f_D(m_+^2,m_{-}^2,t) |^2  \quad \Dz \textrm{tag}\\
\overline{p}_D(m_+^2, m_{-}^2,t) &\equiv& | \overline{f}_D(m_+^2,m_{-}^2,t) |^2 \quad \Dzb \textrm{tag}
\end{eqnarray}
The signatures for interesting processes are the following ones:
\begin{itemize}

\item  Mixing without \CP violation
\begin{eqnarray}
 p_D(m_+^2, m_{-}^2,t) & = & \overline{p}_D(m_-^2, m_{+}^2,t) \quad  \forall\ t \quad  \textrm{but}\\
 p_D(m_+^2, m_{-}^2,0) & \neq & p_D(m_+^2, m_{-}^2,t)
\end{eqnarray}
\item \CP violation in mixing
\begin{eqnarray}
 p_D(m_+^2, m_{-}^2,0) & = & \overline{p}_D(m_-^2, m_{+}^2,0)  \quad  \textrm{and}\\
 p_D(m_+^2, m_{-}^2,t) & \neq & \overline{p}_D(m_-^2, m_{+}^2,t)
\end{eqnarray}
\item Direct \CP violation
\begin{equation}
 p_D(m_+^2, m_{-}^2,0)  \neq  \overline{p}_D(m_-^2, m_{+}^2,0)
\end{equation}
\end{itemize}
and the quantities, to be measured, that enter in the previous Dalitz plot distribution functions, are:
$x$, $y$ (mixing parameters), $|q/p|$ or $\epsilon=\frac{1-|q/p|}{1+|q/p|}$ and $\phi=\arg(\frac{q \bar A_f}{p A_f})$
(\CP-violation parameters).
\par
$x$, $y$, $\epsilon$ and $\phi$ can be extracted in a Dalitz model-dependent
analysis with the isobar or K-matrix approach, using global fits. Examples are
described in references~\cite{Asner:2005sz,Aubert:2005iz}.
For the model-dependent approach,
 we conservatively estimate the \superb\ sensitivity at 75 \invab by extrapolating from the current analyses.
 Statistical errors can be scaled with the square root of luminosity. The result exceeds
 the desired goal of $10^{-3}$, a level not reachable by BES-III.
The second source is from systematic errors due to the experiment. They are mainly due to background parametrization,
 efficiency variation over the Dalitz plot, experimental resolution biases on Dalitz plot variables, decay time
 parametrization, and mistag fractions. Background parametrization is checked with sidebands (according to the
 Monte Carlo, the background does not peak in the \Dz mass signal region), and scales with statistics.
 Efficiency variation studied with Monte Carlo events scales with the Monte Carlo statistics.
Biases on Dalitz plot variable mass resolution are negligible. Decay time parametrization improves with the size of the data sample and due to the
 time resolution at \superb. Mistag fractions can be checked with other final $D$ states;
 their contribution is negligible. It is thus plausible that the errors arising from experimental sources can be
scale with statistics as well, but we prefer to be conservative, and evaluate these systematic errors using an additional safety factor of two.
These errors are shown in Table~VI;   
we can see that they are smaller than the statistical errors.
{
\begin{table}[!htb]
\caption{\label{tab:dalitz_belle}
Current Belle errors with 0.54 \invab on relevant 
mixing and \CP violation parameters.}
\begin{tabular}{lcccc}
\hline\hline
Par. & Stat.         &  Exp. Syst.      &  Model Syst. &  Total \\    \hline
$x$ ($10^{-4}$)        & 30.0 & 8.0  & 12.0 & 33.3 \\
$y$ ($10^{-4}$)        & 24.0 & 10.0 &  7.0 & 26.9 \\
$\epsilon$ ($10^{-4}$) & 15.0 & 2.5  &  4.0 & 15.7 \\
$\phi$ (deg)           & 17.0 & 4.0  &  3.0 & 17.7  \\ \hline\hline
\end{tabular}
\end{table}
}
{\begin{table}[!htb]
\caption{\label{tab:dalitz_sens}
\superb\ errors with 75 \invab on relevant \newline
mixing and \CP violation parameters.}
\begin{tabular}{lcccc}
\hline\hline
Par. & Stat.         &  Exp. Syst.      &  Model Syst. &  Total  \\ \hline
$x$ ($10^{-4}$)        & 2.5  & 1.4   & 4.0 & 4.9  \\
$y$ ($10^{-4}$)        & 2.0 & 1.7  &  2.3 &  3.5  \\
$\epsilon$ ($10^{-4}$) & 1.3 & 0.4   &  1.3 & 1.9  \\
$\phi$ (deg)           & 1.4 & 0.7   &  1.0 & 1.9 \\  \hline\hline
\end{tabular}
\end{table}
}

The last, but not the least important, source of systematic errors, is the
model used, typically isobar or \hbox{K-matrix} models or a partial-wave
analysis. Uncertainties arise from radius parameters, masses and widths of the
resonances, and the choice of resonances included in the fit.
Recent results from CLEO and Belle~\cite{Asner:2005sz,Abe:2007rd} have, however, demonstrated that the mixing and \CP violation parameters
 are not very sensitive to Dalitz model variations.
The sensitivity to models will be checked using two model independent approaches:
\begin{itemize}
\item With a very large data sample, a partial-wave analysis is capable to determine the amplitude and phase
 variation over the phase space directly from data.
\item Data collected at charm threshold will make the \DzDzb relative phase accessible~\cite{White:2007br}.
\end{itemize}
Even if it is extremely difficult to make predictions on the Dalitz model systematics at \superb, it is
 reasonable to assume that these will be substantially reduced with respect to the present errors from Belle~\cite{Abe:2007rd}.   By comparing the CLEO analysis based on 9.0~\invfb with the Belle analysis based on
 540~\invfb, we realize an improvement of the Dalitz model systematic error of more than a factor of four on
 average. This improvement is mainly due to the fact that the larger statistics data sample allows
 a better determination of the Dalitz model parameters.
Contemplating a factor of three improvement for the model error
 at \superb\ seems conservative, since it does not take into account the benefits of partial-wave analysis,
 and the use of data collected at charm threshold.
 Sensitivity predictions for mixing and \CP violation parameters at \superb\ are shown in Table~VII.




\onecolumngrid\twocolumngrid\hbox{}

\setcounter{section}{0}
\rhead[\fancyplain{}{\bf Tau Physics}]%
      {\fancyplain{}{\bf\thepage}}
\clearpage


{\centerline{\rule[0in]{0.9\columnwidth}{2pt}}
\vspace {-0.7cm}
\part*{\centerline{Tau Physics}}
\vskip -8pt
{\centerline{\rule[ 0.15in]{0.9\columnwidth}{2pt}}}


\bigskip
\smallskip

Searches for lepton flavor violation in tau decays constitute one of
the most theoretically and experimentally clean and powerful probes to
extend our knowledge in particle physics.  In this specific area,
\superb has clear advantages over the LHC experiments and SuperKEKB,
and it is complementary to muon LFV searches.
Experimental investigations on
\CP\ violation in tau decay and on the tau EDM and $g\! -\!\! 2$ provide \superb with
additional experimentally clean tools to shed light on unexplored
territories, with the ability to test some specific New Physics  scenarios.
Furthermore, precise tests of lepton universality can
reveal new phenomena, although attaining the required precision is
challenging, \superb is once again the best-positioned project, due
to its very high luminosity.

With an integrated luminosity of 75\invab, \superb will be able to
explore a significant portion of the parameter space of most New Physics
scenarios by searching for LFV in tau decays.  While the MEG
experiment~\cite{Grassi:2005ac} will search for $\mu \to e\gamma$ with
great sensitivity, \superb will uniquely explore transitions between the
third and first or second generations, providing crucial information to
determine the specific New Physics model that produces LFV.
The LHC experiments are, in general, not competitive in LFV searches;
SuperKEKB, with 10\invab, will also be able to explore LFV in tau decay,
but with a sensitivity that does not challenge the majority of New
Physics models. \superb has the
advantage of higher luminosity, which increases its tau LFV
sensitivity by a factor 2.7 in the worst hypothesis of background-dominated
analyses, even assuming no improvement in analysis
techniques. For analyses which are background-free, \superb
will have a sensitivity at least 7.5 times better, and will also profit from
reduced machine background. Furthermore, \superb can have a 85\%
linearly polarized electron beam, which will produce tau leptons with
known and well-defined polarization that can be exploited either to
improve the selection of LFV final states, given a specific LFV
interaction, or to better determine the features of the LFV
interaction, once they are found.

Experimental studies on \CP\ violation in tau decay and on the tau EDM and
$g\! - \!\! 2$ are especially clean tools, because they rely on measurement of
asymmetries with relatively small systematic uncertainties from the
experiment. The beam polarization also improves the experimental sensitivity
for tau EDM and $g\! - \!\! 2$ determinations, by allowing measurements of the polarization
of a single tau, rather than measurements of correlations between
two taus produced in the same events. with this technique,  \superb can test whether
supersymmetry is a viable explanation for the present discrepancy on
the muon $g\! - \!\! 2$.  Although the most plausible New Physics  models constrained
with the available experimental results predict \CP\ violation in tau decay
and the tau EDM in a range that is not measurable, \superb can test
specific models that enhance those effects to measurable levels.

\section{Lepton Flavor Violation}

\subsection*{Predictions from New Physics  models}

In the following, we discuss the size of $\tau$ LFV effects on decays and correlations
that are expected in supersymmetric extensions of the Standard Model and, in particular,
in the so-called constrained MSSM,
The flavor-conserving phenomenology of this framework is
characterized by five parameters: $M_{1/2}$, $M_0$, $A_0$, $\tan\beta$, $\text{sgn}\ \mu$.
We will discuss a subset of the ``Snowmass Points and Slopes'' (SPS)~\cite{Allanach:2002nj}, listed
in Table~\ref{tab:SPS:def:15}, in this
five-dimensional parameter space  to illustrate the main
distinctive features of the model as they relate to lepton flavor violation.

Specifying one such point
is sufficient to determine the phenomenology of the model
relevant for the LHC, but it is not sufficient to unambiguously
compute LFV rates. The amount of
flavor-violation is controlled by other parameters,
which play no role in high-$p_T$ physics.
Nonetheless, specifying the flavor-conserving parameters
allows us to simplify the description of LFV decays
and, in particular, to establish clear correlations
among different processes.

\begin{table}[!htb]
\caption{\label{tab:SPS:def:15}
Values of $M_{1/2}$, $M_0$, $A_0$, $\tan \beta$,
and sign of $\mu$ for the SPS points considered in the analysis.}
\begin{tabular}{cccccc}
\hline\hline
SPS & $M_{1/2}$ (GeV) & $M_0$ (GeV) & $A_0$ (GeV) & $\tan \beta$ &
 $\mu$ \\\hline
 1\,a & 250 & 100 & -100 & 10 &  $>\,0 $ \\
 1\,b & 400 & 200 & 0 & 30 &   $>\,0 $ \\
 2 &  300 & 1450 & 0 & 10 &  $>\,0 $ \\
 3 &  400 & 90 & 0 & 10 &    $>\,0 $\\
 4 &  300 & 400 & 0 & 50 &   $>\,0 $ \\
 5 &  300 & 150 & -1000 & 5 &   $>\,0 $\\\hline\hline
\end{tabular}
\end{table}

At all the SPS points, LFV decays are dominated by the contribution
of dipole-type effective operators of the form ($\bar l_i \sigma_{\mu\nu} l_j F^{\mu\nu}$).
Defining $\mathcal{R}^{(a)}_{(b)}=\BR(\tau \to a)/\BR(\tau \to b)$,
The dipole dominance allows us to establish the
following relations,
\begin{eqnarray}
\mathcal{R}^{(\mu ee)}_{(\mu \gamma)}\approx
1.0 \times 10^{-2} & \to & \BR(\tau \to \mu e^+ e^-) <  5 \times 10^{-10} \nonumber \\
\mathcal{R}^{(\mu \rho^0)}_{(\mu\gamma)}
\approx  2.5 \times 10^{-3} & \to &  \BR(\tau \to \mu \rho^0) < 10^{-10} \nonumber \\
\mathcal{R}^{(3\mu)}_{(\mu \gamma)}
\approx  2.2 \times 10^{-3}  & \to &  \BR(\tau \to 3\mu) < 10^{-10} \nonumber \\
\mathcal{R}^{(\mu \eta)}_{(\mu\gamma)} <  10^{-3} & \to &  \BR(\tau \to \mu \eta)
< 5 \times 10^{-11}, \nonumber
\end{eqnarray}
where the bounds correspond to the present limit
\hbox{$\BR(\tau\to\mu\gamma)< 4.5 \times 10^{-8}$.}  Similar relations hold for
\hbox{$\tau \to e$} transitions. As a result, in such a framework only
$\tau \to \mu \gamma$ and  $\tau \to e \gamma$ decays
are within experimental reach.

To estimate the overall scale of  $\tau \to (\mu,e) \gamma$ rates,
we must specify the value of the LFV couplings, since they are not
determined by the SPS conditions.
In the mass-insertion and leading-log approximation, assuming
that the leading LFV couplings appear in the left-handed
slepton sector, we can write
\begin{equation}
\frac{{\rm \BR}(l_j\to l_i\gamma)}{{\rm \BR}(l_j\to l_i \bar{\nu}_i\nu_{j})}
\approx
\frac{\alpha^3}{G_{F}^2}
\frac{\left|\left(m^2_{\widetilde L} \right)_{ji}\right|^2}{M_{S}^8}
\tan^2\beta \nonumber, \label{eq:min}
\end{equation}
where, to a good approximation,
$M_{S}^8\simeq 0.5 M_0^2 M_{1/2}^2 \times (M_0^2 + 0.6 M_{1/2}^2 )^2$.
In a Grand Unified Theory (GUT) with heavy right-handed neutrinos,
the off-diagonal entries of the slepton mass matrix $m^2_{\widetilde L}$
are likely to be dominated by the flavor mixing in
the (s)neutrino sector.
These terms can be expressed as
\begin{equation}
\label{eq:leadinglog}
\left(m^2_{\widetilde {L}}\right)_{ji} \approx
-\frac{6M_0^2+2A_0^2}{16\pi^2}\ \delta_{ij},
\end{equation}
where $\delta_{ij} = \left(Y^\dagger_\nu Y_\nu\right)_{ji}
\log(M_{GUT}/M_{R})$ in terms of the neutrino
Yukawa couplings ($Y_\nu$), the average
heavy right-handed neutrino mass ($M_R$)
and the GUT scale ($M_{GUT} \sim 10^{15}$--$10^{16}$~GeV).
Given the large phenomenological value
of the 2--3 mixing in the neutrino sector
(and the corresponding suppression of the 1--3 mixing)
we expect $|\delta_{32}| \gg |\delta_{31}|$
hence $\BR(\tau \to \mu \gamma) \gg \BR(\tau \to e \gamma)$.
For sufficiently heavy right-handed neutrinos, the
normalization of $Y_\nu$ is such that
$\BR(\tau \to \mu \gamma)$ can reach
values in the $10^{-9}$ range.
In particular, $\BR(\tau \to \mu \gamma) \gsim 10^{-9}$
if at least one heavy right-handed neutrino has a mass around or above
$10^{13}$~GeV (in SPS 4) or $10^{14}$~GeV (in SPS 1a,1b,2,3,5).

A key issue that must be addressed is the role of
$\BR(\mu \to e \gamma)$ in
constraining  the LFV couplings and,
more generally, the correlations between
$\BR(\tau \to (\mu,e) \gamma)$ and $\BR(\mu \to e \gamma)$
in this framework. An extensive analysis of such questions
has been presented in Ref.~\cite{Antusch:2006vw,Arganda:2005ji},
under the hypothesis of a hierarchical spectrum
for the heavy right-handed neutrinos.

The overall structure of the
$\BR(\tau \to \mu \gamma)$ vs.~$\BR(\mu \to e \gamma)$
correlation in SPS~1a is shown in~Fig.~\ref{fig:herrero1}.
As anticipated, $\BR(\tau \to \mu \gamma)\sim 10^{-9}$
requires a heavy right-handed neutrino around or
above  $10^{14}$~GeV. This possibility is not excluded by
$\BR(\mu \to e \gamma)$ only if the 1--3 mixing in the
lepton sector (the $\theta_{13}$ angle of the neutrino
mixing matrix) is sufficiently small. This is a general
feature, valid at all SPS points, as illustrated in
Fig.~\ref{fig:SPS:t13:ad}.
In Table \ref{tab:SPS:BR} we show the predictions for $\BR(\tau\to\mu\,\gamma)$
and $\BR(\tau\to 3\,\mu)$ corresponding to the neutrino
mass parameters chosen in Fig.~\ref{fig:SPS:t13:ad}
(in particular $M_{N_3} = 10^{14}$~GeV), for the various SPS points.
Note that this case contains points that
are within the \superb sensitivity range,
yet are not excluded by $\BR(\mu \to e \gamma)$
(as illustrated in Fig.~\ref{fig:SPS:t13:ad}).

\begin{figure}[!htb]
\includegraphics[width=\linewidth,angle=0]{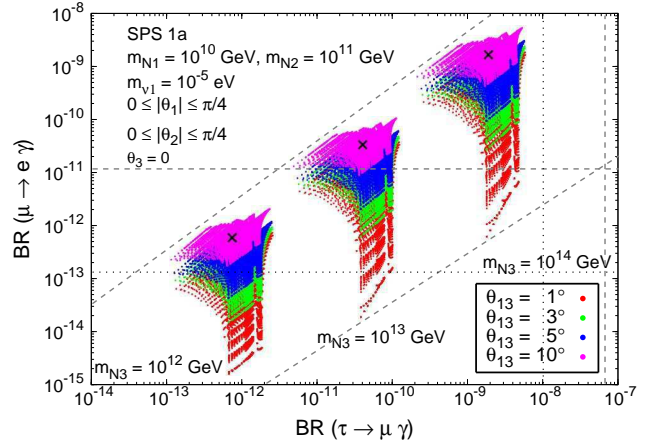}
\caption{\label{fig:herrero1}
$\BR(\tau \to \mu \gamma)$ vs.~$\BR(\mu \to e \gamma)$
in SPS~1a,
for three reference values of the heavy right-handed neutrino
mass and several values of $\theta_{13}$.
The horizontal dashed (dotted) line denotes the present experimental
    bound (future sensitivity) on $\BR(\mu \to e \gamma)$.
    All other relevant parameters are set to the values specified in
    Ref.~\protect{\cite{Antusch:2006vw}}.
     }
\end{figure}

\begin{figure}[!htb]
\includegraphics[angle=-90,width=\linewidth]{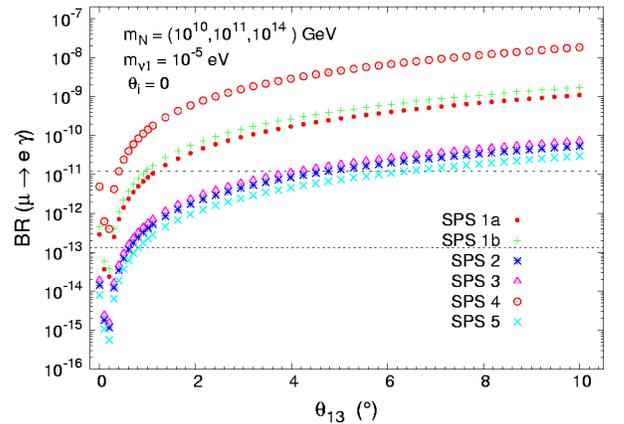}
\caption{\label{fig:SPS:t13:ad}
$\BR(\mu \to e\, \gamma$) as a function of $\theta_{13}$ (in degrees)
for various SPS points. The dashed (dotted) horizontal line denotes the present experimental
    bound (future sensitivity). All other relevant parameters are
    set to the values specified in
    Ref.~\cite{Antusch:2006vw}.
           }
\end{figure}

\begin{table}[!htb]
\caption{\label{tab:SPS:BR}Predictions for  $\BR(\tau \to \mu\, \gamma$) and \hbox{$\BR(\tau\to 3\,\mu$)}
         corresponding to the SPS points. The values of $m_{N_i}$ and $m_{\nu_1}$
         are as specified in Fig.~\ref{fig:SPS:t13:ad}~\cite{Antusch:2006vw}.}

\begin{tabular}{ccccccc}
\hline\hline
SPS               &  1\,a   &  1\,b   &  2  &  3  &  4  &  5   \\ \hline
$\BR(\tau\to\mu\gamma)\times 10^{-9}$ & 4.2 & 7.9 & 0.18 & 0.26 & 97 & 0.019 \\
$\BR(\tau\to 3\mu)\times 10^{-12}$ & 9.4 & 18 & 0.41 & 0.59 & 220 & 0.043 \\ \hline\hline
\end{tabular}
\end{table}

\subsubsection*{LFV in the NUHM scenario}

At large $\tan\beta$ and not too heavy Higgs masses,
another class of LFV interactions is relevant,
the effective coupling between a  $\mu$--$\tau$ pair
and the heavy (scalar and pseudoscalar) Higgs bosons.
This coupling can
overcome the constraints on $\BR(\tau\to \mu\mu\mu)$
and $\BR(\tau\to\mu\eta)$ dictated by $\BR(\tau\to\mu\gamma)$
in the dipole-dominance scenario.
Such a configuration cannot be realized in the CMSSM, but
it could be realized in the so-called NUHM SUSY scenario,
which is also theoretically well-motivated and rather general.
In such a framework, there are specific regions of the
parameter space in which $\tau\to\mu\eta$ could have
a branching ratio in the $10^{-9}$--$10^{-10}$ range,
comparable or even slightly
larger than $\BR(\tau\to\mu\gamma)$~\cite{Paradisi:2005tk}.

Finally, in more exotic New Physics frameworks, such as SUSY
without R parity, Little Higgs Models with T parity (LHT)
or $Z^{'}$ models with non-vanishing LFV couplings ($Z^{'}\ell_{i}\ell_{j}$),
the $\tau\to \mu\mu\mu$ rate could be as large as,
or even larger than $\tau\to\mu\gamma$ (see {\it e.g.},~\cite{Raidal:2008jk}).
In this respect, an improvement of $\BR(\tau\to \mu\mu\mu)$
at the $10^{-10}$ level would be interesting even
with $\BR(\tau\to\mu\gamma)\lsim 10^{-9}$.


\subsection*{\superb experimental reach}{SuperB experimental reach}

A sensitive search for lepton flavor-violating $\tau$ decays at \superb
requires signal to be selected with as high an efficiency as possible, while allowing
minimal, and preferably zero, background.
A  candidate $e^+e^- \to \tau^+\tau^-$ events obtained from an initial screening selection
is divided into hemispheres in the center-of-mass frame,
each containing the decay products of one $\tau$ lepton.
Unlike Standard Model $\tau$~decays, which contain at least one neutrino,
the decay products from a LFV decay have a combined energy in the center-of-mass frame
equal to $\sqrt{s}/2$ and a mass equal to that of the $\tau$.
A requirement on the two dimensional signal region in the $E_{\ell X}$--$M_{\ell X}$ plane
therefore provides a powerful tool to reject backgrounds,
which arise from well-understood Standard Model $\tau$ decays.
Consequently, residual background rates and distributions
are reliably estimated from Monte Carlo simulations and validated
using quantitative comparisons with data as various selection requirements are applied.
Global event properties and an explicit identification of the non-signal
$\tau$ decay can be applied to suppress non-$\tau$ backgrounds with only marginal
loss of efficiency.

The considerable experience developed in searching for these decays
in the ${\sim}0.5$~ab$^{-1}$ data set at \babar enables
 us confidently to estimate background levels to be expected with 75~ab$^{-1}$
 for selection strategies similar to those of the existing experiments.
These lead us to classify the LFV decay modes into two categories for the
purposes of estimating the experimental $\tau$ LFV discovery reach of
\superb: (i) modes having ``irreducible backgrounds'' and (ii)
modes that do not have irreducible backgrounds.
For luminosities of $10^{36}\,\mathrm{cm^{-2}s^{-1}}$,
\taulg decays fall into category (i), whereas $\tau\ra\ell\ell\ell$ and \taulh generally fall
 into category (ii),
where $\ell$ is either a muon or electron and h$^0$ is a hadronic system.
The hadronic system may be identified as a pseudoscalar or vector meson
 ($\pi^0$, $\eta$, $\eta'$, $K^0_S$, $\omega$,$\phi$,$K^{*}$ {\it etc.}) or
a non-resonant system of two pions, two kaons or a pion and kaon.

The category (ii) decay modes have the property that with perfect particle identification
no known process or combination of processes can mimic the signal at rates
relevant to \superb.
The challenge in searching for these decays
is thus to remove all non-$\tau$ backgrounds and to provide as powerful
a particle identification as possible.
For category (i) modes, however, even with perfect particle identification, there
exist backgrounds that limit the discovery sensitivity.
In fact, there are no \taulg Standard Model processes expected
at these luminosities, but there are combinations
of processes that can mimic this signal, even with perfect measurements.
In the case of $\taumg\!\!,$ for example, the irreducible background arises from
events having a $\tau\ra\mu\nu\bar{\nu}$ decay
and a $\gamma$ from initial state radiation (ISR)
in which the photon combines with the muon to form a candidate that accidentally
falls into the signal region in the $E_{\ell X}$--$M_{\ell X}$ plane.
At sufficiently high rates, $\tau\ra\ell\pi^0$ and  $\tau\ra\ell\eta \, (\eta\ra\gamma\gamma)$
searches will suffer the same problems when two hard ISR
photons accidentally reconstruct to a $\pi^0$ or $\eta$ mass,
 but the rate for two hard-photon ISR emission will be roughly 100 times lower
 than  the rate for a signal hard photon emission and lower still
 when requiring a $\gamma\gamma$ mass to match that of a $\pi^0$ or $\eta$.
 Consequently, this  is not expected to  be an issue at \superb luminosities.
Similarly, $\tau\ra ee^+e^-$ and $\tau\ra \mu e^+e^-$ can, in principle,
suffer a background from $\tau\ra\ell\nu\bar{\nu}e^+e^-$ events where
the ISR photon undergoes internal pair production. Such background events are expected
 to start to just become measurable for luminosities roughly 100 times higher than
current experiments, and so might just begin to impact the experimental
bounds placed on those modes at \superb.

The experimental reach is expressed here in terms of
``the expected $90\%$ CL upper limit'' assuming no signal,
as well as in terms of a 4$\sigma$ discovery branching fraction
 in the presence of projected backgrounds.
In the absence of signal,
for large numbers of background events $N_{\rm bkd}$,
the $90\%$ CL upper limit for the number of signal events can be given as
$N^{UL}_{90} \sim 1.64 \sqrt{N_{\rm bkg}}$,
whereas for small $N_{\rm bkg}$ a value for $N^{UL}_{90}$
is obtained using the method described in~\cite{Cousins:1991qz},
which gives, for  $N_{\rm bkg} \sim 0$, $N^{UL}_{90} \sim 2.4$.
If a signal is determined from counting events within a signal region
rather than from a fit, the $90\%$ CL branching ratio upper limit is:
\begin{equation}
  \B^{UL}_{90} =
  \frac{N^{UL}_{90}}{2 N_{\tau\tau} \epsilon} =
  \frac{N^{UL}_{90}}{2 \cal{L}\sigma_{\tau\tau} \epsilon}\,,
\end{equation}
where  $N_{\tau\tau}=\cal{L}\sigma_{\tau\tau}$
is the number of $\tau$-pairs produced in $e^+e^-$ collisions;
$\cal{L}$ is the integrated luminosity,
$\sigma_{\tau\tau}$=0.919~nb~\cite{Banerjee:2007is}
 is the $\tau$-pair production cross section, and
$\epsilon$ is the signal efficiency.

The \taumg projected sensitivity is based on the published
\babar\ analysis~\cite{Aubert:2005ye},
but incorporating changes designed for a very high luminosity
data set and using the improved muon particle identification efficiencies
that became available with a hardware upgrade to the \babar\ muon system.
The published analysis explicitly identifies  the non-signal $\tau$ decays
as specific Standard Model decay modes. In the published analysis, this
 set of tag modes includes $\tau\ra\mu\nu\bar{\nu}$, which has a disproportionate amount
 of $\mu$-pair background compared to the other tag modes. For \superb\ luminosities
it would appear that a more optimal analysis would not include this mode.
The consequence is that the efficiency for a 2$\sigma$ signal ellipse region suffers
a decrease from dropping the $\mu$-tag, but increases from the other improvements
 to both the analysis and the hardware, so that the net efficiency is 7.4\%.
The background levels for 75~ab$^{-1}$ are projected from the Monte Carlo
to be $200\pm50$ events from the $\tau\ra\mu\nu\bar{\nu}(\gamma)$ irreducible background.
This leads to an expected 90\%CL upper limit of $2.3\times 10^{-9}$ and 4$\sigma$ discovery
reach of $5.6\times 10^{-9}$. It is important to note that further improvements can be obtained using the \superb\ polarized electron beam. For a 100\% polarized electron beam,
the polar angles of the signal decay products provide additional background suppression,
as is evident from Figure~\ref{fig:tau-lfv-exp-1}. The ``irreducible background'' would be cut
by 70\% for a 39\% loss in signal efficiency. This would result in approximately a
10\% improvement in the sensitivity:
an expected upper limit of $2.1\times 10^{-9}$ and 4$\sigma$ discovery
level of $5.0\times 10^{-9}$. However, by far the most important aspect
of having the polarization is the possibility to determine the helicity
structure of the LFV coupling from the final state momenta
distributions (see for instance Ref.\cite{Matsuzaki:2007hh} for the
$\tau\to\mu\mu\mu$ process).
Note that for a data sample of  15~ab$^{-1}$ using a
 machine with no polarization, the same analysis and detector
can be expected to yield an expected upper limit of
$5.2\times 10^{-9}$ with a discovery potential of $1.3\times 10^{-8}$.
Similar analyses can be expected to yield comparable sensitivities for the
\taueg LFV decay mode, based on the published \babar\ analysis~\cite{Aubert:2005wa}.

The situation for the other LFV decays,
$\tau\to\ell_1\ell_2\ell_3$\ and $\tau\to\ell h$, is different,
as these modes do not suffer the problem of accidental photons
 with which the \taulg searches must contend.
In these cases, one can project sensitivities assuming $N_{\rm bkg}$
comparable to backgrounds in existing analyses
for approximately the same efficiencies.
For illustrative purposes, we demonstrate how this is accomplished
for the \taummm based on modifications to the published \babar\ analysis~\cite{:2007pw}.
The published analysis managed to suppress the backgrounds for the
data set without explicitly identifying the  Standard Model $\tau$ decays for
 the non-signal $\tau$ and using the loosest muon identification
algorithms.

\begin{figure}[!htb]
  \begin{center}
    \begin{overpic}[width=\linewidth,trim=0 20 0 0,clip]{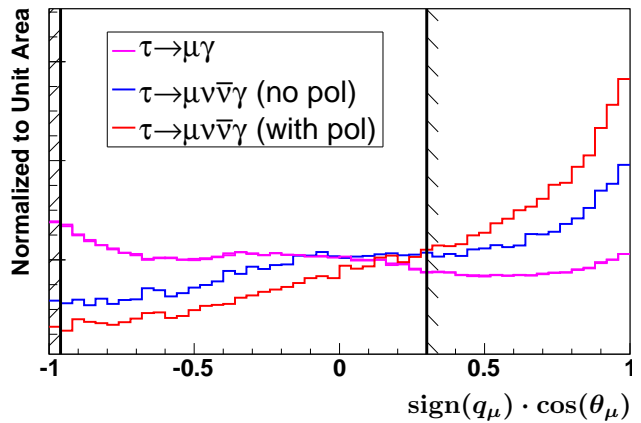}
      \put(0,-4){\makebox[\linewidth][r]{\larger\textbf{$\text{sign}(q_\mu)\cdot\cos(\theta_\mu)$}}}
    \end{overpic}
  \end{center}
  \vspace{1ex}
  \caption{\label{fig:tau-lfv-exp-1}%
    Distribution of the cosine of the signal-side muon multiplied by the
    muon charge for signal and background events with and without electron
    beam polarization in the \taumg search analysis at \superb.}
\end{figure}

Table~\ref{tab:LFVExptSensitivities} summarizes the sensitivities
for various LFV decays.

\begin{table}[!htb]
  \caption{
    \label{tab:LFVExptSensitivities}
    Expected $90\%$ CL upper limits and 4$\sigma$ discovery reach
    on \taumg and \taummm LFV decays with $75 \ {\rm ab}^{-1}$
    with  a polarized electron beam.
  }
  \begin{center}
    \begin{tabular}{lll}
      \hline \hline
      Process &  Expected 90\%CL & 4$\sigma$ Discovery \\
              &  upper limited   &  Reach  \\
      \hline
      $\BR(\tau \to \mu\,\gamma)$          &  $2 \times 10^{-9}$ &  $5 \times 10^{-9}$  \\
      $\BR(\tau \to \mu\, \mu\, \mu)$      &  $2 \times 10^{-10}$ & $8.8 \times 10^{-10}$  \\
      \hline\hline
    \end{tabular}
  \end{center}
\end{table}

\section{Lepton universality}

Tree-level Higgs exchanges in supersymmetric new physics models can
induce modifications of lepton universality of order
0.1\%~\cite{Krawczyk:1987zj}, smaller but close to the present
experimental accuracy of $\approx 0.2\%$~\cite{Pich:2006nt}. As
discussed in Ref.~\cite{Bona:2007qt}, \superb can probably measure
lepton universality to $0.1\%$ or better. However the measurement is
limited by experimental systematic uncertainties on the measurement of
the tau leptonic branching fractions and the tau lifetime, as the
modest progress provided by the existing $B$ Factories also
confirms~\cite{Lusiani:2007cb}.  Therefore it cannot be advocated that
the \superb advantages in terms of luminosity are crucial and
necessary for the advancement of this particular sector, although
large statistical samples will be an advantage to reduce experimental
systematic uncertainties.

\section{{Tau \CP$\!\!$V, EDM and $g\! -\!\! 2$}}

\subsection*{Predictions from New Physics models}

\subsubsection*{\CP violation and $T$-odd observables in tau decay}

\CP violation in the quark sector has been observed both
in the $K$ and in the $B$ systems; the experimental results are thus far
fully explained by the complex phase of the CKM matrix.
On the contrary, $\CP$ violation in the lepton sector has yet not been
observed. Within the Standard Model, $\CP$-violating effects in charged-lepton
decays are  predicted to be vanishingly small.
For instance, the $\CP$ asymmetry rate of $\tau^{\pm}\to K^{\pm}\pi^0\nu$
is estimated to be of order ${\cal O}(10^{-12})$~\cite{Delepine:2005tw}.
Evidence for \CP violation in tau decay would therefore be a clear signal of
New Physics.
In one instance, the $\tau^{\pm}\to K_S \pi^\pm \nu$
rate asymmetry, a small \CP\ asymmetry of $3.3 \times 10^{-3}$
is induced by the known \CP-violating phase of the $\KzKzb$
mixing amplitude~\cite{Bigi:2005ts}. This asymmetry is known to 2\% precision. Thus, this mode can serve as a calibration, and in addition, any deviation from the expected asymmetry would be a sign of New Physics.

Most of the known New Physics models cannot generate observable
\CP-violating effects in $\tau$ decays (see {\it e.g.},~\cite{Raidal:2008jk}).
The only known exceptions are R parity-violating supersymmetry~\cite{Delepine:2007qg}
or specific non-supersymmetric multi-Higgs models.
In such a framework, the $\CP$ asymmetries of various $\tau$-decay channels
can be enhanced up to the $10^{-1}$ level, without conflicting with
other observables, and saturating the experimental limits
obtained by CLEO~\cite{Bonvicini:2001xz}.
Similar comments also apply to $T$-odd  \CP-violating asymmetries in
the angular distribution of $\tau$ decays.

\subsubsection*{Tau electric dipole moment}

In natural SUSY frameworks, lepton EDMs ($d_{\ell}$) scale linearly with the
lepton mass. As a result, the existing limits on the electron EDM generally
preclude any visible effect in the $\tau$ and $\mu$ cases.
In multi-Higgs models, however, EDMs scale with the cube of the lepton
masses~\cite{Barger:1996jc}, $d_{\tau}$ can thus be substantially enhanced.
However, in this case the electron and muon EDMs receive sizable
two-loop effects via Barr-Zee diagrams, which again scale linearly with the lepton masses.  As a result, one can derive an approximate bound
 $d_{\tau}\lsim 0.1\times (m_{\tau}/m_{\mu})^3 (m_{\mu}/m_e)d_{e}$
which is still very strong.  From the
present experimental upper bound on the electron EDM,
$d_{e}\lsim 10^{-27} e\,{\rm cm}$, it follows that $d_{\tau}\lsim 10^{-22} e\,{\rm cm}$.

\subsubsection*{Tau $g\! - \!\! 2$}

The Standard Model prediction for the muon anomalous magnetic
moment is not in perfect agreement with recent experimental results. In particular,
 $\Delta a_{\mu} = a_{\mu}^{\rm exp} - a_{\mu}^{\rm SM}
\approx(3 \pm 1)\times 10^{-9}$.
Within the MSSM, this discrepancy can naturally
be accommodated, provided $\tan\beta \gsim 10$ and $\mu >0$.

A measurement of the $\tau$ anomalous magnetic moment
could be very useful to confirm or disprove the
interpretation of $\Delta a_{\mu}$ as due to New Physics contributions.
The natural scaling of heavy-particle effects
on lepton magnetic dipole moments,
implies $\Delta a_\tau/\Delta a_\mu \sim m^{2}_{\tau}/m^{2}_{\mu}$.
Thus, if we interpret the present muon discrepancy
$\Delta a_{\mu} = a_{\mu}^{\rm exp} - a_{\mu}^{\rm SM} \approx(3 \pm 1)\times 10^{-9}$
as a signal of New Physics, we should expect
$\Delta a_{\tau}\approx 10^{-6}$.

In the supersymmetric case, such an estimate holds for all
the SPS points (see Table~\ref{tab:SPS:gm2}) and, more generally,
in the limit of almost
degenerate slepton masses. If $m_{\tilde{\nu}_{\tau}}^2<<m_{\tilde{\nu}_{\mu}}^2$
(as happens, for instance, in the so-called effective-SUSY scenario),
$\Delta a_{\tau}$ could be enhanced up to the $10^{-5}$ level.

\begin{table}[!htb]
\caption{\label{tab:SPS:gm2}
Values of $\Delta a_{\mu}$ and $\Delta a_{\tau}$ for various SPS points.}
\begin{tabular}{ccccccc}
\hline\hline
SPS               &  1\,a   &  1\,b   &  2  &  3  &  4  &  5   \\ \hline
$\Delta a_{\mu}\times 10^{-9}$ &  3.1 & 3.2 & 1.6  & 1.4  & 4.8  & 1.1 \\
$\Delta a_{\tau}\times 10^{-6}$ &  0.9 & 0.9 & 0.5  & 0.4  & 1.4  & 0.3 \\ \hline\hline
\end{tabular}
\end{table}


\subsection*{\superb experimental reach}

\subsubsection*{\CP violation and $T$-odd observables in tau decay}

A first search for \CP violation in tau decay has been conducted by the CLEO
collaboration~\cite{Bonvicini:2001xz}, looking for a
tau-charge-dependent asymmetry of the angular distribution of the
hadronic system produced in $\tau \to K_S\pi\nu$.  In multi-Higgs
doublet New Physics, the \CP-violating asymmetry arises from the Higgs coupling
and the interference between $S$ wave scalar exchange and $P$ wave
vector exchange.  The Cabibbo-suppressed decay mode into $K_S\pi\nu$
has a larger mass-dependent Higgs coupling;  the events in the
sidebands of the $K_S$ mass distributions can thus be used to calibrate the
detector response.  With a data sample of 13.3\invfb ($12.2\EE{6}$ tau
pairs), the mean of the optimal asymmetry observable is $\left<\xi\right> =
(-2.0 \pm 1.8)\EE{-3}$.
As the above measurement relies on detector calibration with side-band
events, it is conceivable that \superb with 75\invab would not be
limited by systematics and would therefore reach an experimental
resolution $\Delta\left<\xi\right> \approx 2.4\EE{-5}$.


\subsubsection*{Tau electric dipole moment}

The tau electric dipole moment (EDM) influences both the angular
distributions and the polarization of the tau produced in \epem
annihilation. With a polarized beam, it is possible to
construct observables from the angular distribution of the products of
a single tau decay that unambiguously discriminate between the
contribution due to the tau EDM and other
effects~\cite{GonzalezSprinberg:2007qj,Bernabeu:2006wf}.
Recent work has provided
an estimate of the \superb
upper limit sensitivity for the real part of the tau EDM
$\left|\text{Re}\{d_\tau^\gamma\}\right| \le
7.2\EE{-20}\,e\,\text{cm}$ with
75\invab~\cite{GonzalezSprinberg:2007qj}. The result
assumes a $100\%$ polarized electron beam colliding with unpolarized
positrons at the \FourS peak, no uncertainty on the polarization, and
perfect reconstruction of the tau decays $\tau \to \pi\nu$. Studies
have been done assuming more realistic conditions:
\begin{itemize}
\item an electron beam with a linear polarization of $80\% \pm 1\%$;
\item $80\%$ geometric acceptance;
\item track reconstruction efficiency $97.5\%\pm 0.1\%$ (similarly to
  what has been achieved in LEP analyses~\cite{Schael:2005am} and
  \babar ISR analyses~\cite{Davier:2008:babarIsrEffSyst}.
\end{itemize}
The process $\epem \to \tautau$ is simulated with the KK
generator~\cite{Jadach:1999vf} and the Tauola package for tau
decay~\cite{Jadach:1999vf}; the simulation includes the
complete spin correlation density matrix of the initial-state beams
and  the final state tau leptons. Tau EDM effects are simulated by
weighting the tau decay product angular distributions.
The studies are not complete, and do not yet include
uncertainties in reconstructing the tau direction. The preliminary
indications are that the tau EDM experimental resolution is $\approx
10\EE{-20} e\,\text{cm}$, corresponding to an angular asymmetry of
$3\EE{-5}$; the uncertainties in track reconstruction give a $\approx
1\EE{-20}$ systematic contribution.  Asymmetries
proportional to the tau EDM depend on events that go into the same
detector regions but arise from tau leptons produced at different angles,
minimizing the impact of efficiency uncertainties. It must be added
that all the hadronic tau channels have at least theoretically the same
statistical power as the $\tau \to \pi\nu$ mode in measuring the tau
polarization~\cite{Kuhn:1995nn}, and can therefore be used to improve
the experimental resolution.

A search for the tau EDM with unpolarized beams has been completed at
Belle~\cite{Inami:2002ah}.  In this case, one
must measure correlations of the angular distributions of both tau
leptons in the same events, thereby losing in both reconstruction efficiency and
statistical precision.  The analysis shows the impact of inefficiency
and uncertainties in the tau direction reconstruction, and also
demonstrates that all tau decays, including leptonic decays with two
neutrinos, provide statistically useful information for measurement of the tau
EDM.  With $29.5\invfb$ of data, the experimental resolution on the
real and imaginary parts of the tau EDM is
$[0.9{-}1.7]\EE{-17}\,e\,\text{cm}$, including systematic effects.  An
optimistic extrapolation to \superb at $75\invab$, assuming systematic
effects can be reduced according to statistics, corresponds to an
experimental resolution of $[17{-}34]\EE{-20}$.


\subsubsection*{Tau $g\! - \!\! 2$}

In a manner similar to an EDM, the tau anomalous moment ($g\! - \!\! 2$) influences both the angular
distribution and the polarization of the tau produced in \epem
annihilation.  Polarized beams allow the measurement of the real part of the
$g\! - \!\! 2$ form factor by statistically measuring the tau polarization with
the angular distributions of its decay
products. Bernab\'eu {\it et al.}~\cite{Bernabeu:2007rr} estimate that \superb
with 75\invab will measure the real and imaginary part of the $g\! - \!\! 2$
form factor at the \FourS with a resolution in the range
$[0.75-1.7]\EE{-6}$. Two measurements of the real part of
$g\! - \!\! 2$ are proposed, one fitting the polar angle distribution of the tau leptons,
and one based on the measurement of the tau transverse and
longitudinal polarization from the angular distribution of its decay
products.  All events with tau leptons decaying either in $\pi\nu$ or
$\rho\nu$ are considered, but no detector
effects are accounted for. For the tau polarization measurements,
electron beams with perfectly known 100\% polarization are assumed.
Studies simulating more realistic experimental conditions are ongoing.
While the polar angle distribution measurement will conceivably suffer
from uncertainties in the tau direction reconstruction, the
preliminary results on the tau EDM measurement, mentioned above,
indicate that asymmetries measuring the tau polarization are least
affected by reconstruction systematics. Transposing  the preliminary results obtained with
simulations for the tau EDM to the real part
of the $g\! - \!\! 2$ form factor, one can estimate that $a_\mu = (g-2)/2$
can be measured with a statistical error of $2.4\EE{-6}$, with
systematic effects from reconstruction uncertainties one order of
magnitude lower.

\bigskip










\onecolumngrid\twocolumngrid\hbox{}
\clearpage 

\setcounter{section}{0}
\rhead[\fancyplain{}{\bf Spectroscopy}]%
      {\fancyplain{}{\bf\thepage}}

{\centerline{\rule[0in]{0.9\columnwidth}{2pt}}
\vspace {-0.7cm}
\part*{\centerline{Spectroscopy and the}}
\vspace {-1cm}
\part*{\centerline{Decays of Quarkonia}}
\vskip -8pt
{\centerline{\rule[ 0.15in]{0.9\columnwidth}{2pt}}




\bigskip
\smallskip

Although the Standard Model is well-established,
QCD, the fundamental theory of strong interactions, provides a quantitative comprehension only of phenomena
at very high energy scales, where perturbation theory is effective due
to asymptotic freedom.
The description of hadron dynamics below the QCD dimensional transmutation scale
is therefore far from being under full theoretical control.

Systems that include heavy quark-antiquark pairs
(quarkonia) are a unique and, in fact, ideal laboratory for probing both the high
energy regimes of QCD, where an expansion in terms of the coupling
constant is possible, and the low energy regimes, where
nonperturbative effects dominate.
For this reason, quarkonia have been studied for decades in great detail.
The detailed level of understanding of the quarkonia mass spectra
is such that a particle mimicking quarkonium properties, but not fitting any quarkonium level,
is most likely to be considered to be of a different nature.

In particular, in the past few years the $B$ Factories and the Tevatron have provided evidence for states that
do not admit the conventional mesonic interpretation and that instead could be
made of a larger number of constituents (see Sec.~\ref{sec:data}). While
this possibility has been considered since the
beginning of the quark model~\cite{ GellMann:1964nj}, the actual identification of such states
would represent a major revolution in our understanding of elementary particles. It would also imply the existence
of a large number of additional states that have not yet been observed.

Finally, the study of the strong bound states could be of relevance to
understanding the Higgs boson, if it turns out to be itself
a bound state, as predicted by several technicolor models (with or without extra dimensions)~\cite{techni}.

The most likely possible states beyond the mesons and the
baryons are:

\begin{itemize}

\item {\bf hybrids:} bound states of a quark-antiquark pair and a number of
constituent gluons. The lowest-lying state is expected to have quantum numbers
$J^{PC}=0^{+-}$. Since a quarkonium state cannot have
these quantum numbers (see below), this a unique signature for
hybrids. An additional signature is the preference for a hybrid to
decay into quarkonium and a state that can be produced by the excited
gluons (\eg, $\pi^+\pi^-$ pairs); see \eg,~Ref.~\cite{ibridi}.
\item {\bf molecules:} bound states of two mesons, usually represented as
$[Q\bar{q}][q^{\prime}\bar{Q}]$, where $Q$ is the heavy quark. The
system would be stable if the binding energy were to set the mass of the
states below the sum of the two meson masses.
While this could be the case for when $Q=b$, this does not apply for
$Q=c$, the case for which most of the current experimental data exist. In this case,
the two mesons can be bound by pion exchange. This means that only
states decaying strongly into pions can bind with other mesons (\eg,
there could be $D^*D$ states), but
that the bound state could decay into its constituents~\cite{others}.
\item {\bf tetraquarks:} a bound quark pair, neutralizing its color with a bound
antiquark pair, usually represented as $[Qq][\bar{q^{\prime}}\bar{Q}]$. A full
nonet of states is predicted for each spin-parity, \ie, a large number
of states are expected. There is no need for these states to be close to
any threshold~\cite{mppr}.

\end{itemize}

In addition, before the panorama of states is fully clarified, there is always the
lurking possibility that some of the observed states are misinterpretations of
threshold effects: a given amplitude might be enhanced when new hadronic final
 states become energetically possible, even in the absence of resonances.

While there are now several good experimental candidates for unconventional states,
the overall picture is not complete and needs confirmation, as well as
discrimination between the alternative explanations.
A much larger dataset than is currently available is needed, at several energies, to pursue this program; this capability is uniquely within the reach  of \superb.

Finally, bottomonium decays also allow direct searches for physics beyond the Standard Model
in regions of the parameters space that have not been reached by LEP.

\section{Light meson spectroscopy}%

 The problem of the interpretation of the light scalar mesons, namely
$f_0, a_0,\kappa,\sigma$, is one of the oldest problems in hadronic physics
\cite{gell}.
For many years the question of the existence of the $\sigma$ meson as a resonance in $\pi\pi$ scattering has been debated~\cite{resu};  only recently has a thorough analysis of $\pi\pi$ scattering amplitudes shown that the $\sigma(500)$ and $\kappa(800)$ can be considered to be
proper resonances~\cite{Caprini}.

Reconsideration of the $\sigma$ was triggered by the E791 analysis of $D\to 3\pi$ data~\cite{e791}; a number of papers have commented on those results, \eg,~Ref.~\cite{torni}.  The role of the scalar mesons in several exclusive $B$ decays could be rather relevant: for example, in the perspective of a high precision measurement of the $\alpha$ angle at the SuperB factory, the hadronic contributions, like the one of the isoscalar $\sigma$ in $B\to rho \pi$, must be properly controlled~\cite{io}.
Also diverse studies on light and heavy scalar mesons could be
performed analyzing the Dalitz plots of exclusive decays like $B\to KKK$ and $B\to K \pi\pi$. In this respect, having sufficient statistics to clearly
assess the presence of a scalar $\kappa (800)$ resonance, would certainly be a major result for hadron spectroscopy.

Beyond the ``taxonomic'' interest in the classification of scalar mesons, the idea that these mesons could play a key role in our understanding of
aspects of non-perturbative QCD has been raised; see, for example, the interesting paper, Ref.~\cite{gribov}.

In what follows we would like to underscore the latter point by observing that:

\begin{itemize}%

\item Light scalar mesons are most likely the lightest particles with an {\it exotic} structure, \ie, they cannot be classified as $q\bar q$ mesons.
\item Their dynamics is tightly connected with instanton physics.  Recent discussions have shown that instanton effects facilitate the creation of a consistent model for the description of light scalar meson dynamics, under the hypothesis
that these particles are diquark-antidiquark mesons.%

\end{itemize}%

Therefore, new modes of aggregation of quark matter could be established by the experimental/theoretical investigation of these particles, further expanding the role of instantons in hadronic physics.

The idea of four-quark mesons dates back to the pioneering papers by Jaffe~\cite{jaffe}, while the discussion of exotic mesons and hadrons in terms of diquarks was introduced in Ref.~\cite{wilc} and then extended in Ref.~\cite{noim} to the scalar meson sector.

In the following, we will assume that the scalar mesons below $1$~GeV are indeed  bound states of  a spin~0 diquark and an anti-diquark (we will often call this a tetraquark).  A spin~0 diquark field can be written as:

\begin{equation}
\qq_{i\alpha}= \epsilon_{ijk}\epsilon_{\alpha\beta\gamma} \bar q^{j\beta}_C\gamma_5 q^{k\gamma},%
\label{defdq}
\end{equation}


\par\noindent
where Latin indices label flavor and Greek letters label color.  The color is saturated, as in a standard $q\bar q$ meson: $\qq^\alpha {\bar \qq}_\alpha$. Therefore, since a spin zero diquark is in a ${\bf \bar 3}$-flavor representation,  nonets of $\qq\bar \qq$ states are allowed (crypto-exotic states). The sub-GeV scalar mesons most likely represent the lowest tetraquark nonet.

The $\qq\bar \qq$ model of light-scalars is very effective at explaining the most striking feature of these particles, namely  their inverted pattern, with respect to that of ordinary $q\bar q$ mesons, in the  mass-versus-$I_3$ diagram~\cite{jaffe}, as shown in Fig.~\ref{fig:i3}.

\begin{figure}[!htb]
\includegraphics[width=40mm]{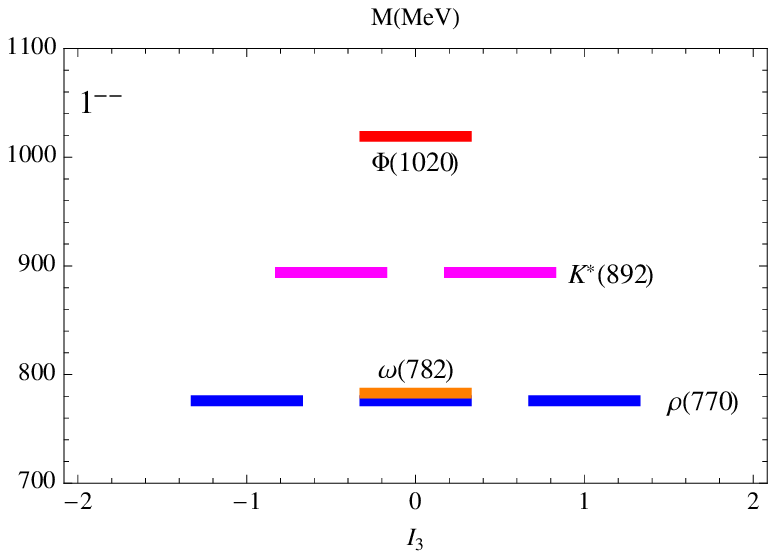}
\includegraphics[width=40mm]{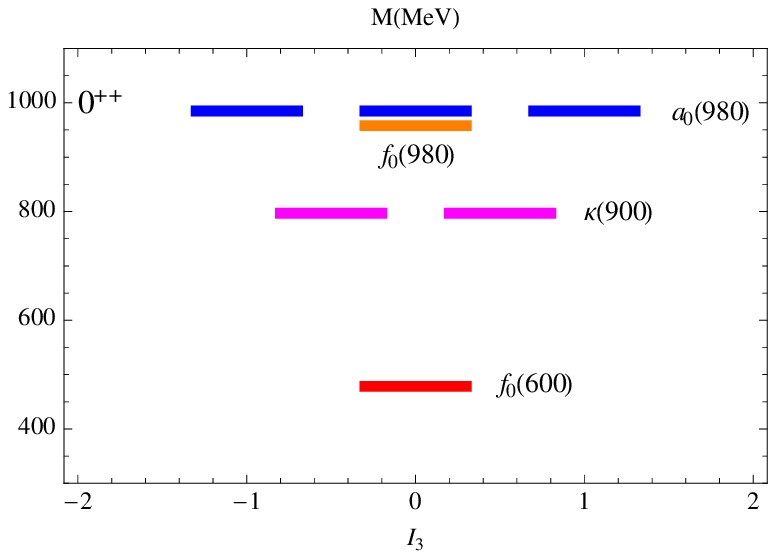}
\caption{\label{fig:i3}
Vector mesons ($q\bar q$ states) and the sub-GeV scalar mesons in the $I_3-m$ plane.}
\end{figure}

Such a pattern is not explained in a $q\bar q$ model,
in which, for example, the $f_0(980)$ would be an $s\bar s$ state~\cite{torni} while the
$I=1$, $a_0(980)$, would be a $u\bar u+d\bar d$ state. If this were the case, the degeneracy of the two particles appears rather unnatural.

Besides a correct description of the mass-$I_3$ pattern, the tetraquark model offers the possibility of explaining the decay rates of scalars at a level never reached by standard $q\bar q$ descriptions.
The effective decay Lagrangian into two pseudoscalar mesons, \eg, $\sigma \to \pi\pi$, is written as:
\begin{equation}
{\cal L}_{\rm exch.}=c_f S^i_j  \epsilon^{j t u}\epsilon_{i r s}  \partial_\mu \Pi^r_t  \partial^\mu \Pi^s_u,
\label{exchange}
\end{equation}
where $i,j$ are the flavor labels of $\qq^i$ and $\bar\qq^j$, while $r,s,t,u$ are the flavor labels of the quarks $\bar q^t,\bar q^u$ and  $ q^r, q^s$. $c_f$ is the effective coupling weighting this interaction term and ${S},{\Pi}$ are the scalar and pseudoscalar matrices. This Lagrangian describes the quark exchange amplitude for  the quarks to tunnel out of their diquark shells to form ordinary mesons~\cite{noim}. Such a mechanism is an alternative to the color string breaking  $\qq\ \gluon q\bar q\gluon\ \bar \qq\to B\bar B$, {\it i.e.}, a baryon-anti-baryon decay, which is phase-space forbidden to sub-GeV scalar mesons.

The main problem with eq.~(\ref{exchange}) is that it is not able to describe the decay $f_0\to \pi\pi$, since $f_0=( \qq^2{\bar\qq}^2+\qq^1\bar \qq^1)/\sqrt{2}$, being $1,2,3$ the $u,d,s$ flavors so that, see equation~(\ref{defdq}),  $\qq^1=[ds]$ and $\qq^2=[su]$. An annihilation diagram would be needed to replace the $s$ quarks, inducing a small rate that does not match the observation.

Alternatively, one can suppose  the mixing between the two isoscalars $f_0$ and $\sigma$ is at work, the $\sigma$ component ($\qq^3{\bar\qq}^3$) providing the $\pi\pi$ decay. However, as discussed in~\cite{winprog}, such mixing is expected to be too small, $<5^\circ$, to account for
the structure of the inverted mass pattern (a precise determination of the $\kappa$ mass would be crucial to fix this point).

A solution that improves the overall  agreement with data  of all light scalar mesons decay rates  has been found~\cite{winprog}. In low energy QCD, instantons  generate a quark interaction term that can be written as:
\begin{equation}
{\cal L}_I= \det (\bar q_L^i q_R^j),
\label{six}
\end{equation}
$i,j=1,2,3$ being flavor indices. Such a left-right mixing interaction is
screened at high energies, the instanton action scaling as $S\sim
\exp(-8\pi^2/g^2)$.  In addition to  the quark-exchange diagrams, described at
the effective theory level by the Lagrangian of eq.~(\ref{exchange}), (see
Fig.~\ref{fig:qdiags}~(a)), there are also contributions such as those in
Fig.~\ref{fig:qdiags}~(b)~\cite{footnote1}.

\begin{figure}[!htb]
\includegraphics[width=80mm]{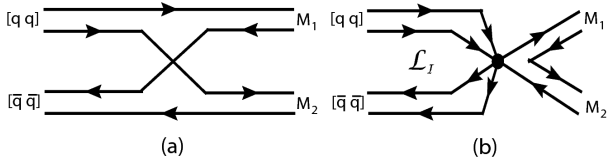}
\caption{\label{fig:qdiags}
Decay of a tetraquark scalar meson $S$ in two $q \bar{q}$ mesons $M_1 M_2$: (a) quark rearrangement (b) instanton-induced process.}
\end{figure}

The quark-level instanton interaction, Fig.~\ref{fig:qdiags}(b), reflects into an effective meson interaction of the kind:
\begin{equation}
{\cal L}_I=c_I {\rm Tr} ({\bf S \times (\partial \Pi)^2} ),
\label{einst}
\end{equation}
$c_I$ being an effective coupling as $c_f$ in~(\ref{exchange}). Assuming that the low energy dynamics of light scalar mesons is described by:
\begin{equation}
{\cal L}={\cal L}_{\rm exch.}+{\cal L}_{I},
\end{equation}
one can reach a remarkably satisfying description of light meson decays~\cite{winprog}. Namely:
\begin{itemize}
\item  Such a good description of decays is possible {\it only} if the assumption is made that sub-GeV light scalars are
diquark-antidiquark mesons (see Table~\ref{tab:res}). In the $q\bar  q$ hypothesis, the agreement of $a_0\to\pi^0\eta$ with data appears very poor.
\item The inverted mass spectrum of super-GeV scalar mesons can be explained by assuming that they form
the lightest $q\bar q$ scalar multiplet, deformed in the mass-$I_3$ pattern by mixing with the lowest exotic multiplet
of sub-GeV scalar mesons (see Fig.~\ref{fig:i3h}~\cite{winprog}).
\end{itemize}

\begin{figure}[!htb]
\includegraphics[width=50mm]{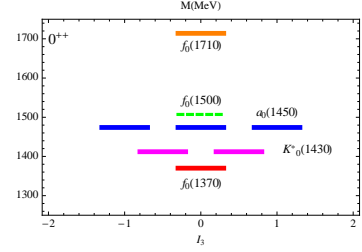}
\caption{\label{fig:i3h}
Super-GeV scalar mesons in the $I_3-m$ plane.}
\end{figure}
One of the isoscalars in the decuplet in Fig.~\ref{fig:i3h} is likely to be the lowest {\it glueball}; there are arguments favoring the $f_0(1500)$ as the most probable glueball candidate.

We quote a table from~\cite{winprog} describing at what level one can fit the decays of the lightest scalar mesons in
a diquark-antidiquark picture:

\begin{table}[!htb]
\begin{center}

\caption{\label{tab:res}
\footnotesize Numerical results, amplitudes in GeV. Second and third columns: results obtained with a decay  Lagrangian including or not including instanton effects, respectively (Labels $I$ and no-$I$ mean that we add or do not add the instanton contribution.).
No $f_0-\sigma$ mixing is assumed in this table. Fourth column: best fit, see text, with instanton effects included. Fifth column: predictions for a $q\bar q$ picture of the light scalars.
The $\eta-\eta^\prime$ singlet-octet mixing angle assumed: $\phi_{_{PS}}=-22^\circ$~\cite{ginop}.
Data for $\sigma$ and $\kappa$ decays are from~\cite{Caprini}, the reported amplitudes correspond to: $\Gamma_{\rm tot}(\sigma) = 272 \pm 6$, $\Gamma_{\rm tot}(\kappa) = 557 \pm 24$.
 }
\begin{tabular}{lccccc}
\hline\hline
Proc. &   \multicolumn{3}{c}{ ${\cal A}_{\rm th}([qq][\bar q \bar q])$} &${\cal A}_{\rm th}(q\bar q)$& ${\cal A}_{\rm expt}$  \\
    & $I$ &  no-$ I$ & best fit & $I$ & \\ \hline
$ \sigma   ( \pi^+\pi^-) $ &
{\rm input}  &   {\rm input}  & 1.7 &
{\rm input}  &$2.27 (0.03)$   \\ \hline
$\kappa^+   (K^0 \pi^+)  $ &   $5.0$ &
5.5 & 3.6 & 4.4&$5.2 (0.1) $   \\  \hline
$f_0  (\pi^+\pi^-) $    &    {\rm input} &{\bf 0}&1.6 &
{\rm input}  &$1.4 (0.6)$  \\
$f_0     (K^+ K^-) $    & $4.8$  & 4.5&3.8&  4.4 &$3.8(1.1)$  \\  \hline
$a_0   (\pi^0 \eta)$    &  $4.5$ & 5.4&3.0& 8.9&$2.8 (0.1)$    \\
$a_0  (K^+ K^-)$        &   $3.4 $  & 3.7&2.4 & 3.0 &$2.16(0.04)$  \\  \hline\hline
\end{tabular}  \\ [2pt]

\end{center}
\end{table}


A relative of the lowest lying scalar mesons may have been found very
 recently by \babar: the $Y(2175)$, a particle first observed in the decay
 $Y\to \phi f_0(980)$~\cite{y2175}.  This object could be a
 radial excitation of the lowest lying scalar mesons, of the kind
 $\qq^1{\bar \qq}^1+\qq^2{\bar\qq}^2$ and could strikingly manifest all
the three tetraquark decay mechanisms: the instanton ($Y\to
\phi(1020)f_0(980)$), the quark rearrangement ($Y\to KK^*$), and the
string breaking ($Y\to KK^*$) mechanisms. It is to be noted that only the
first decay mode has been observed; there are only hints of the other two.

We tend to exclude the possibility of a $Y(2175)$ built as $\qq^3{\bar \qq}^3$ because, though it would contain four $s$ quarks as the observed final state, it would involve spin ~1 diquarks, because of Fermi statistics. Spin~1 diquarks are thought to be energetically disfavoured, but, worse, they are in the ${\bf 6_f}$ representation, thus requiring a large number of exotic particles: ${\bf 6}\otimes {\bf \bar 6}={\bf 1}\oplus{\bf 8}\oplus{\bf 27}$. The search for other decay mechanisms would be quite crucial to test this hypothesis.

Searches of radially excited partners of the scalar mesons in the high statistics data samples from a SuperB factory, would deeply improve
the comprehension of the tetraquark picture. To give an example, consider that
predictions of lighter partners of the $Y(2175)$, to be found in ISR,
are at hand. Are the good, spin zero, diquarks the only relevant building blocks, or bad, spin one, diquarks are also
effective degrees of freedom to describe states at higher mass than the standard scalar nonets?
It is decisive to understand to what extent the actual models for multiquark particles are predictive.

\section{Charmonium}
\label{sec:data}

In the past few years the $B$~Factories have observed several states with clear $c\bar{c}$ content,
which do not behave like standard mesons, and that are therefore an indication of new spectroscopy.

The $X(3872)$ was the first state found that did not easily fit into charmonium spectroscopy.
It was initially observed decaying into $J/\psi\pi^+\pi^-$ with a mass just beyond the open charm
threshold~\cite{Choi:2003ue}. The $\pi^+\pi^-$ invariant mass distribution, the observation of the
$X\to J/\psi\gamma$ and the full angular analysis from CDF~\cite{Abulencia:2006ma} and Belle~\cite{Abe:2005iya}
favor the assignment of $J^{PC}=1^{++}$ for this state, and of $B\to J/\psi\rho$ as its dominant decay.
There are therefore several indications that this is not a charmonium state: the mass assignment does
not match any prediction of long-verified potential models (see
Fig.~\ref{fig:summary}); the
dominant decay would be isospin-violating; and the state is relatively narrow (less than a few MeV)
despite that fact that its mass is above threshold for the production of two charmed mesons.

\begin{figure}[!htb]
\epsfig{file=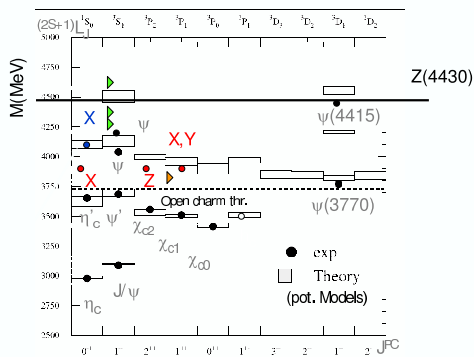, height=5cm}
 \caption{\label{fig:summary}
Measured masses of the newly observed states, positioned
in the spectroscopy according
to their most likely quantum numbers. The charged state ($Z(4430)$) clearly has
no $C$ quantum number.}
\end{figure}

Another aspect of interest of the $X(3872)$ are the measurements of its
mass, the most recent of which
is Ref.~\cite{babar:2007rv}: there
is an indication that there are two different particles, one decaying into $J/\psi\pi\pi$
and one into $D^{*0}D^0$, their masses differing
by about 4.5 standard deviations.
This observation makes the $X(3872)$ a good
tetraquark candidate:
di-quarks with an heavy meson are, in fact, flavor-triplets, and therefore
pairs give rise to the same nonet structure as conventional
mesons. There should therefore be two states with $S=I_3=0$
very close in mass~\cite{mppr}.
Without this evidence, the closeness to
the $D^0D^{*0}$ threshold suggests the hypothesis that this is a molecule composed of these two mesons.

Furthermore, the  $B$~Factories investigate
a large range of masses for particles with $J^{PC}=1^{--}$ by looking for events where the initial state
radiation brings the $e^+e^-$ center-of-mass energy down to the particle's mass. While in principle
only particles already observed in $R=\sigma_{had}/\sigma_{\mu\mu}$ scans could be produced, the high luminosity has
allowed the observation of several new particles: the  $Y(4260)\to J/\psi\pi^+\pi^-$ ~\cite{Aubert:2005rm}, the
$Y(4350)$~\cite{Aubert:2006ge} and the $Y(4660)$~\cite{belle:2007ea}, both
observed in their decay to $\psi(2S)\pi\pi$.

The invariant mass of the two pions in these decays is a critical observable in discerning the nature of these particles,
which are unlikely to be charmonium, since their masses are above the
open-charm threshold, yet they are relatively narrow.  Furthermore, their
decays to two
charmed mesons have not yet been observed, the most stringent limit being~\cite{Aubert:2007pa}
$\BR(Y(4260)\to D\bar{D})/\BR(Y(4260)\to J/\psi\pi^+\pi^-)<1.0 @$ 90\%
confidence level.

\begin{figure*}[!htb]
\begin{center}
\epsfig{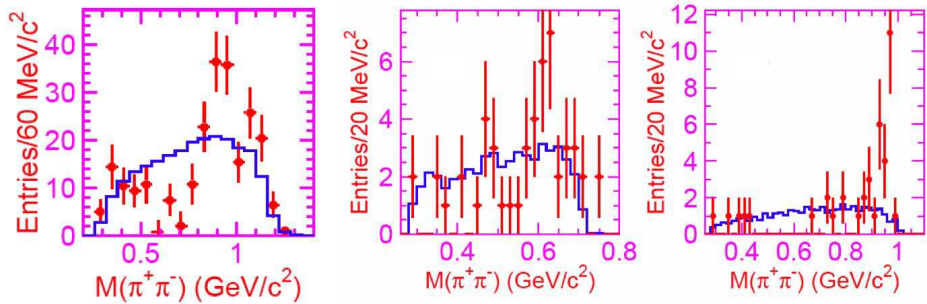}
 \caption{\label{fig:bellepipiinv}
 Di-pion invariant mass distribution in $Y(4260)\to\jpsi \pi^+\pi^-$ (left), $Y(4350)\to \psi(2S) \pi^+\pi^-$ (center),\newline
and $Y(4660)\to \psi(2S) \pi^+\pi^-$ (right) decays.
}

\end{center}
\end{figure*}

Figure~\ref{fig:bellepipiinv}
shows the di-pion invariant mass spectra for all
regions in which new resonances have been observed. There is some
indication that only the $Y(4660)$ has a well-defined intermediate
state (most likely an $f_0$), while others have a more complex
structure.

These observations make the $Y(4260)$ a good hybrid candidate, and the $Y(4350)$ and  $Y(4660)$ good candidates for
$[cd][\bar{c}\bar{d}]$ and $[cs][\bar{c}\bar{s}]$ tetraquarks, respectively. The latter would, in fact, prefer to decay into an
$f_0$, while the mass difference between the two states is consistent with the hypothesis that the two belong to the same nonet.

The turning point in the query for states beyond charmonium
was therefore the observation by the Belle Collaboration of a charged
state
decaying into $\psi(2S)\pi^\pm$~\cite{belle:2007wg}.
Figure~\ref{fig:belleZ4430}
shows the fit to the  $\psi(2S)\pi$ invariant mass distribution in
$B\to \psi(2S)\pi K$ decays, returning a mass $M=4433\pm4\rm{MeV/c}^2$
and a width $\Gamma=44^{+17}_{-13}$ MeV.

\begin{figure}[!htb]
\epsfig{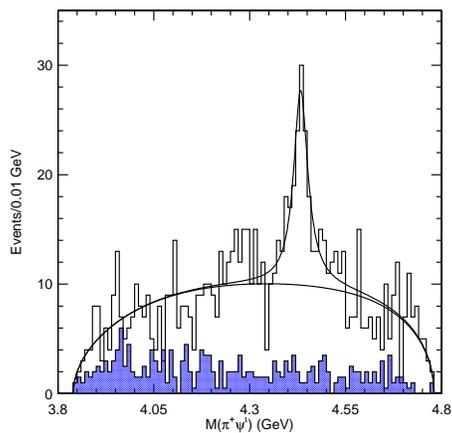}
 \caption{\label{fig:belleZ4430}
 The $\psi(2S)\pi$ invariant mass distribution in
\hbox{$B\to \psi(2S)\pi K$ decays.}}
\end{figure}

In terms of quarks, such a state must
contain a $c$ and a $\bar{c}$, but given its charge
it must also contain at least a $u$ and a $\bar{d}$. The only open
options are the tetraquark, the molecule or threshold effects. The
latter two options are viable due to the closeness of the $D_1D^*$
threshold.

Finding the corresponding neutral state, observing a decay mode of the
same state, or at least having a confirmation of its existence, are
critical before a complete picture can be drawn.

There are several reasons why a run at fifty to one hundred times the existing integrated luminosity is
critical to convert these of hints into a complete, solid picture:
\begin{itemize}
\item all the new states, apart from the $X(3872)$, have been observed in only
a single decay channel, with significance that are barely above 5$\sigma$.
a hundredfold increase in statistics would allow searches in several other
modes. It is in particular critical to observe both the decay to
charmonium and to $D$-meson pairs and/or $D_s$ meson pairs. Since the
branching fraction of observable final states for the $D$ and especially
for the $D_s$ mesons are particularly low, current experiments do not have
the sensitivity to observe all the decays.
\item the models predict several other states, such as the neutral
partners of the $Z(4430)$ and the nonet partners, for instance
 $[cd][\bar{c}\bar{s}]$ candidates decaying into a charmonium state and a kaon,
 at a significantly lower rate
(see \eg, Ref.~\cite{Maiani:2007vr}) than the observed modes. Furthermore ,
several of these states decay into particles (in particular neutral pions
and kaons) that have a low detection efficiency.
 \end{itemize}

\section{Bottomonium}

Exotic states with two bottom quarks, analogous to those with two charm quarks, could also exist.
In this respect, bottomonium spectroscopy is a
very good testbench for
speculations advanced to explain the charmonium states.
On the other side, searching for new
bottomonium states
is more challenging, since they tend to be broader and there are
more possible decay channels. This explains why there are still eight unobserved
states with masses
below open bottomonium threshold.

Among the known states, there is already one with unusual behavior: there
has been a recent
observation~\cite{Abe:2007tk} of an anomalous enhancement, by two orders
of magnitude, of
the rate of $\FiveS$ decays to the $\OneS$ or a $\TwoS$ and two pions. This
indicates that either the $\FiveS$ itself or a state very close by in
mass has a
decay mechanism that enhances the amplitudes for these processes.

In order to understand whether the exotic state coincides with the
$\FiveS)$ or not, a high luminosity
(at least 20 fb$^{-1}$ per point to have a 10\% error) scan of the
resonance region is needed.

In any case, the presence of two
decay channels to other bottomonium states excludes the possibility of
this state being a molecular aggregate, but all other models are possible, and would predict
a large variety of not yet observed states.

As an example, one can estimate possible resonant states with the tetraquark
model, by assuming that  the masses of states with two $b$ quarks can be obtained
from one with two $c$ quarks by adding the mass difference between
the $\OneS)$ and the $\jpsi$. Under this assumption, which
works approximately for the known bottomonium states, we could expect three nonets
that could be produced by the $\ThreeS$ and decaying into $\OneS$ and
pions. Assuming that the production and decay rates of these new states
are comparable to the charmonium states, and assuming
a data sample of $\ThreeS$ events comparable in size to the current $\FourS$
sample is needed to clarify the picture, we would need about $10^9\ \ThreeS$
mesons, corresponding to an integrated luminosity of \hbox{0.3 ab$^{-1}$.}

As already mentioned, searching for bottomonium-like states would require
higher statistics than the corresponding charmonium ones; this therefore
represents an even stronger case for \superb.

\section{Search for Physics Beyond the Standard Model in Bottomonium Decays}

In spite of intensive searches performed at LEP \cite{Schael:2006cr},
the possibility of
a rather light non-standard Higgs boson has not been ruled out in
several scenarios beyond the Standard Model
\cite{Dermisek:2005gg,Dermisek:2006py,SanchisLozano:2007wv},
due to the fact that a new scalar may be uncharged under the
gauge symmetries, similar to a sterile neutrino in the fermion
case.  These studies indicate that its mass could be less than twice
the $b$ mass, placing it within the reach of \superb$\!\!$.
Moreover, the
LHC might not be able to unravel a signal from a
light Higgs boson whose mass is
below $B\bar{B}$ threshold, since it will be difficult for the soft decay products
to pass the LHC triggers.  Dark matter may also be
light, evading LEP searches if it does not couple strongly to the
$Z^0$~\cite{Gunion:2005rw,McElrath:2005bp,Fayet:2007ua,Bird:2006jd}.
\superb\ will be {\it required} in most of these cases to precisely determine
its masses and couplings, and will play an important discovery role.

\subsection*{Light Higgses}

A Higgs $h$ with $M_h < M_\Upsilon$ can be
produced in $\Upsilon(nS)$ decays via the Wilczek mechanism
with a branching ratio
approximately given by the leading-order formula \cite{Wilczek:1977zn}
\[ \frac{\Gamma(\Upsilon(nS) \to \gamma h)}{\Gamma(\Upsilon(nS) \to \mu
\mu)} =
    \frac{\sqrt{2} G_F m_b^2}{\alpha \pi M_{\NS}} E_\gamma X_d^2
\]
where $X_d$ is a model-dependent quantity containing the coupling of the
Higgs
to bottom quarks, $m_b$ is the bottom quark mass, $\alpha$ and $G_F$ are
the
electroweak parameters, and $E_\gamma =
(M_{\NS}/2)(1-M_h^2/M_{\NS)}^2)$
is the photon energy.

From a theoretical viewpoint, the existence of a light
pseudoscalar Higgs is not unexpected in many
extensions of the SM. As an especially appealing example, the
Next-to-Minimal Supersymmetric Standard
Model (NMSSM) has a gauge singlet added to the
MSSM two-doublet Higgs sector (see \cite{Fullana:2007uq} and references therein for a short
summary of other scenarios leading to a light Higgs
boson) leading
to seven physical Higgs bosons, five of them neutral, including
two pseudoscalars.

In the limit of either
slightly broken $R$ or Peccei-Quinn (PQ) symmetries,
the lightest \CP-odd Higgs boson (denoted by $A_1$)
can be much lighter than the other Higgs bosons.
Interestingly, the authors of \cite{Dermisek:2005gg}
interpret the excess of $Z^0$+$b$-jet events found at LEP
as a signal, in this formalism, of a
Standard Model-like Higgs decaying partly into $b \bar{b}$, but
dominantly into $\tau$'s via two light pseudoscalars.

Let us write the physical Higgs boson $A_1$ as a mixture of
singlet ($A_{s}$)
and non-singlet ($A_{MSSM}$) fractions parametrized by the
angle $\theta_A$, according to
\[ A_1 = \cos{\theta_A} A_{MSSM}+\sin{\theta_A} A_s \]
The $A_1$ coupling to down-type fermions turns out to be proportional to
$X_d = \cos{\theta_A}$ $\tan{\beta}$, where $\tan{\beta}$ denotes
the ratio of the vevs of the up- and down-type Higgs bosons. For
$\cos{\theta_A}$ close to zero, the $A_1$ almost completely
decouples from flavor physics.
However, if $\cos{\theta_A} \sim 0.1-0.5$, present LEP and $B$ physics
bounds can be simultaneously satisfied~\cite{Domingo:2007dx}, while
a light Higgs could still show up
in $\Upsilon$ radiative decays into tauonic pairs:
\[ \NS \to \gamma A_1(\to \tau^+\tau^-)\ ;\ \ \ n=1,2,3. \]

\begin{figure}[!htb]
\begin{center}
\includegraphics[width=15pc,angle=270]{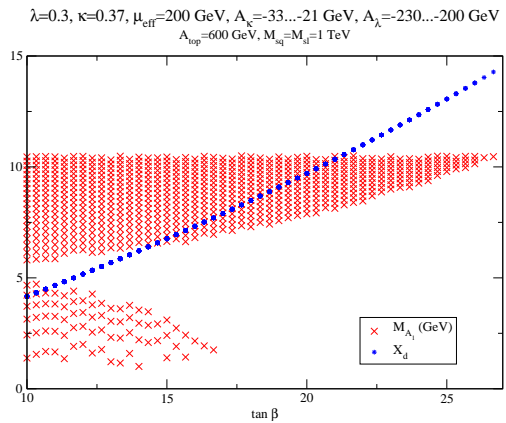}
\caption{\label{fig:ups_gamma_a_univers}
Plot of $X_d=\cos{\theta_A}$ $\tan{\beta}$ (blue points)
and $A_1$ mass in GeV (red crosses) versus $\tan{\beta}$.
All points were generated using the NMHDECAY code \protect{\cite{Ellwanger:2005dv}}
satisfying both LEP and$B$ physics constraints
using a particular set of NMSSM parameters \protect{\cite{SanchisLozano:2008}.}}
\end{center}
\end{figure}

As this light Higgs acquires its couplings to Standard Model fermions via mixing
with the Standard Model Higgs, it therefore couples to mass, and will decay to the
heaviest available Standard Model fermion.  In the region $M_{A_1} > 2 M_\tau$, there
are
two measurements which have sensitivity: lepton universality of $\Upsilon$
decays, and searches for a monochromatic photon peak in tauonic $\Upsilon$
decays. \\

{\bf{The measurement of lepton universality}}
 compares the branching ratios of $\Upsilon$ to $e^+e^-$,
$\mu^+\mu^-$ and $\tau^+\tau^-$
\cite{SanchisLozano:2003ha,SanchisLozano:2006gx},
which should all be identical up to kinematic
factors in the Standard Model, due to the gauge symmetry.
It is relevant especially when the $A_1$ mass is
within about 500 MeV of an $\Upsilon$ mass, so that the monochromatic photon
signal is buried under backgrounds.  It is also the best measurement
when $M_{A_1} > M_\Upsilon$, which causes there to be a photon spectrum,
rather than monochromatic line.

Using the NMHDECAY code \cite{Ellwanger:2005dv}, we have
randomly generated masses and couplings for the $A_1$ Higgs
below the $B\bar{B}$ threshold, under the condition of passing
all current LEP and $B$ physics bounds built into the
NMHDECAY \cite{Domingo:2007dx}.
We actually chose a physically-motivated
set of NMSSM parameters favoring the existence of a scenario
with of a light $A_1$ \cite{SanchisLozano:2007wv,SanchisLozano:2008}.

In Fig.~\ref{fig:ups_gamma_a_univers} we plot the resulting points
of our scan
for the $A_1$ mass and $X_d$ values as a function of $\tan{\beta}$.
Let us stress that, in view of the available large $X_d$ values,
such a light \CP-odd Higgs could provide a signal in
$\Upsilon$ leptonic decays, whose first hint
would be an apparent breaking of lepton universality, {\it e.g.}
at the few percent level. Indeed, the tauonic mode would be (slightly)
enhanced by the New Physics channel with respect
to the electronic and muonic modes, because of the large leptonic
mass difference
\cite{SanchisLozano:2003ha,SanchisLozano:2006gx, Fullana:2007uq}.
The degree of enhancement of the tauonic channel (\ie, of the
New Physics contribution) obviously depends on the
assumed set of the NMSSM parameters (notably $\tan{\beta}$) but seems
sizeable for reasonable values of them, as can be seen from
Fig.~\ref{fig:ups_gamma_a_univers}.

Moreover, the observation (non-observation) of a monochromatic photon
from the radiative process
would become the smoking gun pointing out (excluding)
the existence of such a light
non-standard Higgs boson.

\begin{figure}[!htb]
\begin{center}
\includegraphics[width=16pc]{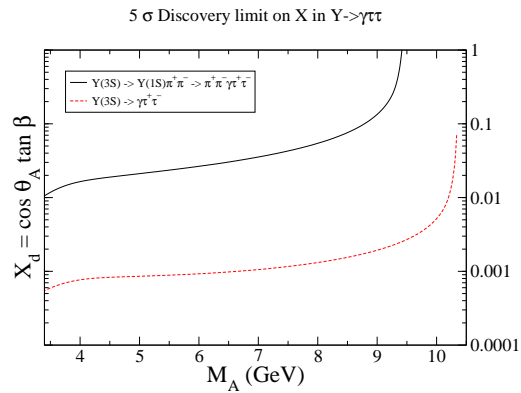}
\caption{\label{fig:ups_gamma_a_mono}
Plot of the $5 \sigma$ discovery potential of \superb\
with $\Upsilon(3S)$ data, in the mode $\Upsilon(3S) \to \pi^+ \pi^-
\Upsilon(1S) \to \pi^+ \pi^- \tau^+ \tau^- \gamma$ (solid black) and
$\Upsilon(3S) \to \tau^+ \tau^- \gamma$ (dashed red). An integrated
luminosity of 1 ab$^{-1}$ was assumed.}
\end{center}
\end{figure}

{\bf{In the search for monochromatic photons} }
the first relevant decay mode is $\ThreeS \to \OneS \pi^+
\pi^-$
first, followed by $\OneS \to \gamma \tau^+ \tau^-$, which has only a
4.5\%
branching fraction, but has low background.
The second decay mode is $\ThreeS \to \gamma
\tau^+ \tau^-$, which suffers from much worse backgrounds
from $e^+ e^- \to \tau^+\tau^- \gamma$ events, but also has a rate that is
more than a factor of ten higher.
The corresponding exclusion plots are in Fig.~\ref{fig:ups_gamma_a_mono}.

\subsection*{Invisible decays and light dark matter}

Finally, if Dark Matter is lighter than 5 GeV, it will require a Super $B$
Factory to determine its properties.  Generally, in this mass region one
needs
two particles, the dark matter particle $\chi$, and a boson that couples
it to
the Standard Model $U$.  The most promising searches are in
invisible and radiative decays of the $\Upsilon$, which can be measured in
the mode $\ThreeS \to \pi^+ \pi^- \OneS \to \pi^+ \pi^- +
invisible$, which is sensitive to a vector $U$.  However, to substantially
improve on existing measurements from Belle and CLEO, far-forward tagging
must
be incorporated into the design of the detector.  This is needed to veto
events
in which the $\OneS$ decays to a two-body state, with decay
products that disappear down the beampipe~\cite{McElrath:2005bp}.

The second most promising signature is radiative decays $\Upsilon \to
\gamma + invisible$.  This is probably the most
favored mode theoretically,
and is
sensitive to a scalar or pseudoscalar $U$.  The  mediator
coupling the Standard Model particles to final-state $\chi$'s
can be a pseudoscalar Higgs, $U=A_1$, which can be naturally
light, and would appear in this mode~\cite{Gunion:2005rw}.
In such models the Dark Matter can be naturally be a bino-like neutralino.

\section{Summary}

\superb\ will open a unique window on this physics because it
allows a high statistics study of the current hints of new aggregations
of quarks and gluons. Besides the physics one can study in running at the
$\FourS$ resonance, the following alternative energies are of interest:
$\ThreeS$ (at least 0.3 ab$^{-1}$) and a high luminosity scan between 4-5
GeV (5 MeV steps of 0.2 fb$^{-1}$ each would require a total of
40 fb$^{-1}$)~\cite{footnote2}.
While this is not huge statistics, this scan is only feasible with
\superb. The only possible competitor, BES-III, is not planning to scan
above 4 GeV, since their data sample would, in any case, be lower than that
 of the $B$ Factories alone.

Finally, the search for exotic particles among the decay products of the
bottomonia can probe regions of the parameters space of non-minimal
supersymmetric models that cannot be otherwise explored directly, for
instance at LHC. These studies are particularly efficient when producing
$\NS$ mesons with $n<4$.

The superiority of \superb\  with respect to the planned upgrade of Belle
lies both in the ten times higher statistics, which broadens the range of
cross sections the experiment is sensitive to, but also in the flexibility
to change center of mass energy.

\onecolumngrid\twocolumngrid\hbox{}

\setcounter{section}{0}
\rhead[\fancyplain{}{\bf Appendix: Tools}]%
      {\fancyplain{}{\bf\thepage}}
\clearpage
{\centerline{\rule[0in]{0.9\columnwidth}{2pt}}
\vspace {-0.7cm}
\part*{\centerline{Appendix:}}
\vspace {-1cm}
\part*{\centerline{Physics Tools}}
\vskip -8pt
{\centerline{\rule[ 0.15in]{0.9\columnwidth}{2pt}}


\bigskip
\smallskip

We describe herein the tools used to simulate physics events and evaluate detector performance at the
\superb flavor factory. The simulation should meet two main
requirements. First, since the design of the subsystems is evolving, the user
should be able to perform optimization studies and modify the detector description in a simple way. Second, the program should be very fast, to simulate
very large numbers of physics events.
Table~\ref{tab:exp_rates} shows the event rate expected at a luminosity of
$1.0\times 10^{36}$~cm$^{-2}s^{-1}$. Over one year it translates to $1.1\times
10^{10}$ $\Upsilon(4S)$ decays and a total of about $5.4\times 10^{10}$
$e^+e^-\rightarrow q\bar{q}$ ($q=u,d,s,c,b$) and $\tau^+\tau^-$ decays.\\

\begin{table}[!htb]
\begin{center}
\caption{Physics rates at $1.0\times 10^{36}$~cm$^{-2}s^{-1}$.}
\begin{tabular}{lc}
\hline
\hline
Process & Rate at $\mathcal{L}=1\times 10^{36}$~cm$^{-2}s^{-1}$\\
 & (kHz)\\
\hline
$\Upsilon(4S)\rightarrow B\bar{B}$ & 1.1\\
$udsc$ continuum & 3.4 \\
$\tau^+\tau^-$ & 0.94 \\
$\mu^+\mu^-$ & 1.16 \\
$e^+e^-$ for $|\cos\theta_{\textrm{Lab}}|<0.95$& 30 \\
\hline
\end{tabular}
\label{tab:exp_rates}
\end{center}
\end{table}

At this stage, a single tool cannot fulfill completely both
requirements. Therefore the development of the simulation tools moves along parallel paths. A very fast
and relatively simple simulation program has been already developed and is
operational. It can simulate large amounts of both hadronic and $\tau^+\tau^-$
events while allowing to some extent the modification of the detector
configuration.
An upgrade schedule has been defined to increase the
accuracy of the simulation without sacrificing the speed.
More details are provided in the next section.

In parallel, a project is planned where the detailed description of both the detector
and the interaction region are done within the Geant4~\cite{geant4} framework.

Finally, the \babar\ simulation and reconstruction packages are being used to
perform \superb\ subdetector optimization studies. Although some aspects of the
\babar\ simulation make its evolution towards \superb\ not attractive, there are
good reasons why the possibility of exploring it for \superb\ can continue to be
particularly important. Detailed performance evaluations for \superb\ can in
fact be carried out by introducing minor modifications to the \babar\
detector. This will represent for a while the main option available to extract
the parameters needed as input by the \superb\ fast simulation. Negotiations
with \babar\ management are currently underway to extend access to
non-\babar\ members.

\section*{The parametric fast simulation}
\label{sec:app_fastsimu}
The simplest fast simulation program we have, named PravdaMC~\cite{pravda}, is
a very fast Monte Carlo which uses parametrization to simulate the detector
response. The radius, thickness and material of the beam pipe is configurable.
The tracking system can be modified by changing the number of active
layers of the silicon detector, the intrinsic spatial resolutions and the
amount of interaction length, as well as the number and dimension of the drift
chamber cells and their spatial resolutions. The current tracking algorithm is
TRACKERR~\cite{trackerr} which starts from the truth Monte Carlo charged
particle to produce the track and evaluate the error matrix of its parameters
taking into account the energy loss and the multiple scattering. The main
limitation is that the trajectory is not modified by the energy loss and
therefore it is a perfect helix. This approximation is poor for very low
momentum tracks, like soft pions from $D^{*\pm}$.

The response of the electromagnetic calorimeter is analytic.
In the current version of the program, the response of the DIRC and IFR to the
passage of a charged particle is implemented as an efficiency map of a
particle identification algorithm provided externally.

PravdaMC uses the same generators-framework interface as used by the \babar\
simulation code. In particular it can generate both hadronic
$e^+e^-\rightarrow q\bar{q}$ events (including obviously
$e^+e^-\rightarrow\Upsilon(4S)$) and $e^+e^-\rightarrow\tau^+\tau^-$ events.
In the latter case it is possible to generate events where the
$e^-$ or $e^+$ beams are polarized, which is a unique and important
aspect of the $\tau$ physics program at the \superb\ flavor factory.


Activity is ongoing to develop an improved fast simulation. It uses PravdaMC
as a basis but eventually it will become a completely different program.
First, TRACKERR is replaced by a more accurate track fitting
algorithm based on the \babar\ track reconstruction and taking into account
all the effects of the interaction between particles and materials.
Second, the response of the DIRC, EMC and IFR is simulated through the
parametrization of the physics quantities measured by each subsystem and
used to perform the analysis of the physics events. Several sources can be used
to tune the parametrization of the detectors output: the real data collected
by the \babar\ detector, the Geant4 simulation of the \babar\ detector and the
standalone detailed simulation of the \superb\ subsystems.

\subsection*{Readout and analysis of simulated data}
The analysis of simulated events requires several specific tools.
Composition and
vertexing algorithms for the reconstruction of the signal decay trees,
the algorithms to determine the flavor and vertex position of the recoil $B$,
and an extensive set of utilities for signal/background separation are
inherited from the \babar\ experiment and therefore are mature and fully
functional. The output of the simulation with the information of the simulated
tracks and neutral clusters together with the reconstructed composite
particles are stored in ROOT files~\cite{root}.
Effort is ongoing to make the existing tools independent of the \babar\
framework.  

\section*{Simulation with Geant4}\label{sec:app_geant4}

A medium-term plan for the development of a detailed simulation of the
\superb\ detector has been defined.
The simulation of the machine-induced backgrounds is at present accomplished
with a Geant4 application that incorporates a preliminary description of
the \superb\  detector volumes.
This initial effort of describing the \superb\  detector in Geant4
can represent the basis for the future development of a
detailed detector simulation.
At present some work is needed to improve the usability and
maintainability
of the tool for background studies.  The most important improvement consists
in decoupling
the geometry description from the code. The "technology" is available, since
using a markup
language to allow definition of geometry  data in XML format is now
implemented in Geant4
through GDML files.
Input from the subdetectors is needed to refine the current initial models.
When the detailed simulation of the \superb\ detector will be available, it
will be used to tune the output of the fast simulation including the effects
of the machine backgrounds.

\section*{Simulation of tau pair production with polarized beams}

The \superb project includes the ability to operate with an 85\%
longitudinally polarized electron beam, which is especially relevant
for tau physics studies. For this document, tau pairs produced with
polarized beams have been simulated with the KK
generator~\cite{Jadach:1999vf} and Tauola~\cite{Jadach:1999vf}.
That simulation framework includes all QED effects up to the second
order. Tau decays are simulated taking into account spin polarization
effects as well., and the complete spin correlations density matrix of
the initial-state beams and final state is incorporated in an exact
manner.

\onecolumngrid\twocolumngrid

\end{document}